\documentclass[11pt,letterpaper]{article}

\newif\ifanonymous
\anonymousfalse

\usepackage[margin=1in]{geometry}
\IfFileExists{lmodern.sty}{\usepackage[T1]{fontenc}\usepackage{lmodern}}{}
\usepackage[utf8]{inputenc}
\usepackage{microtype}
\usepackage[dvipsnames]{xcolor}

\usepackage{amsmath,amssymb,amsthm,mathtools,thmtools,thm-restate}

\makeatletter
\@ifundefined{newcounteralias}{}{%
    \renewcommand\thmt@autorefsetup{%
        \@xa\def\csname\thmt@envname autorefname\@xa\endcsname\@xa{\thmt@thmname}%
    }%
}
\makeatother

\declaretheorem[numberwithin=section]{theorem}
\declaretheorem[sibling=theorem]{lemma,proposition,corollary,claim,observation}
\declaretheorem[sibling=theorem,style=definition]{definition,example,construction}

\newcommand{\TEQ}{\operatorname{TEQ}}

\newcommand{\Out}{\operatorname{Out}}

\newcommand{\PSPACE}{\mathsf{PSPACE}}
\newcommand{\NP}{\mathsf{NP}}

\newcommand{\mD}{\mathcal{D}}   
\newcommand{\mC}{\mathcal{C}}   

\newcommand{\R}{\mathrel{R}}
\newcommand{\RX}[1]{\mathrel{R_{#1}}}

\newcommand{\iverson}[1]{[#1]}

\newcommand{\dout}{d}      
\newcommand{\eout}{o}      
\newcommand{\cout}{c}      
\newcommand{\orole}{o}     
\newcommand{\Gad}{G}       
\newcommand{\Filt}{F}      
\newcommand{\test}{\tau}      
\newcommand{\relay}{\rho}     
\newcommand{\Rd}[1]{B^{#1}}
\newcommand{\Rdi}[2]{B_{#1}^{#2}}
\newcommand{\rone}{B^{1}}      
\newcommand{\rzero}{B^{0}}     
\newcommand{\broot}{r}     
\newcommand{\tri}{t}       
\newcommand{\shw}{w}       
\newcommand{\shz}{z}       
\newcommand{\groot}{B^{1}}     
\newcommand{\slink}{s}     
\newcommand{\king}{g}      
\newcommand{\wanc}{w}      
\newcommand{\anc}{a}       
\newcommand{\ax}[1]{\kappa_{#1}}        
\newcommand{\axl}[2]{\kappa_{#1,#2}}    
\newcommand{\oc}[1]{\lambda_{#1}}       
\newcommand{\ocl}[2]{\lambda_{#1,#2}}   
\newcommand{\Cle}[1]{C_{\le #1}}      
\newcommand{\Hs}[1]{H_{#1}}       
\newcommand{\bet}[1]{\beta_{#1}}  

\newcommand{\Wex}{W_{\exists}}
\newcommand{\Wall}{W_{\forall}}

\newcommand{\prob}[1]{\textup{\textsc{#1}}}

\usepackage{booktabs}
\usepackage{array}
\usepackage{enumitem}
\setlist{itemsep=2pt,topsep=4pt}
\usepackage{tikz}
\usetikzlibrary{arrows.meta,positioning,fit,calc,shapes.geometric,backgrounds,decorations.pathreplacing}
\usepackage{float}
\usepackage{caption}
\newfloat{algorithm}{htbp}{loa}
\floatname{algorithm}{Algorithm}

\usepackage{url}
\usepackage[colorlinks=true,linkcolor=blue!50!black,citecolor=green!35!black,urlcolor=blue!50!black]{hyperref}
\usepackage[nameinlink,noabbrev,capitalise]{cleveref}
\Crefname{appendix}{Appendix}{Appendices}
\Crefname{subappendix}{Appendix}{Appendices}
\Crefname{subsubappendix}{Appendix}{Appendices}
\crefname{claim}{Claim}{Claims}
\crefname{observation}{Observation}{Observations}
\Crefname{observation}{Observation}{Observations}
\crefname{construction}{Construction}{Constructions}
\Crefname{construction}{Construction}{Constructions}

\hypersetup{hypertexnames=false}

\title{The Tournament Equilibrium Set Is PSPACE-Complete,\\
       Even under Hereditary Uniqueness}

\ifanonymous
  \author{Anonymous submission}
\else
  \author{Ethan Dickey \qquad Athina Terzoglou\\[2pt]
          {\normalsize Department of Computer Science, Purdue University}}
\fi
\date{}

\hypersetup{pdftitle={The Tournament Equilibrium Set Is PSPACE-Complete Even under Hereditary Uniqueness}}
\ifanonymous\else\hypersetup{pdfauthor={Ethan Dickey, Athina Terzoglou}}\fi

\begin{document}

\begin{titlepage}
\maketitle
\thispagestyle{empty}
\begin{abstract}

\noindent
The tournament equilibrium set (TEQ), proposed by Schwartz in 1990, is a classical voting rule based on pairwise majority comparisons. Its definition is recursive: an alternative is judged by the alternatives that beat it, called its dominators, and by which of those dominators TEQ itself selects. Deciding whether a given alternative belongs to TEQ was known to be NP-hard and to lie in PSPACE. We close this gap by proving that the problem is PSPACE-complete. To our knowledge, this is the first PSPACE-completeness result for membership in a tournament solution.

The hardness persists even when every nonempty subtournament has a unique minimal retentive set, the property conjectured by Schwartz to hold for all tournaments. Moreover, every such subtournament in our hard instances contains an alternative with at most three dominators. We prove that a single alternative with at most four dominators suffices both to guarantee uniqueness and to find a TEQ winner in $O(N^2)$ time. Thus, on our hard instances, finding one winner is efficient even though recognizing a specified winner is PSPACE-complete.

We also study how the complexity changes with the dominator bound. If every nonempty subtournament has an alternative with at most one dominator, the entire TEQ can be computed in quadratic time; allowing two dominators makes membership NP-hard. Under the same condition with four dominators, all $t$ winners can be computed in $O(2^t N^2)$ time, where $N$ is the number of alternatives.

\end{abstract}

\end{titlepage}

\thispagestyle{empty}
\begingroup
  \linespread{0.95}\small
  \tableofcontents
\endgroup
\clearpage
\setcounter{page}{1}

\section{Introduction}\label{sec:intro}

A \emph{tournament} is a complete directed graph: a set of alternatives in which every pair has been compared and one member of each pair has won. Tournaments arise wherever a choice must be made from pairwise comparisons. In an election, alternative $a$ beats $b$ if a majority of the voters prefers $a$ to $b$. In a round-robin competition, the comparisons are the results of the matches. Increasingly, they arise in machine learning, where systems are trained and evaluated from relative judgments rather than absolute scores: a user clicks on one of two search rankings~\cite{YBKJ2012,Yan2022}, or an annotator prefers one of two responses of a language model~\cite{Munos2024}, and the aggregate of many such judgments is, again, a tournament. Tournaments are also a familiar object in algorithms, through the problems of ranking a tournament with few backward edges~\cite{Alon2006,CTY2007,KMS2007} and of fixing a knockout tournament~\cite{VW2010}.

The most basic question about a tournament is which alternative(s) should win. When one alternative beats every other, the answer is clear: that alternative, the \emph{Condorcet winner}. But pairwise comparisons need not be transitive. A majority may prefer $a$ to $b$, $b$ to $c$, and $c$ to $a$, and comparison data collected in practice often has no Condorcet winner at all~\cite{ZKWR2015,Ramamohan2016}. A \emph{tournament solution}~\cite{Fishburn1977,Laslier1997} answers the question in general. It is a rule that selects from every tournament a nonempty set of winners, selects the Condorcet winner whenever there is one, and does not depend on the names of the alternatives. The classical tournament solutions each make precise a different idea of what it means to win in the presence of cycles: the \emph{Copeland set} collects the alternatives with the most wins, the \emph{uncovered set} those that beat every other alternative directly or in two steps, the \emph{Banks set} the winners of maximal transitive subtournaments, and the \emph{bipartisan set} the support of the optimal mixed strategy of the tournament viewed as a zero-sum game~\cite{LLL1993b}. The same concepts have been rediscovered in machine learning. The dueling bandits literature, which studies learning from pairwise comparisons, first assumed a Condorcet winner~\cite{YBKJ2012}, then moved to the Copeland winner~\cite{ZKWR2015}, to the \emph{von Neumann winner}~\cite{DHSSZ2015} (a lottery over alternatives, related to the bipartisan set), and to general tournament solutions~\cite{Ramamohan2016}; recent work on learning from human preferences targets the same equilibrium~\cite{Munos2024}.

\paragraph{The tournament equilibrium set.}
This paper concerns the tournament equilibrium set of Schwartz~\cite{Schwartz1990}, or TEQ, which is among the most selective of the classical tournament solutions and has one of the most intricate definitions. Its guiding idea is that not every defeat should count the same. An alternative that loses only to opponents who are themselves weak has a better claim than one that loses to a strong opponent.

To make this precise, call the alternatives that beat $y$ the \emph{dominators} of $y$, and evaluate them by the same rule, applied to the subtournament they induce: the \emph{serious challengers} of $y$ are the alternatives that TEQ selects among the dominators of $y$. A set of alternatives is \emph{retentive} if it contains the serious challengers of each of its members. A retentive set is \emph{minimal} if it contains no smaller nonempty retentive set. Distinct minimal retentive sets are disjoint, and TEQ selects all alternatives in their union. The definition is recursive, since the serious challengers of $y$ are computed by TEQ itself on a smaller tournament, and it is well founded, since the dominators of $y$ exclude $y$. Intuitively, TEQ computes the serious challengers of each alternative, finds all nonempty sets that contain the serious challengers of each of their members, discards any such retentive set that contains a smaller one, and selects every alternative in the remaining sets.

TEQ occupies a distinguished position among tournament solutions. Starting with any tournament rule, we can identify the alternatives it selects among each alternative's dominators, then select the union of the minimal sets that are retentive with respect to these choices. This gives a new tournament rule, to which we can apply the same operation again. Remarkably, no matter which rule we start with, at most $\lfloor N/2 \rfloor$ iterations produce TEQ on every tournament with $N$ alternatives~\cite{BBFH2014}.
Schwartz conjectured that every tournament has a unique minimal retentive set, and under that conjecture TEQ satisfies monotonicity, stability, and a number of other standard axioms~\cite{LLL1993,BBH2016}. The conjecture was eventually refuted, first nonconstructively~\cite{BCKLNSST2013} and then by an explicit tournament on $24$ vertices~\cite{BS2013}, while it is known to hold for all tournaments with at most $14$ vertices~\cite{Yang2016}. Nevertheless, uniqueness remains important for understanding TEQ on restricted classes of tournaments, even though it does not hold in general.

\paragraph{The complexity question.}
The computational complexity of the classical tournament solutions is by now well understood~\cite{Hudry2009,BBH2016}. The Copeland set and the uncovered set can be computed in polynomial time, as can the minimal covering set~\cite{BF2008}. Deciding membership in the Banks set is NP-complete~\cite{Woeginger2003}, although some Banks winner can be found in polynomial time~\cite{Hudry2004}, while deciding membership in the Slater set is $\Theta_2^p$-complete~\cite{Lampis2022}. For TEQ, Brandt, Fischer, Harrenstein, and Mair~\cite{BFHM2008} proved in 2008 that deciding whether a given alternative is selected is NP-hard. The recursive definition gives a polynomial-space algorithm, but the gap between these bounds remained open. We close this gap.

\begin{restatable}[Main theorem]{theorem}{thmmain}\label{thm:main}
    Deciding whether a designated vertex of a tournament belongs to TEQ is PSPACE-complete under polynomial-time many-one reductions.
\end{restatable}

Thus, the recursion in the definition of TEQ is as expressive as the alternation of quantifiers in a quantified Boolean formula, and TEQ membership has no polynomial-size certificates verifiable in polynomial time unless $\NP=\PSPACE$. Computing the entire TEQ set is thus also PSPACE-hard. To our knowledge, this is the first PSPACE-completeness result for membership in a tournament solution, that is, a rule that uses only the outcome of each pairwise comparison, not the margin of victory. The only comparable result concerns Stable Voting~\cite{SV2026}, whose recursion is driven by the margins of pairwise victories; its lower bound does not transfer directly to TEQ, which never sees those margins.

\subsection{Where does the difficulty come from?}\label{sec:results}

A natural first guess is that TEQ is hard because of the competing minimal retentive sets that refuted Schwartz's conjecture: a rule that has to reconcile several such sets might well be intractable. Our second result rules this out. The reduction behind \Cref{thm:main} produces tournaments in which every nonempty set of alternatives, with the comparisons among them unchanged, has a unique minimal retentive set (\Cref{thm:structured-hardness}). Since the tournament on such a subset is an \emph{induced subtournament}, the uniqueness promise holds at every level of the recursion. The promise does not say that only one alternative is selected; in a three-cycle the unique minimal retentive set contains all three vertices. Nor does it reveal which alternatives are selected, and this is where the difficulty lies: their membership can still depend on the answers to many smaller TEQ queries.

The hard instances have a further local property, which the reduction gives us for free: in every nonempty induced subtournament, some vertex has at most three dominators. This turns out to be a strong condition. We prove that whenever a tournament has a vertex with at most four dominators, its minimal retentive set is unique (\Cref{thm:local}) and some member of TEQ can be found in quadratic time (\Cref{thm:select}). Under the hereditary form of this condition (that is, the recursive form: every nonempty induced subtournament also has a vertex with at most four dominators), these conclusions also hold for every nonempty induced subtournament. The bound concerns some vertex of each subtournament, not every vertex, so these tournaments can be arbitrarily large and complex elsewhere.

Put together, these results separate two questions that are usually treated as one. On every tournament with a vertex of at most four dominators, and in particular on the hard instances, \emph{finding} a member of TEQ takes quadratic time, while \emph{recognizing} whether a given alternative is a member is PSPACE-complete. A gap of this kind was known for the Banks set, where finding a member is easy and recognizing one is NP-complete~\cite{Woeginger2003,Hudry2004}; for TEQ, recognition is PSPACE-complete. For the Banks set, a member can be found in every tournament. Our procedure needs a vertex with at most four dominators, and whether a member of TEQ can be found in polynomial time in general remains open.

Finding one member of TEQ also helps us find the rest. Brandt et al.~\cite{BFHM2010} sped up the recursion in practice by starting from the alternatives with the most pairwise wins, although their worst-case running time remained exponential in the number $N$ of alternatives. Brandt, Brill, and Seedig~\cite{BBS2011} gave a decomposition algorithm running in $O(N^2+N f(\delta))$ time, where $\delta$ is the size of the largest indecomposable subproblem and $f(k)$ is the time needed by the underlying TEQ algorithm on at most $k$ alternatives. With direct recursion, $f(k)$ is exponential in $k$, and $\delta$ can be as large as $N$. Under the aforementioned hereditary four-dominator condition, starting from a known winner and repeatedly adding serious challengers lets us find all $t$ winners in $O(2^t N^2)$ time (\Cref{thm:select}).
For any $k\ge 1$, the same algorithm finds $k$ distinct winners in $O(2^{k-1}N^2)$ time, returning the entire TEQ within this bound if there are fewer than $k$ winners (\Cref{cor:partial-teq}). For fixed $k$, this gives quadratic time regardless of the total number of winners, even when the tournament admits no nontrivial decomposition. Taking $k=\lfloor\log_2 N\rfloor$, we can find $k$ winners in $O(N^3)$ time, or all of them if there are fewer.

Finally, we locate the dominator bound on the complexity map. If every induced subtournament has a vertex with at most one dominator, TEQ is computed in quadratic time; with two dominators, membership is already NP-hard; with three, it is PSPACE-complete (\Cref{thm:boundary,thm:structured-hardness}). Whether two dominators suffice for PSPACE-hardness is the main question we leave open.

\subsection{Proof ideas}\label{sec:overview}

\paragraph{Membership as reachability.}
To use the recursive queries in our reduction, we record the recursive answers in a graph on the alternatives called the \emph{contestation graph}~\cite{LLL1993}. This graph has an edge $x\to y$ exactly when $x$ is a serious challenger of $y$, that is, when TEQ selects $x$ among the dominators of $y$. A set is retentive exactly when every edge in the contestation graph pointing to a member of the set starts inside the set. The minimal retentive sets are therefore precisely the strongly connected components with no incoming edge from another component (i.e. the source components). If a vertex $r$ can reach every alternative in the contestation graph, its component is the unique source component. TEQ then selects exactly the alternatives that have a path back to $r$ (\Cref{lem:return}). This turns membership into a question about paths, which the reduction can control.

\paragraph{Simulating quantifiers.}
To evaluate a quantified Boolean formula, we give each occurrence of a literal (in a clause) its own vertex, and we represent the value $0$ of a variable by deleting its vertices and the value $1$ by keeping them. A clause gadget then recognizes the one pattern of surviving vertices that falsifies its clause. Each quantifier must combine two answers, one with the variable's vertices present and one with them deleted, and copying the tournament for both answers at every quantifier would blow up its size exponentially. Instead, each quantifier adds a constant number of vertices around a single copy. Two of them, the \emph{readers}, have exactly the two relevant subtournaments among their dominators, so that the recursive queries at the readers evaluate both states of the same copy. The remaining vertices create the conditional return paths mentioned above that combine the two answers, in parallel for an existential quantifier and in series for a universal one (\Cref{sec:hardness}). The difficulty is that a tournament has an edge between every pair of vertices, so each new vertex is compared with everything, and these comparisons could create unintended paths or disturb the readers' queries. A two-vertex \emph{shield} isolates the parts of a query we want to read from the parts we want to ignore, and the gadgets are designed so that deleting vertices inside them leaves a gadget of the same kind, which is what the induction over nested quantifiers requires.

\paragraph{Uniqueness from a vertex with few dominators.}
The uniqueness argument begins with two hypothetical minimal retentive sets. Neither can be a singleton, and each must then contain two dominators of every vertex outside it, so a vertex with at most four dominators leaves only a few configurations. Counting alone does not rule them out. The remaining cases are eliminated with \emph{captain edges}, in the terminology of Yang~\cite{Yang2016}: an edge from $a$ to $b$ is a captain edge if no vertex beats both $a$ and $b$. Then $a$ beats every other dominator of $b$, so $a$ is the only serious challenger of $b$, and every retentive set containing $b$ contains $a$ (\Cref{sec:uniqueness}). The same local constraints yield the procedure for finding a member (\Cref{sec:algorithms}). Proofs omitted from the body appear in the appendices, in the order of the sections.

\subsection{Related work}\label{sec:related}

\paragraph{TEQ and retentive sets.}
Tournament solutions are surveyed by Laslier~\cite{Laslier1997}, by Brandt, Brill, and Harrenstein~\cite{BBH2016}, and by Suksompong~\cite{Suksompong2021}. Laffond, Laslier, and Le Breton~\cite{LLL1993} characterized TEQ through the source components of the contestation graph. Minimal retentive sets with respect to general tournament solutions were studied by Brandt et al.~\cite{BBFH2014}, and the structure of stable tournament solutions by Brandt et al.~\cite{BBSS2018}. Yang~\cite{Yang2016} verified Schwartz's conjecture for tournaments of order at most $14$, and proved it for locally transitive tournaments and for tournaments in which the captain edges contain a directed cycle through all vertices. \Cref{thm:local} does not follow from these results: it asks for a single vertex with at most four dominators and places no condition on the rest of the tournament. Our use of captain edges builds on Yang's work. Mnich, Shrestha, and Yang~\cite{MSY2015} established structural properties of minimal retentive sets, including that every nonsingleton minimal retentive set induces a strongly connected tournament.

\paragraph{Complexity of TEQ.}
Brandt et al.~\cite{BFHM2008,BFHM2010} proved NP-hardness of TEQ membership and studied exact algorithms based on recursive expansion with memoization. Their hard instances have a unique minimal retentive set, and their argument gives uniqueness in every induced subtournament containing the designated vertex~\cite[Lemma~1]{BFHM2010}, but not in every induced subtournament, which is what \Cref{thm:structured-hardness} requires. Brandt, Brill, and Seedig~\cite{BBS2011} gave a fixed-parameter algorithm for TEQ based on decomposing the tournament into blocks whose vertices have identical comparisons with every outside vertex. Bachmeier et al.~\cite{BBGHKPS2019} proved NP-hardness on the majority tournaments of seven voters, and Constantinescu and Wattenhofer~\cite{CW2022} showed that two-crossing preference profiles, in which the voters can be ordered so that the relative preference between any two alternatives changes at most twice, still realize every tournament, which yields NP-hardness on that domain. Our reduction transfers to the same domain (\Cref{cor:two-crossing}).

\paragraph{PSPACE-hardness in voting.}
Dickey, Psomas, and Terzoglou~\cite{SV2026} show PSPACE-completeness of winner determination for Stable Voting and Simple Stable Voting, recursive rules defined on weighted tournaments. Earlier PSPACE-hardness results in computational social choice rely on succinct or combinatorial representations of the preferences or on sequential processes; see the discussion in~\cite{SV2026}. \Cref{thm:main} is the first such result for a rule that depends only on the orientation of each pairwise comparison.
\section{Preliminaries}\label{sec:prelim}

\paragraph{Tournaments.}
A tournament $T$ on a finite vertex set $V$ is an irreflexive relation $\succ$ such that for all distinct $x,y\in V$ exactly one of $x\succ y$ and $y\succ x$ holds. We say that $x$ \emph{beats} $y$, or that the edge $xy$ points from $x$ to $y$. For sets $A,B$ we write $A\succ B$ if $a\succ b$ for all $a\in A$ and $b\in B$, and we omit braces around singletons.

TEQ evaluates an alternative by looking at those that beat it. The \emph{dominators} of $y$ in $X\subseteq V$ are $D_X(y)=\{x\in X:x\succ y\}$, and $D(y)=D_V(y)$. The \emph{indegree} of $y$ is $|D(y)|$. A vertex with no dominators is a \emph{Condorcet winner}. The recursive queries are themselves tournaments: $T[X]$ is the \emph{induced subtournament} on $X$, with the comparisons between vertices of $X$ unchanged. We identify $X$ with $T[X]$ when the edges are clear, and write $T-Z$ for $T[V\setminus Z]$.

\begin{definition}[TEQ]\label{def:teq}
Set $\TEQ(\varnothing)=\varnothing$. For a nonempty tournament $X$, a nonempty set $S\subseteq X$ is \emph{TEQ-retentive} (or simply \emph{retentive}) in $X$ if $\TEQ(D_X(y))\subseteq S$ for every $y\in S$. The \emph{tournament equilibrium set} $\TEQ(X)$ is the union of the inclusion-minimal retentive sets of $X$.
\end{definition}

We say that a vertex $x$ is \emph{selected} in $X$ if $x\in\TEQ(X)$.

The definition is well founded because $D_X(y)\subsetneq X$. The set $X$ itself is retentive, so minimal retentive sets exist and $\TEQ(X)\ne\varnothing$. In a three-vertex cycle $a\succ b\succ c\succ a$, each vertex has one dominator. A retentive set containing $a$ must therefore also contain $c$, then $b$. The same reasoning applies from either of the other vertices, so the whole cycle is the unique minimal retentive set and TEQ selects all three vertices.

We study the following problem. The full-set version asks to output $\TEQ(T)$, with the original vertex labels. Throughout, $N=|V|$, and a tournament is given by its $\binom N2$ orientation bits, for instance as an adjacency matrix.

\begin{center}
\fbox{\begin{minipage}{0.9\linewidth}
\prob{TEQ Membership}\\
\emph{Input:} a tournament $T$ and a designated vertex $d$.\\
\emph{Question:} is $d\in\TEQ(T)$?
\end{minipage}}
\end{center}

\paragraph{The contestation graph.}
To see how the answers to the smaller dominator queries constrain a retentive set, we record each selected dominator as an edge into the alternative it challenges. The resulting graph lets us express retentiveness as closure under predecessors.

\begin{definition}[Contestation graph]\label{def:relation}
For a nonempty tournament $X$, the \emph{contestation graph} $R_X$ on $X$ is defined by
\begin{equation}\label{eq:recursive-relation}
x\RX{X}y\quad\Longleftrightarrow\quad x\in\TEQ(D_X(y)).
\end{equation}
\end{definition}

We call the set $\TEQ(D_X(y))$ the \emph{query} at $y$ in $X$. Thus $x\RX{X}y$ holds exactly when $x$ is selected in the query at $y$, and the definition of TEQ uses nothing about $X$ beyond the queries at its vertices.

Every edge of $R_X$ is an edge of the tournament, but only the selected dominators of $y$ contribute edges into $y$. We drop the subscript when $X$ is the whole tournament under consideration.

The following example shows how the contestation graph records the selected dominators and why it must be recomputed for a different ambient tournament. Deleting a vertex can change a dominator query even when an edge between surviving vertices stays the same.

\begin{figure}[H]
\centering
\begin{tikzpicture}[>=Stealth, every node/.style={circle,inner sep=1.5pt,minimum size=13pt}, x=1cm, y=1cm]
\tikzset{arc/.style={->,thick},rel/.style={->,very thick,blue!60!black}}
\begin{scope}
\node (a) at (0,1.5) {$a$}; \node (b) at (1.8,1.5) {$b$};
\node (c) at (0,0) {$c$};   \node (d) at (1.8,0) {$d$};
\draw[arc] (a)--(b); \draw[arc] (b)--(c); \draw[arc] (c)--(a);
\draw[arc] (a)--(d); \draw[arc] (c)--(d); \draw[arc] (d)--(b);
\node[draw=none,rectangle] at (0.9,-0.8) {\small (a) $T$ on $X=\{a,b,c,d\}$};
\end{scope}
\begin{scope}[xshift=4.2cm]
\node (a) at (0,1.5) {$a$}; \node (b) at (1.8,1.5) {$b$};
\node (c) at (0,0) {$c$};   \node (d) at (1.8,0) {$d$};
\draw[rel] (a)--(b); \draw[rel] (b)--(c); \draw[rel] (c)--(a); \draw[rel] (c)--(d);
\node[draw=none,rectangle] at (0.9,-0.8) {\small (b) $R_X$};
\end{scope}
\begin{scope}[xshift=8.4cm]
\node (a) at (0,1.5) {$a$}; \node (b) at (1.8,1.5) {$b$};
\node (d) at (1.8,0) {$d$};
\draw[rel] (a)--(b); \draw[rel] (a)--(d);
\node[draw=none,rectangle] at (0.9,-0.8) {\small (c) $R_Y$ for $Y=X\setminus\{c\}$};
\end{scope}
\end{tikzpicture}
\caption{\Cref{ex:four}. Black arrows are tournament edges and blue arrows are edges of the contestation graph. Deleting $c$ creates the contestation edge $a\RX{Y}d$, although the edge $a\succ d$ is unchanged.}
\label{fig:example}
\end{figure}
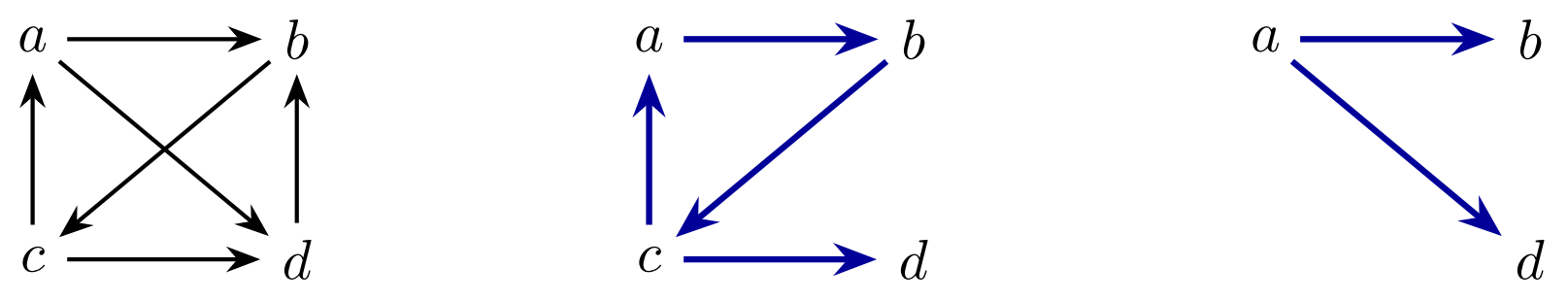

\begin{example}\label{ex:four}
Let $X=\{a,b,c,d\}$ with $a\succ b\succ c\succ a$ and $\{a,c\}\succ d\succ b$ (\Cref{fig:example}). The four dominator queries and their answers are
\[
\begin{array}{c|cccc}
y&a&b&c&d\\\hline
D_X(y)&\{c\}&\{a,d\}&\{b\}&\{a,c\}\\
\TEQ(D_X(y))&\{c\}&\{a\}&\{b\}&\{c\}.
\end{array}
\]
Thus $R_X$ consists of the cycle $a\RX{X}b\RX{X}c\RX{X}a$ and the edge $c\RX{X}d$. Any retentive set containing a vertex of the cycle must contain the whole cycle, and a retentive set containing $d$ must contain $c$ and therefore the cycle. The cycle is itself retentive, so it is the unique minimal retentive set and $\TEQ(X)=\{a,b,c\}$. The edge $a\succ d$ is not a contestation edge. Now delete $c$. The tournament $Y=\{a,b,d\}$ is transitive with Condorcet winner $a$, and $R_Y$ has the edges $a\RX{Y}b$ and $a\RX{Y}d$. The unchanged edge $a\succ d$ has become a contestation edge: $R_Y$ is not the restriction of $R_X$ to $Y$.
\end{example}

A source strongly connected component has no incoming contestation edge from another component. In the example, $\{a,b,c\}$ is the only such component. This is no coincidence: the predecessor-closure requirement makes the source components precisely the minimal retentive sets.

\begin{proposition}[\cite{LLL1993}]\label{prop:source-components}
Let $X$ be a nonempty tournament and $R_X$ its contestation graph. The minimal retentive sets of $X$ are exactly the source strongly connected components of $R_X$. Hence $\TEQ(X)$ is the union of the source components of $R_X$.
\end{proposition}

\begin{proof}
A nonempty set is retentive exactly when it contains every $R_X$-predecessor of each of its members, that is, when it is closed under $R_X$-predecessors. A closed set containing one vertex of a strongly connected component contains the whole component. A source component is closed. Conversely, starting at any component contained in a nonempty closed set and following incoming edges must eventually reach a source component, which is also contained in the set. Hence the minimal closed sets are exactly the source components.
\end{proof}

The source-component characterization suggests a way to prove membership: arrange for one vertex $r$ to reach the entire contestation graph. Any source component must then contain $r$, and a vertex is selected exactly when it can return to $r$. This criterion will let the reduction encode Boolean answers by the existence of paths.

\begin{lemma}[Return-path criterion]\label{lem:return}
Suppose some vertex $r$ of a nonempty tournament $X$ reaches every vertex of $X$ in $R_X$. Then $X$ has a unique minimal retentive set, namely the strongly connected component of $r$, and a vertex $d\in X$ belongs to $\TEQ(X)$ if and only if $d$ reaches $r$ in $R_X$.
\end{lemma}

\begin{proof}
Any vertex that reaches the component of $r$ is reached by $r$ and hence lies in that component, so the component is a source. A second source component would have to be reached by $r$, which is impossible. Since $r$ reaches $d$, the vertex $d$ lies in the component of $r$ exactly when it reaches $r$.
\end{proof}

\begin{proposition}\label{prop:pspace}
\prob{TEQ Membership} can be decided in polynomial space.
\end{proposition}

\begin{proof}
Compute $R_X$ by answering the query at each vertex recursively, and then compute its source components. The recursion depth is at most $N$, and each level stores one contestation graph and the current domain, so $O(N^3)$ bits suffice when completed queries are recomputed rather than stored.
\end{proof}

\section{The Tournament Construction}\label{sec:hardness}

We now prove that deciding whether a given vertex belongs to the tournament equilibrium set is PSPACE-hard, which is the hardness half of our main theorem. The proof is a reduction from TQBF, the problem of deciding whether a quantified Boolean formula is true. Given such a formula $\Phi$, we construct in polynomial time a tournament $T_1$ together with a vertex $o_1$ such that
\[
\Phi\text{ is true}\quad\Longleftrightarrow\quad o_1\in\TEQ(T_1).
\]
The index records that $T_1$ is the outermost of $n+1$ nested tournaments $T_1\supseteq T_2\supseteq\cdots\supseteq T_{n+1}$, one for each quantifier and one for the quantifier-free part. The innermost tournament $T_{n+1}$ is built first and $T_1$ last. Each $T_i$ has a designated vertex $o_i$, its \emph{output}; this is the vertex whose membership in $\TEQ(T_i)$ will carry a truth value, and the output of $T_1$ is the vertex $o_1$ above. This section gives the construction in full (\Cref{sec:hard-construction}); it is short, and it can be read without knowing why it works. We then explain in broad terms how the resulting tournament evaluates the formula (\Cref{sec:hard-ideas}). \Cref{sec:correctness} proves that the construction is correct, and explains the role of each gadget where it proves the corresponding lemma.

\paragraph{The formula.}\label{sec:hard-formula}
An instance of TQBF is a closed formula
\[
\Phi=Q_1x_1\,Q_2x_2\cdots Q_nx_n\ (C_1\land\cdots\land C_m),
\]
where each quantifier $Q_i$ is $\exists$ or $\forall$, each clause $C_\ell$ is a disjunction of literals, and a literal is a variable $x_i$ or its negation $\neg x_i$. We write $\psi=C_1\land\cdots\land C_m$ for the quantifier-free part and $\psi(\alpha)$ for its truth value under an assignment $\alpha\in\{0,1\}^n$, with $1$ standing for true. We assume that $\Phi$ has the following form, which simplifies the construction:
\begin{enumerate}[label=\textup{(F\arabic*)},leftmargin=3.2em]
\item\label{F1} every clause has exactly three literals, and their three variables are distinct;
\item\label{F2} every clause has at least one positive literal, and its negative literals are listed first: $C_\ell=(\ocl{\ell}{1}\lor\ocl{\ell}{2}\lor\ocl{\ell}{3})$, where the first $j_\ell\in\{0,1,2\}$ literals are negative and the remaining ones are positive;
\item\label{F3} every variable occurs in some clause.
\end{enumerate}
TQBF remains PSPACE-complete under these assumptions: it is PSPACE-complete for clauses with at most three literals~\cite{SM1973}, and every such formula can be brought into this form in linear time (\Cref{lem:normal} in \Cref{hard:normal}). Throughout this section and the next, we illustrate the construction on the running example
\begin{equation}\label{eq:running}
\Phi_{\mathrm{ex}}=\forall x_1\,\exists x_2\,\exists x_3\ \bigl[(x_1\lor x_2\lor x_3)\land(\neg x_1\lor\neg x_2\lor x_3)\bigr],
\end{equation}
whose two clauses have $j_1=0$ and $j_2=2$ negative literals. The formula is true: whatever the value of $x_1$, setting $x_3=1$ satisfies both clauses.

\subsection{Main Construction}\label{sec:hard-construction}

We now present the construction, which takes the formula $\Phi$ to the tournament $T_1$. This subsection only specifies the vertices and the edges of $T_1$. What each part does, and why the construction is correct, is explained afterwards, in \Cref{sec:hard-ideas,sec:correctness}.

Let $\Phi$ have $n$ variables and $m$ clauses. The tournament $T_1$ is built from the inside out, in three steps. First, every clause receives six vertices: one for each of its three literals, an output, and two auxiliary vertices. Second, every clause after the first receives two more vertices, a \emph{filter}, which sits between its six vertices and those of the earlier clauses; the result is one tournament for the quantifier-free part $\psi$, which we call $T_{n+1}$. Third, every quantifier $Q_ix_i$ receives six vertices, which are added around everything built so far, from the innermost quantifier $Q_nx_n$ to the outermost quantifier $Q_1x_1$; the result after adding the vertices of $Q_ix_i$ is the tournament $T_i$. Altogether $T_1$ has $6m+2(m-1)+6n=8m+6n-2$ vertices.

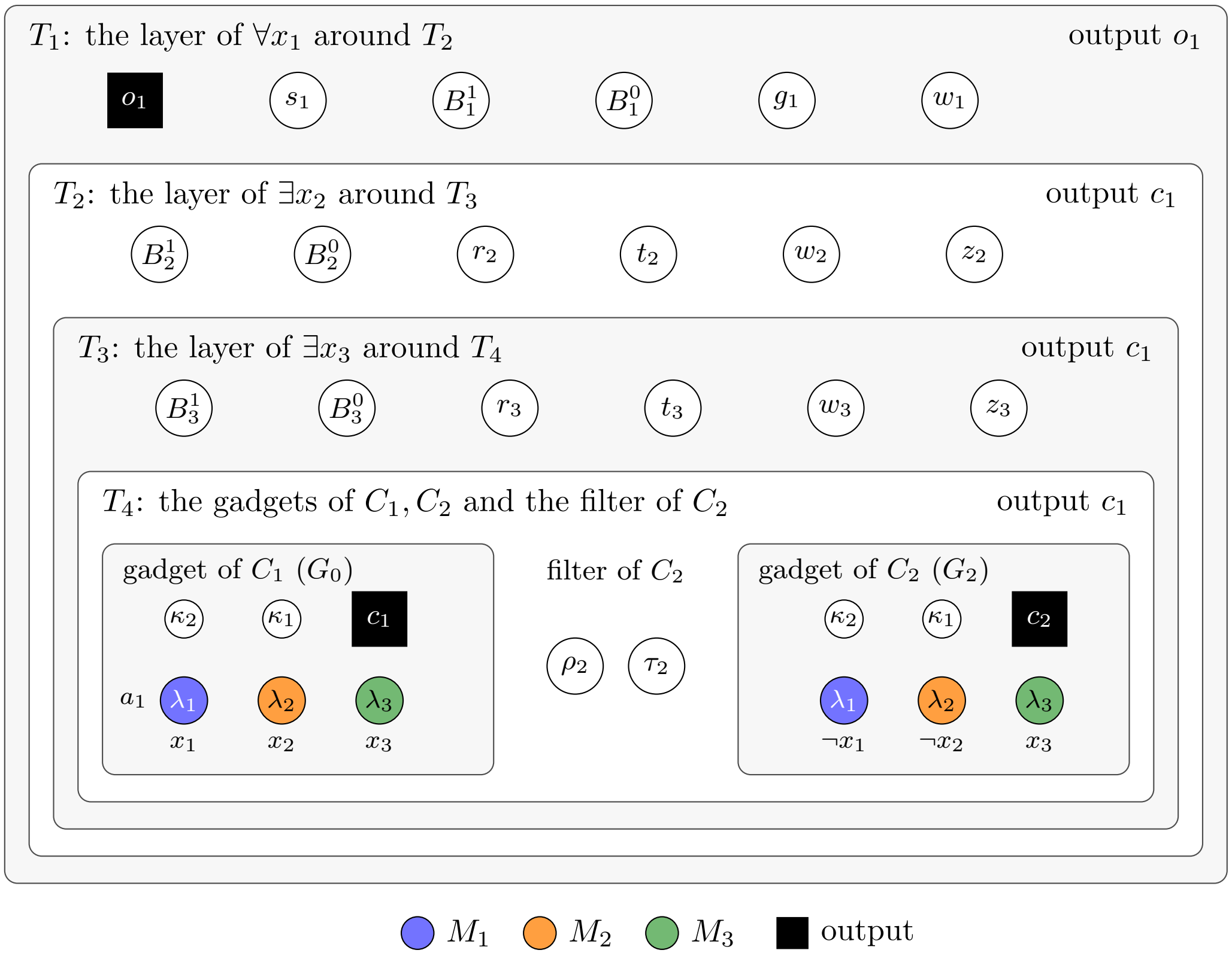
\begin{figure}[ht]
\centering
\begin{tikzpicture}[x=0.9cm,y=1cm,font=\small,>=Stealth]
\tikzset{
  family/.style={draw=black!70,rounded corners=4pt},
  gv/.style={circle,draw,fill=white,inner sep=0pt,minimum size=6.2mm,font=\footnotesize},
  outv/.style={rectangle,draw,fill=black,text=white,inner sep=0pt,minimum size=6mm,font=\footnotesize},
  aux/.style={circle,draw,fill=white,inner sep=0pt,minimum size=4.2mm,font=\scriptsize},
  occ/.style={circle,draw,inner sep=0pt,minimum size=5.2mm,font=\scriptsize},
  m1/.style={occ,fill=blue!55,text=white},
  m2/.style={occ,fill=orange!75},
  m3/.style={occ,fill=green!50!black!55}
}
\draw[family,fill=black!3] (0,0) rectangle (15,9.7);
\node[anchor=west] at (0.15,9.35) {$T_1$: the layer of $\forall x_1$ around $T_2$};
\node[anchor=east] at (14.85,9.35) {output $\eout_1$};
\node[outv] at (1.6,8.65) {$\eout_1$};
\foreach \n/\xx in {\slink_1/3.6,\groot_1/5.6,\rzero_1/7.6,\king_1/9.6,\wanc_1/11.6} \node[gv] at (\xx,8.65) {$\n$};
\draw[family,fill=white] (0.3,0.3) rectangle (14.7,7.95);
\node[anchor=west] at (0.45,7.6) {$T_2$: the layer of $\exists x_2$ around $T_3$};
\node[anchor=east] at (14.55,7.6) {output $\cout_1$};
\foreach \n/\xx in {\rone_2/1.9,\rzero_2/3.9,\broot_2/5.9,\tri_2/7.9,\shw_2/9.9,\shz_2/11.9} \node[gv] at (\xx,6.95) {$\n$};
\draw[family,fill=black!3] (0.6,0.6) rectangle (14.4,6.25);
\node[anchor=west] at (0.75,5.9) {$T_3$: the layer of $\exists x_3$ around $T_4$};
\node[anchor=east] at (14.25,5.9) {output $\cout_1$};
\foreach \n/\xx in {\rone_3/2.2,\rzero_3/4.2,\broot_3/6.2,\tri_3/8.2,\shw_3/10.2,\shz_3/12.2} \node[gv] at (\xx,5.25) {$\n$};
\draw[family,fill=white] (0.9,0.9) rectangle (14.1,4.55);
\node[anchor=west] at (1.05,4.2) {$T_4$: the gadgets of $C_1,C_2$ and the filter of $C_2$};
\node[anchor=east] at (13.95,4.2) {output $\cout_1$};
\draw[family,fill=black!3] (1.2,1.2) rectangle (6.0,3.75);
\node[anchor=west,font=\footnotesize] at (1.3,3.45) {gadget of $C_1$ ($\Gad_0$)};
\node[aux] at (2.2,2.92) {\scriptsize$\ax{2}$}; \node[aux] at (3.4,2.92) {\scriptsize$\ax{1}$};
\node[outv] at (4.6,2.92) {$\cout_1$};
\node[m1,label={[font=\scriptsize]below:$x_1$},label={[font=\scriptsize]left:$\anc_1$}] at (2.2,2.02) {\scriptsize$\oc{1}$};
\node[m2,label={[font=\scriptsize]below:$x_2$}] at (3.4,2.02) {\scriptsize$\oc{2}$};
\node[m3,label={[font=\scriptsize]below:$x_3$}] at (4.6,2.02) {\scriptsize$\oc{3}$};
\node[font=\footnotesize] at (7.5,3.45) {filter of $C_2$};
\node[gv] at (7.0,2.4) {$\relay_2$};
\node[gv] at (8.0,2.4) {$\test_2$};
\draw[family,fill=black!3] (9.0,1.2) rectangle (13.8,3.75);
\node[anchor=west,font=\footnotesize] at (9.1,3.45) {gadget of $C_2$ ($\Gad_2$)};
\node[aux] at (10.3,2.92) {\scriptsize$\ax{2}$}; \node[aux] at (11.5,2.92) {\scriptsize$\ax{1}$};
\node[outv] at (12.7,2.92) {$\cout_2$};
\node[m1,label={[font=\scriptsize]below:$\neg x_1$}] at (10.3,2.02) {\scriptsize$\oc{1}$};
\node[m2,label={[font=\scriptsize]below:$\neg x_2$}] at (11.5,2.02) {\scriptsize$\oc{2}$};
\node[m3,label={[font=\scriptsize]below:$x_3$}] at (12.7,2.02) {\scriptsize$\oc{3}$};
\begin{scope}[yshift=-0.55cm,xshift=3.75cm]
\node[m1,minimum size=3.6mm] at (0.9,0) {}; \node[anchor=west] at (1.1,0) {$M_1$};
\node[m2,minimum size=3.6mm] at (2.4,0) {}; \node[anchor=west] at (2.6,0) {$M_2$};
\node[m3,minimum size=3.6mm] at (3.9,0) {}; \node[anchor=west] at (4.1,0) {$M_3$};
\node[outv,minimum size=3.4mm] at (5.5,0) {}; \node[anchor=west] at (5.7,0) {output};
\end{scope}
\end{tikzpicture}
\caption{The tournament $T_1$ for $\Phi_{\mathrm{ex}}=\forall x_1\,\exists x_2\,\exists x_3\,[(x_1\lor x_2\lor x_3)\land(\neg x_1\lor\neg x_2\lor x_3)]$. It has $32$ vertices: $6+2+6$ in $T_4$ and $6$ in each layer. Every vertex appears once. Each box is the vertex set of the tournament $T_i$ built so far, so the boxes are nested induced subtournaments, not copies. The output of each box is marked in its top right corner, and the designated vertex is the output $\orole_1$ of the outermost box. Colors mark the occurrence sets $M_1,M_2,M_3$, and $\anc_1$ is the anchor of the $\forall x_1$-layer. The edges are those of \Cref{def:clause,def:filter,def:or,def:and}.}
\label{fig:reduction-example}
\end{figure}

We now describe the three steps in detail. \Cref{def:reduction} then puts them together, and \Cref{fig:reduction-example} shows the result for the running example.

\paragraph{Step 1: clause gadgets.}\label{sec:hard-gadgets}
The six vertices of a clause are an \emph{output}, two \emph{auxiliary} vertices, and three \emph{occurrence vertices}, one for each literal of the clause. How these six vertices are connected depends only on the number $j\in\{0,1,2\}$ of negative literals of the clause. This gives three tournaments on six vertices, one for each value of $j$, which we call the \emph{clause gadgets}.

\begin{definition}[Clause gadgets]\label{def:clause}
A clause gadget has six vertices: the output $\cout$, the auxiliary vertices $\ax{1}$ and $\ax{2}$, and the occurrence vertices $\oc{1},\oc{2},\oc{3}$ of the three literals of the clause, in order. Let $j\in\{0,1,2\}$ be the number of negative literals of the clause; these are its first $j$ literals. The clause gadget $\Gad_j$ is the tournament on the six vertices obtained from the transitive tournament
\[
\oc{3}\succ\oc{2}\succ\oc{1}\succ\ax{2}\succ\ax{1}\succ\cout,
\]
in which every vertex beats all the vertices to its right, by reversing the edges between the following pairs. The pairs are listed as they are oriented in $\Gad_j$, that is, after the reversal:
\[
\begin{array}{ll}
\Gad_0: & \cout\succ\oc{1},\ \cout\succ\oc{2},\ \cout\succ\oc{3};\\
\Gad_1: & \cout\succ\ax{2},\ \cout\succ\oc{2},\ \cout\succ\oc{3};\\
\Gad_2: & \cout\succ\ax{1},\ \cout\succ\ax{2},\ \cout\succ\oc{3},\ \ax{1}\succ\ax{2},\ \ax{1}\succ\oc{2},\ \ax{2}\succ\oc{1}.
\end{array}
\]
\end{definition}

\begin{figure}[H]
\centering
\begin{tikzpicture}[>=Stealth, font=\small,
  outv/.style={draw,rectangle,thick,inner sep=0pt,minimum size=5.5mm},
  aux/.style={draw,circle,inner sep=0pt,minimum size=4.6mm},
  occ/.style={draw,circle,thick,inner sep=0pt,minimum size=5.5mm},
  neg/.style={occ,fill=black!18},
  exc/.style={->,thick,shorten >=0.5pt,shorten <=0.5pt}]
\foreach \j/\shift in {0/0,1/4.3,2/8.6} {
  \begin{scope}[xshift=\shift cm]
  \node[font=\normalsize] at (0.9,2.05) {$G_{\j}$};
  \node[aux]  (v2) at (0,1.15) {$\ax{2}$};
  \node[aux]  (v1) at (0.9,1.15) {$\ax{1}$};
  \node[outv] (v0) at (1.8,1.15) {$\cout$};
  \ifnum\j>0 \node[neg] (v3) at (0,0) {$\oc{1}$}; \else \node[occ] (v3) at (0,0) {$\oc{1}$}; \fi
  \ifnum\j>1 \node[neg] (v4) at (0.9,0) {$\oc{2}$}; \else \node[occ] (v4) at (0.9,0) {$\oc{2}$}; \fi
  \node[occ] (v5) at (1.8,0) {$\oc{3}$};
  \foreach \k/\x in {3/0,4/0.9,5/1.8} {
    \pgfmathtruncatemacro{\isneg}{(\k-2)<=\j ? 1 : 0}
    \ifnum\isneg=1 \node[font=\scriptsize] at (\x,-0.5) {neg.}; \else \node[font=\scriptsize] at (\x,-0.5) {pos.}; \fi
  }
  \ifnum\j=0
    \draw[exc] (v0) -- (v3); \draw[exc] (v0) -- (v4); \draw[exc] (v0) -- (v5);
  \fi
  \ifnum\j=1
    \draw[exc] (v0) to[bend right=40] (v2); \draw[exc] (v0) -- (v4); \draw[exc] (v0) -- (v5);
  \fi
  \ifnum\j=2
    \draw[exc] (v0) -- (v1); \draw[exc] (v1) -- (v2); \draw[exc] (v0) to[bend right=40] (v2);
    \draw[exc] (v2) -- (v3); \draw[exc] (v1) -- (v4); \draw[exc] (v0) -- (v5);
  \fi
  \end{scope}
}
\end{tikzpicture}
\caption{The three clause gadgets of \Cref{def:clause}; negative occurrence vertices are shaded. Only the reversed edges are drawn, and all other edges follow the order $\oc{3}\succ\oc{2}\succ\oc{1}\succ\ax{2}\succ\ax{1}\succ\cout$.}
\label{fig:clause-gadgets}
\end{figure}
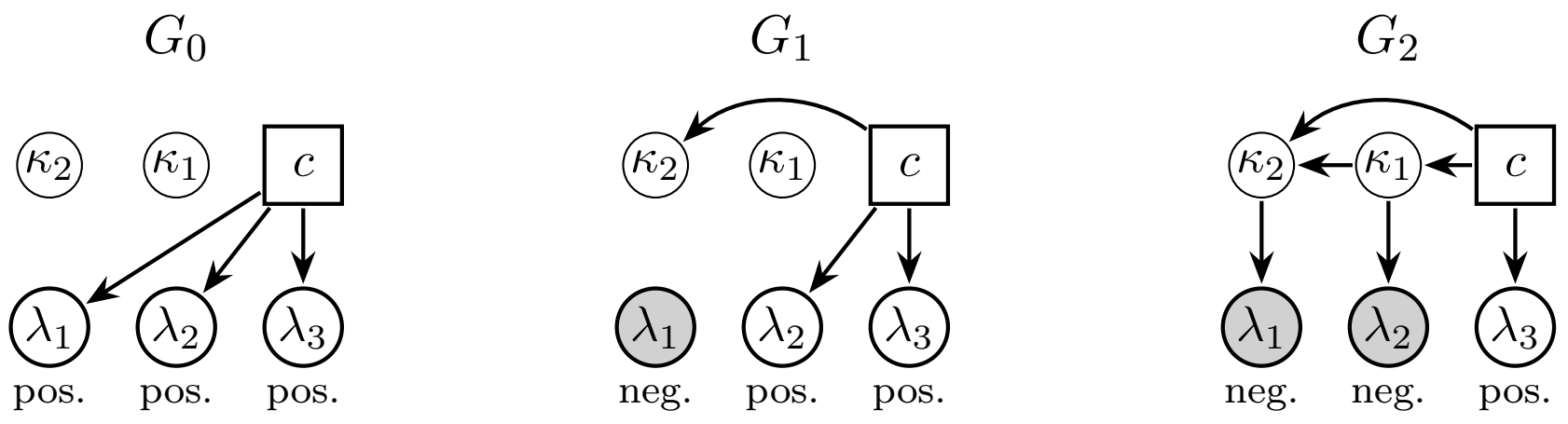

The gadget of a clause $C_\ell$ is a copy of $\Gad_{j_\ell}$, where $j_\ell$ is the number of negative literals of $C_\ell$. We write $\cout_\ell$, $\axl{\ell}{1},\axl{\ell}{2}$ and $\ocl{\ell}{1},\ocl{\ell}{2},\ocl{\ell}{3}$ for its six vertices, so that the occurrence vertex of the literal $\ocl{\ell}{k}$ is denoted by the same symbol $\ocl{\ell}{k}$. \Cref{fig:clause-gadgets} shows all three gadgets.

A variable $x_i$ usually occurs in several clauses, and so has one occurrence vertex in each of them, whether it occurs there as $x_i$ or as $\neg x_i$. We write $M_i$ for the set of all occurrence vertices of $x_i$:
\begin{equation}\label{eq:Mi}
M_i=\bigl\{\ocl{\ell}{k}\ :\ 1\le\ell\le m,\ k\in\{1,2,3\},\ \text{and the literal }\ocl{\ell}{k}\text{ is }x_i\text{ or }\neg x_i\bigr\}.
\end{equation}

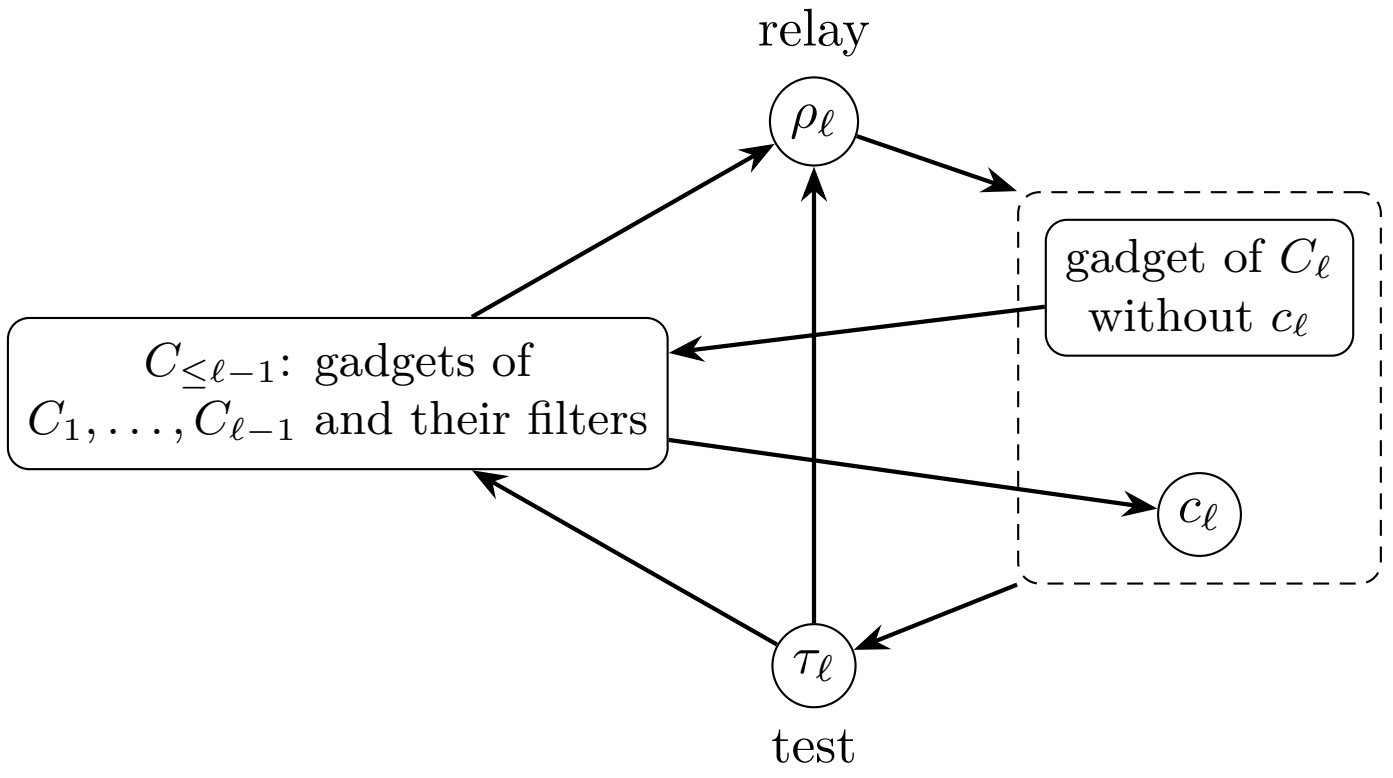
\begin{figure}[H]
\centering
\begin{tikzpicture}[>=Stealth, font=\small]
\node[draw,rounded corners,minimum width=3.3cm,minimum height=1.0cm,align=center,font=\footnotesize] (H) at (-1.0,0) {$\Cle{\ell-1}$: gadgets of\\$C_1,\ldots,C_{\ell-1}$ and their filters};
\node[draw,rounded corners,minimum width=1.5cm,minimum height=.65cm,align=center,font=\footnotesize] (B) at (4.7,.7) {gadget of $C_\ell$\\without $\cout_\ell$};
\node[draw,circle,inner sep=2pt] (b) at (4.7,-.8) {$\cout_\ell$};
\node[draw,circle,inner sep=2pt,label=above:{relay}] (x) at (2.15,1.8) {$\relay_\ell$};
\node[draw,circle,inner sep=2pt,label=below:{test}] (y) at (2.15,-1.8) {$\test_\ell$};
\begin{scope}[on background layer]
\node[draw,densely dashed,rounded corners,fit=(B)(b),inner sep=5pt,label={[font=\scriptsize]right:{}}] (K) {};
\end{scope}
\draw[->,thick] (B) -- (H);
\draw[->,thick] (H) -- (b);
\draw[->,thick] (H) -- (x);
\draw[->,thick] (x) -- (K.north west);
\draw[->,thick] (K.south west) -- (y);
\draw[->,thick] (y) -- (H);
\draw[->,thick] (y) -- (x);
\end{tikzpicture}
\caption{The filter of $C_\ell$ (\Cref{def:filter}): the two vertices $\test_\ell$ and $\relay_\ell$ between $\Cle{\ell-1}$ and the gadget of $C_\ell$. An arrow between two parts means that every vertex of the first beats every vertex of the second.}
\label{fig:filter}
\end{figure}

\paragraph{Step 2: the filter.}
After Step~1 we have $m$ separate clause gadgets, one for each of $C_1,\ldots,C_m$. Step~2 joins them into a single tournament by going through the clauses from left to right, starting with $C_2$ and ending with $C_m$. When it reaches $C_\ell$, the clauses $C_1,\ldots,C_{\ell-1}$ have already been joined into one tournament, which we call $\Cle{\ell-1}$. The step then adds two new vertices, the \emph{test} $\test_\ell$ and the \emph{relay} $\relay_\ell$, which together form the \emph{filter} of $C_\ell$. They sit between $\Cle{\ell-1}$ and the gadget of $C_\ell$, and all edges between these two parts and the two new vertices are oriented as shown in \Cref{fig:filter}. The result is $\Cle{\ell}$, and the process continues with $C_{\ell+1}$. We call the final tournament $\Cle{m}$ the \emph{clause tournament}.

\begin{definition}[Filter]\label{def:filter}
Let $\Cle{1}$ be the gadget of $C_1$. For $\ell=2,\ldots,m$, the tournament $\Cle{\ell}$ is obtained from $\Cle{\ell-1}$ and the gadget of $C_\ell$ by adding two vertices, the \emph{test} $\test_\ell$ and the \emph{relay} $\relay_\ell$, which form the \emph{filter} of $C_\ell$, and orienting the new edges as follows:
\begin{itemize}[leftmargin=*,itemsep=1pt,topsep=3pt]
\item between the two parts: every vertex of the gadget of $C_\ell$ other than its output $\cout_\ell$ beats every vertex of $\Cle{\ell-1}$, and every vertex of $\Cle{\ell-1}$ beats $\cout_\ell$;
\item the relay: every vertex of $\Cle{\ell-1}$ beats $\relay_\ell$, and $\relay_\ell$ beats every vertex of the gadget of $C_\ell$;
\item the test: every vertex of the gadget of $C_\ell$ beats $\test_\ell$, and $\test_\ell$ beats every vertex of $\Cle{\ell-1}$;
\item $\test_\ell$ beats $\relay_\ell$.
\end{itemize}
The output of $\Cle{\ell}$ is $\cout_1$, for every $\ell$.
\end{definition}

\paragraph{Step 3: the $\exists$-layer and the $\forall$-layer.}
After Step~2 we have the tournament $T_{n+1}=\Cle{m}$, whose output is $\orole_{n+1}=\cout_1$. Step~3 goes through the quantifiers from the innermost to the outermost, that is, for $i=n,n-1,\ldots,1$. When it reaches $Q_ix_i$, the tournament $T_{i+1}$ and its output $\orole_{i+1}$ have already been built. The step adds six new vertices, the \emph{layer} of $Q_ix_i$, around $T_{i+1}$, and the result is the tournament $T_i$ with output $\orole_i$. We call the vertices of $T_{i+1}$ the \emph{old} vertices. Which six vertices are added, and how they are connected, depends on the quantifier: \Cref{def:or} is for $Q_i=\exists$ and \Cref{def:and} for $Q_i=\forall$. In both, the only old vertices that are treated specially are the output $\orole_{i+1}$ and the occurrence vertices $M_i$ of the variable $x_i$, defined in~\eqref{eq:Mi}.

In both layers, two of the six vertices are called \emph{readers} and are written $\Rdi{i}{1}$ and $\Rdi{i}{0}$. The superscripts refer to the two values $1$ and $0$ of the variable $x_i$: the reader $\Rdi{i}{b}$ will be the vertex that handles the case $x_i=b$, as \Cref{sec:hard-ideas} explains. The other four vertices have plain letters. The words \emph{root}, \emph{guard}, \emph{link} and \emph{anchor} in the two definitions, like \emph{test} and \emph{relay} above, name the roles that these vertices play in the correctness proof; the root is the vertex to which the return-path criterion (\Cref{lem:return}) will be applied.

Each definition has two parts: how the six new vertices are connected among themselves, and how they are connected to the old vertices. Each definition is followed by a figure that shows both parts.

\begin{definition}[$\exists$-layer]\label{def:or}
Let $Q_i=\exists$. The tournament $T_i$ is obtained from $T_{i+1}$ by adding six vertices: the two \emph{readers} $\rone_i$ and $\rzero_i$, the \emph{root} $\broot_i$, and the three \emph{guards} $\tri_i,\shw_i,\shz_i$. Its output is $\orole_i=\orole_{i+1}$. The new edges are oriented as follows.
\begin{itemize}[leftmargin=*,itemsep=1pt,topsep=3pt]
\item Among the new vertices: $\rone_i,\rzero_i,\tri_i$ form a cycle $\rone_i\succ\rzero_i\succ\tri_i\succ\rone_i$, and $\shz_i$ beats $\shw_i$. Moreover, $\shz_i$ and $\shw_i$ beat each of $\rone_i,\rzero_i,\tri_i$; each of $\rone_i,\rzero_i,\tri_i$ beats $\broot_i$; and $\broot_i$ beats $\shz_i$ and $\shw_i$.
\item Against the old vertices: $\broot_i,\tri_i,\shw_i$ beat every old vertex, and $\rone_i,\shz_i$ lose to every old vertex. The reader $\rzero_i$ loses to every old vertex outside $M_i$ and beats every vertex of $M_i$.
\end{itemize}
\end{definition}

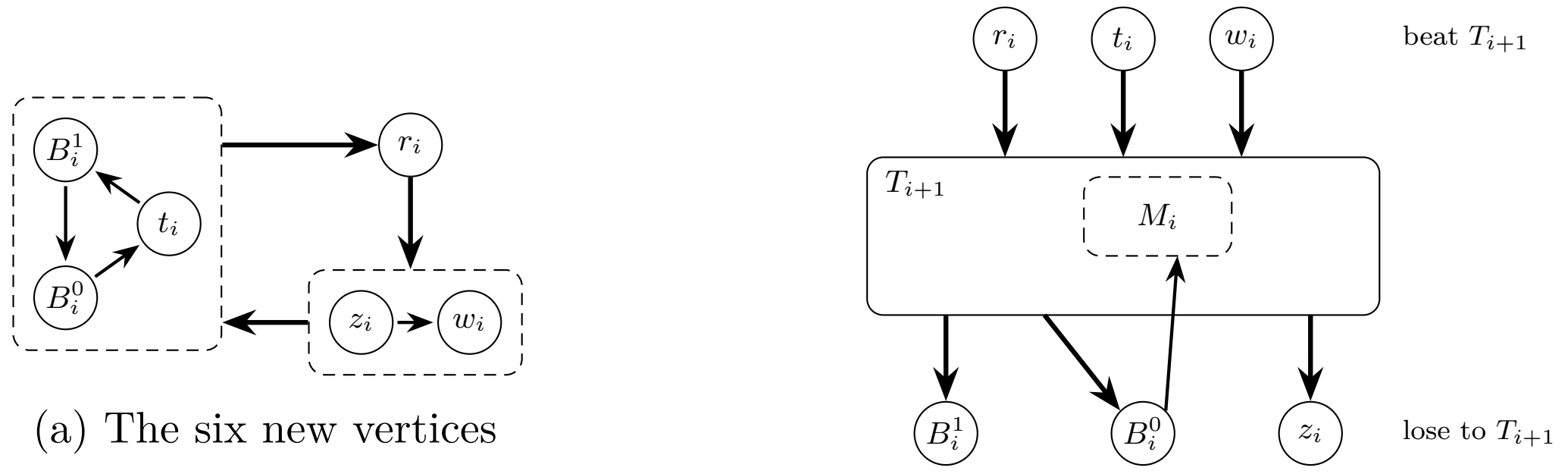
\begin{figure}[H]
\centering
\begin{minipage}[c]{0.44\linewidth}\centering
\begin{tikzpicture}[>=Stealth, scale=0.86, transform shape, v/.style={circle,draw,inner sep=0pt,minimum size=6.4mm,font=\footnotesize}]
\tikzset{arc/.style={->,thick,shorten >=1pt,shorten <=1pt},blk/.style={->,very thick}}
\node[v] (p) at (0,1.5) {$\rone_i$};
\node[v] (q) at (0,0) {$\rzero_i$};
\node[v] (t) at (1.05,0.75) {$\tri_i$};
\draw[arc] (p)--(q); \draw[arc] (q)--(t); \draw[arc] (t)--(p);
\node[draw,densely dashed,rounded corners,fit=(p)(q)(t),inner sep=5pt] (tri) {};
\node[v] (b) at (3.5,1.55) {$\broot_i$};
\node[v] (z) at (3.0,-0.25) {$\shz_i$};
\node[v] (w) at (4.1,-0.25) {$\shw_i$};
\draw[arc] (z)--(w);
\node[draw,densely dashed,rounded corners,fit=(z)(w),inner sep=5pt] (sh) {};
\draw[blk] (tri.east |- b) -- (b);
\draw[blk] (b) -- (b |- sh.north);
\draw[blk] (sh.west) -- (tri.east |- sh.west);
\end{tikzpicture}

\smallskip
(a) The six new vertices
\end{minipage}\hfill
\begin{minipage}[c]{0.54\linewidth}\centering
\begin{tikzpicture}[>=Stealth, font=\small, scale=0.86, transform shape,
  v/.style={circle,draw,inner sep=0pt,minimum size=6.4mm,font=\footnotesize},
  inn/.style={draw,densely dashed,rounded corners},
  blk/.style={->,very thick},
  arc/.style={->,thick}]
\draw[rounded corners] (-.2,-.9) rectangle (5.0,.7);
\node[anchor=north west,font=\footnotesize] at (-.15,.7) {$T_{i+1}$};
\node[v] (b) at (1.2,1.9) {$\broot_i$};
\node[v] (t) at (2.4,1.9) {$\tri_i$};
\node[v] (w) at (3.6,1.9) {$\shw_i$};
\node[v] (p) at (.6,-2.1) {$\rone_i$};
\node[v] (q) at (2.6,-2.1) {$\rzero_i$};
\node[v] (z) at (4.3,-2.1) {$\shz_i$};
\draw[inn] (2.0,-.3) rectangle (3.5,.5);
\node[font=\footnotesize] at (2.75,.1) {$M_i$};
\foreach \u in {b,t,w} \draw[blk] (\u.south) -- (\u.south |- 0,.7);
\draw[blk] (.6,-.9) -- (p.north);
\draw[blk] (4.3,-.9) -- (z.north);
\draw[blk] (1.6,-.9) -- (q.north west);
\draw[arc] (q.north east) -- (2.95,-.3);
\node[font=\scriptsize,anchor=west] at (5.1,1.9) {beat $T_{i+1}$};
\node[font=\scriptsize,anchor=west] at (5.1,-2.1) {lose to $T_{i+1}$};
\end{tikzpicture}

\smallskip
(b) New against old
\end{minipage}
\caption{The $\exists$-layer (\Cref{def:or}). (a) The edges among the six new vertices; a thick arrow between two groups means that every vertex of the first beats every vertex of the second. In the dominator set of either reader, $w_i$ and $z_i$ play the roles of $u$ and $v$, respectively, in the two-vertex shield of \Cref{lem:shield}. (b) The edges between the new vertices and the old vertices, that is, $T_{i+1}$. The vertices drawn above the box beat every old vertex, and those drawn below it lose to every old vertex; the thin arrow, which points upward, marks the exception.}
\label{fig:exists-layer}
\end{figure}

\begin{definition}[$\forall$-layer]\label{def:and}
Let $Q_i=\forall$, and let $\anc_i$ be the occurrence vertex of $x_i$ in the first clause that contains $x_i$; we call it the \emph{anchor} of the layer. It exists by~\ref{F3}, and it belongs to $M_i$. The tournament $T_i$ is obtained from $T_{i+1}$ by adding six vertices: the two \emph{readers} $\groot_i$ and $\rzero_i$, the new \emph{output} $\orole_i$, the \emph{link} $\slink_i$, and the two \emph{guards} $\king_i,\wanc_i$. The reader $\groot_i$ also serves as the \emph{root} of the layer. The new edges are oriented as follows.
\begin{itemize}[leftmargin=*,itemsep=1pt,topsep=3pt]
\item Among the new vertices: $\groot_i$ beats the other five. The other five are arranged in a circle, in the order $\eout_i,\king_i,\rzero_i,\wanc_i,\slink_i$, and each of them beats the next two vertices of the circle. For instance, $\eout_i$ beats $\king_i$ and $\rzero_i$, and $\slink_i$ beats $\eout_i$ and $\king_i$.
\item Against the old vertices: $\king_i$ beats every old vertex, and $\groot_i$ loses to every old vertex. The guard $\wanc_i$ beats every old vertex except the anchor $\anc_i$, which beats $\wanc_i$. The vertices $\eout_i$ and $\slink_i$ lose to every old vertex except the output $\orole_{i+1}$, which they beat. The reader $\rzero_i$ loses to every old vertex outside $M_i$ and beats every vertex of $M_i$.
\end{itemize}
\end{definition}

\begin{figure}[H]
\centering
\begin{minipage}[c]{0.44\linewidth}\centering
\begin{tikzpicture}[>=Stealth, scale=0.86, transform shape, v/.style={circle,draw,inner sep=0pt,minimum size=6.4mm,font=\footnotesize}]
\tikzset{arc/.style={->,thick,shorten >=1pt,shorten <=1pt}}
\node[v] (e) at (90:1.45)  {$\eout_i$};
\node[v] (r) at (18:1.45)  {$\king_i$};
\node[v] (q) at (-54:1.45) {$\rzero_i$};
\node[v] (w) at (-126:1.45){$\wanc_i$};
\node[v] (s) at (162:1.45) {$\slink_i$};
\draw[arc] (e)--(r); \draw[arc] (r)--(q); \draw[arc] (q)--(w); \draw[arc] (w)--(s); \draw[arc] (s)--(e);
\draw[arc] (e)--(q); \draw[arc] (r)--(w); \draw[arc] (q)--(s); \draw[arc] (w)--(e); \draw[arc] (s)--(r);
\draw[densely dashed,rounded corners] (0,0) circle (1.95);
\node[v] (g) at (-3.1,0) {$\groot_i$};
\draw[->,very thick] (g)--(-1.95,0);
\end{tikzpicture}

\smallskip
(a) The six new vertices
\end{minipage}\hfill
\begin{minipage}[c]{0.54\linewidth}\centering
\begin{tikzpicture}[>=Stealth, font=\small, scale=0.86, transform shape,
  v/.style={circle,draw,inner sep=0pt,minimum size=6.4mm,font=\footnotesize},
  inn/.style={draw,densely dashed,rounded corners},
  blk/.style={->,very thick},
  arc/.style={->,thick}]
\draw[rounded corners] (-.2,-.9) rectangle (5.0,.7);
\node[anchor=north west,font=\footnotesize] at (-.15,.7) {$T_{i+1}$};
\node[v] (r) at (1.6,1.9) {$\king_i$};
\node[v] (w) at (3.7,1.9) {$\wanc_i$};
\node[v] (e) at (.3,-2.1) {$\eout_i$};
\node[v] (s) at (1.4,-2.1) {$\slink_i$};
\node[v] (g) at (2.6,-2.1) {$\groot_i$};
\node[v] (q) at (4.1,-2.1) {$\rzero_i$};
\node[circle,fill,inner sep=1.3pt,label={[font=\scriptsize]above:$\orole_{i+1}$}] (o) at (.85,-.35) {};
\draw[inn] (3.3,-.3) rectangle (4.85,.55);
\node[font=\footnotesize] at (4.5,-.05) {$M_i$};
\node[circle,fill,inner sep=1.3pt,label={[font=\scriptsize]right:$\anc_i$}] (a) at (3.75,.22) {};
\draw[blk] (r.south) -- (1.6,.7);
\draw[blk] (w.south west) -- (3.45,.7);
\draw[arc] (a) -- (w.south);
\draw[blk] (.1,-.9) -- (e.north west);
\draw[arc] (e.north east) -- (o);
\draw[blk] (1.65,-.9) -- (s.north east);
\draw[arc] (s.north west) -- (o);
\draw[blk] (2.6,-.9) -- (g.north);
\draw[blk] (3.1,-.9) -- (q.north west);
\draw[arc] (q.north) -- (4.1,-.3);
\node[font=\scriptsize,anchor=west] at (5.1,1.9) {beat $T_{i+1}$};
\node[font=\scriptsize,anchor=west] at (5.1,-2.1) {lose to $T_{i+1}$};
\end{tikzpicture}

\smallskip
(b) New against old
\end{minipage}
\caption{The $\forall$-layer (\Cref{def:and}), drawn as in \Cref{fig:exists-layer}. (a) The vertex $\groot_i$ beats the five vertices on the circle, and each of these beats the next two clockwise. (b) The thin arrows mark the three exceptions: $\eout_i$ and $\slink_i$ beat the output $\orole_{i+1}$, the reader $\rzero_i$ beats $M_i$, and the anchor $\anc_i$ beats $\wanc_i$.}
\label{fig:forall-layer}
\end{figure}

\paragraph{The three steps together.}
For reference, we collect the three steps in one statement.

\begin{construction}[The tournament $T_1$ and its output $\orole_1$]\label{def:reduction}
Let $\Phi=Q_1x_1\cdots Q_nx_n\,(C_1\land\cdots\land C_m)$ satisfy~\ref{F1}--\ref{F3}.
\begin{enumerate}[label=\arabic*.,leftmargin=*,itemsep=3pt]
\item For each clause $C_\ell$, take its gadget, a fresh copy of $\Gad_{j_\ell}$ (\Cref{def:clause}).
\item Join the gadgets by the filters of $C_2,\ldots,C_m$ (\Cref{def:filter}), and set $T_{n+1}=\Cle{m}$ and $\orole_{n+1}=\cout_1$.
\item For $i=n,n-1,\ldots,1$, obtain $T_i$ and $\orole_i$ from $T_{i+1}$ and $\orole_{i+1}$ by adding the $\exists$-layer of \Cref{def:or} if $Q_i=\exists$, and the $\forall$-layer of \Cref{def:and} if $Q_i=\forall$.
\item Return $T_1$ and its output $\orole_1$.
\end{enumerate}
\end{construction}

Every edge is fixed directly by the syntax of $\Phi$; no TEQ computation is involved, and $T_1$ is computed in $O(N^2)$ time, where $N=8m+6n-2$ is its number of vertices. The construction only adds vertices at each step, so each $T_{i+1}$ is an induced subtournament of $T_i$. The output changes only at the $\forall$-layers: $\orole_i$ is a new vertex if $Q_i=\forall$, and $\orole_i=\orole_{i+1}$ if $Q_i=\exists$. The designated vertex $\orole_1$ is therefore the new output of the outermost $\forall$-layer, or $\cout_1$ if $\Phi$ has no universal quantifier. \Cref{fig:reduction-example} shows the vertices of the tournament for the running example~\eqref{eq:running}; its edges are those prescribed by the four definitions.

\subsection{How the tournament evaluates the formula}\label{sec:hard-ideas}

This subsection explains, without proofs, how $T_1$ evaluates $\Phi$. Recall that TEQ is defined recursively. To decide which vertices of a tournament $T$ are selected, TEQ calls itself on the dominator set $D(y)$ of every vertex $y$, that is, on the set of vertices that beat $y$. This call is the query $\TEQ(D(y))$ at $y$, and nothing else about $T$ is used. The edges of $T$ therefore decide which smaller tournaments the recursion visits. This is the only tool of the reduction. By choosing who beats whom, we choose these smaller tournaments, and we choose them so that each one corresponds to $\Phi$ under a partial assignment of its variables.

\paragraph{Assigning variables by deleting vertices.}
The set $M_i$ of \eqref{eq:Mi} contains the occurrence vertices of the variable $x_i$, positive and negative alike, and it represents $x_i$. Keeping $M_i$ stands for $x_i=1$, and deleting $M_i$ stands for $x_i=0$. An assignment $\alpha\in\{0,1\}^k$ to the first $k\le n$ variables thus corresponds to deleting
\[
Z_\alpha=\bigcup_{i\le k:\ \alpha_i=0}M_i.
\]
Each quantifier layer has two dedicated vertices, its readers, whose dominator sets control the value of its variable. In the layer of $x_i$, the dominator set of one reader contains $M_i$, and the dominator set of the other does not.

\paragraph{The two readers.}\label{sec:hard-reading}
Consider the layer of $Q_ix_i$. The reader $\Rdi{i}{1}$ loses to every vertex of $T_{i+1}$, so its dominator set $D(\Rdi{i}{1})$ contains all of $T_{i+1}$. The query at $\Rdi{i}{1}$ is thus a question about $T_{i+1}$, that is, about the rest of the formula with $x_i=1$. The reader $\Rdi{i}{0}$ beats the vertices of $M_i$ and loses to all other vertices of $T_{i+1}$, so $D(\Rdi{i}{0})$ contains $T_{i+1}-M_i$ and no vertex of $M_i$. The query at $\Rdi{i}{0}$ is a question about the rest of the formula with $x_i=0$. In this way the recursion of TEQ tries both values of $x_i$. 


For each of the two values, whether the rest of the formula, $Q_{i+1}x_{i+1}\cdots Q_nx_n\,\psi$, is true is encoded by whether the output $\orole_{i+1}$ belongs to TEQ. The rest of the formula is true with $x_i=1$ exactly when $\orole_{i+1}\in\TEQ(T_{i+1})$, and it is true with $x_i=0$ exactly when $\orole_{i+1}\in\TEQ(T_{i+1}-M_i)$. The other four vertices of the layer combine the two answers. If $Q_i=\exists$, then $\orole_i\in\TEQ(T_i)$ exactly when at least one of the two memberships holds. If $Q_i=\forall$, then $\orole_i\in\TEQ(T_i)$ exactly when both hold (\Cref{lem:or,lem:and}). All of this remains true after the occurrence vertices of outer variables have been deleted, that is, for every assignment to $x_1,\dots,x_{i-1}$.

\paragraph{The quantifier-free part.}
After the last quantifier, the rest of the formula is $\psi=C_1\land\cdots\land C_m$ and every variable has a value. The tournament is then the clause tournament $T_{n+1}$ with the vertices of $Z_\alpha$ deleted, for a complete assignment $\alpha\in\{0,1\}^n$, and its output is $\cout_1$. Keeping an occurrence vertex of $x_i$ means $x_i=1$, whether the literal is positive or negative. A positive literal is therefore true when its vertex is kept, and a negative literal is true when its vertex is deleted. So a clause is false exactly when the vertices of its positive literals are deleted and those of its negative literals are kept. For example, the clause $(\neg x_1\lor\neg x_2\lor x_3)$ of the running example is false only for $x_1=x_2=1$ and $x_3=0$, which keeps its first two occurrence vertices and deletes the third. The gadget of $C_\ell$ is designed so that its output $\cout_\ell$ is not in TEQ of the gadget for exactly this pattern (\Cref{lem:clause}). The filter of $C_\ell$ lets the common output $\cout_1$ stay in TEQ only if $\cout_\ell$ is in TEQ of its gadget (\Cref{lem:filter}). As a result, $\cout_1\in\TEQ(T_{n+1}-Z_\alpha)$ exactly when $\alpha$ satisfies every clause.

\begin{figure}[H]
\centering
\begin{tikzpicture}[x=1cm,y=1cm,>=Stealth,
  nd/.style={draw,rounded corners=2pt,fill=white,inner sep=2.5pt,font=\footnotesize},
  leaf/.style={nd,font=\scriptsize,inner sep=2pt},
  val/.style={font=\scriptsize,text=blue!60!black},
  rowlab/.style={font=\footnotesize,anchor=east,align=right},
  ed/.style={thick},
  el/.style={font=\scriptsize,fill=white,inner sep=1pt}]
\def\ya{0}\def\yb{-1.45}\def\yc{-2.9}\def\yd{-4.35}
\node[rowlab] at (-0.75,\ya) {layer $\forall x_1$:\\ $\eout_1\in\TEQ(\,\cdot\,)$? $=\land$};
\node[rowlab] at (-0.75,\yb) {layer $\exists x_2$:\\ $\cout_1\in\TEQ(\,\cdot\,)$? $=\lor$};
\node[rowlab] at (-0.75,\yc) {layer $\exists x_3$:\\ $\cout_1\in\TEQ(\,\cdot\,)$? $=\lor$};
\node[rowlab] at (-0.75,\yd) {clause tournament:\\ $\cout_1\in\TEQ(\,\cdot\,)$? $=\psi(\alpha)$};
\def\sp{1.58}
\foreach \k/\a/\v in {0/111/1,1/110/0,2/101/1,3/100/1,4/011/1,5/010/1,6/001/1,7/000/0} {
  \node[leaf] (L\k) at ({\k*\sp},\yd) {$T_4-Z_{\a}$};
  \node[val,below=1pt of L\k] {$\psi(\a)=\v$};
}
\foreach \k/\a in {0/11,1/10,2/01,3/00} {
  \pgfmathsetmacro{\xx}{(2*\k+0.5)*\sp}
  \node[nd] (C\k) at (\xx,\yc) {$T_3-Z_{\a}$};
  \node[val,anchor=west] at ([xshift=1pt]C\k.east) {$1$};
}
\foreach \k/\a in {0/1,1/0} {
  \pgfmathsetmacro{\xx}{(4*\k+1.5)*\sp}
  \node[nd] (B\k) at (\xx,\yb) {$T_2-Z_{\a}$};
  \node[val,anchor=west] at ([xshift=1pt]B\k.east) {$1$};
}
\pgfmathsetmacro{\xx}{3.5*\sp}
\node[nd] (A) at (\xx,\ya) {$T_1$};
\node[val,anchor=west] at ([xshift=1pt]A.east) {$1$};
\draw[ed] (A) -- node[el,pos=0.45] {$x_1{=}1$} (B0);
\draw[ed] (A) -- node[el,pos=0.45] {$x_1{=}0$} (B1);
\foreach \k in {0,1} {
  \pgfmathtruncatemacro{\l}{2*\k}\pgfmathtruncatemacro{\r}{2*\k+1}
  \draw[ed] (B\k) -- node[el,pos=0.5] {$x_2{=}1$} (C\l);
  \draw[ed] (B\k) -- node[el,pos=0.5] {$x_2{=}0$} (C\r);
}
\foreach \k in {0,1,2,3} {
  \pgfmathtruncatemacro{\l}{2*\k}\pgfmathtruncatemacro{\r}{2*\k+1}
  \draw[ed] (C\k) -- node[el,pos=0.5] {$1$} (L\l);
  \draw[ed] (C\k) -- node[el,pos=0.5] {$0$} (L\r);
}
\end{tikzpicture}
\caption{The evaluation tree of $\Phi_{\mathrm{ex}}$ in the tournament $T_1$ of \Cref{fig:reduction-example}. Each node asks whether the output of the tournament written in it belongs to TEQ, with the answer in blue. The left child is the question asked by the reader $\Rdi{i}{1}$ of the layer, and the right child is the one asked by $\Rdi{i}{0}$. The answers are combined by $\land$ at the layer of $\forall x_1$ and by $\lor$ at the layers of $\exists x_2$ and $\exists x_3$.}
\label{fig:recursion-tree}
\end{figure}
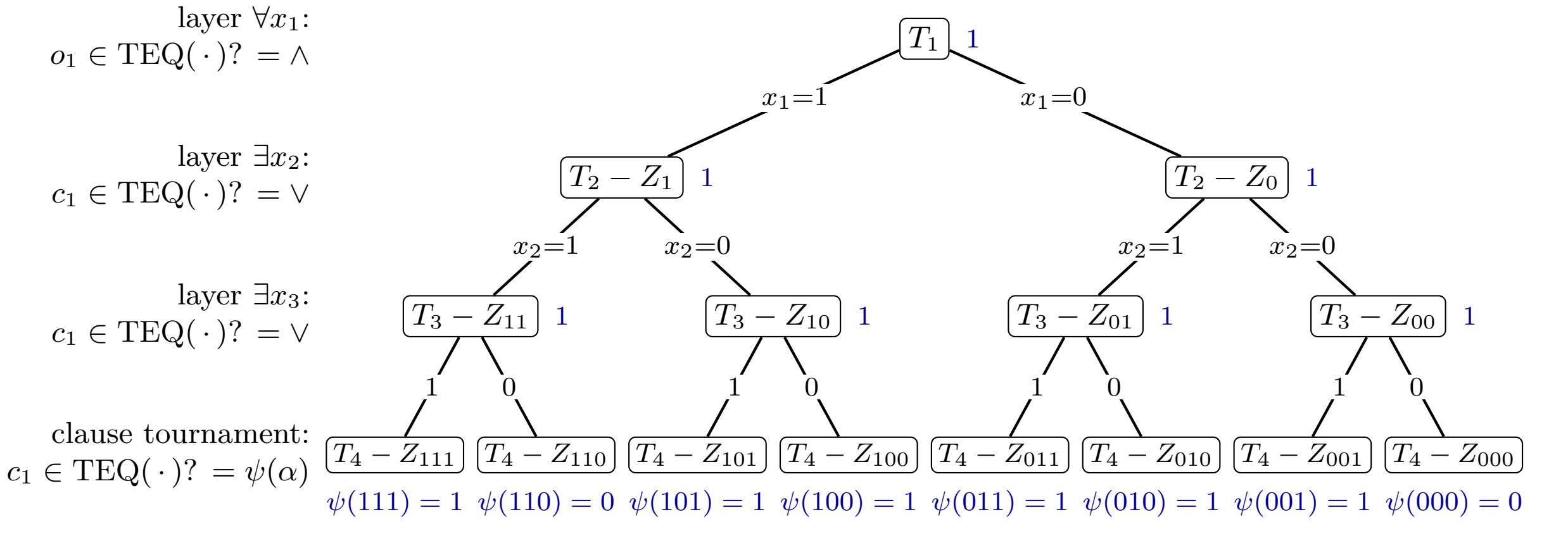

\paragraph{The evaluation tree.}
In summary, following the readers from the outermost layer inward evaluates $\Phi$ one variable at a time. The readers of the first layer ask about $T_2$ and $T_2-M_1$. The readers of the second layer ask about each of these two tournaments with and without $M_2$, and so on inward. After $i-1$ steps the question is
\[
\text{is }\orole_i\in\TEQ(T_i-Z_\alpha)\,?
\]
where $\alpha\in\{0,1\}^{i-1}$ records which readers were followed, and after $n$ steps it is the question about the quantifier-free part (\Cref{prop:matrix}). These questions form the \emph{evaluation tree} of $\Phi$, shown in \Cref{fig:recursion-tree} for the running example. The tree has $2^n$ leaves, whereas $T_1$ has only $8m+6n-2$ vertices, because every leaf is a subtournament of the same tournament $T_1$.

The tree shows only two queries per layer, whereas TEQ has a query at every vertex of every tournament it visits. \Cref{sec:correctness} shows that the other queries do not interfere, and that the two answers of a layer combine to exactly $\lor$ or $\land$.
\section{Correctness of the reduction}\label{sec:correctness}

In this section, we prove that the construction of \Cref{sec:hardness} is correct.

\begin{theorem}[Correctness of the reduction]\label{thm:correct}
Let $\Phi$ be a quantified Boolean formula satisfying~\ref{F1}--\ref{F3}, and let $T_1$ and $\orole_1$ be as in \Cref{def:reduction}. Then $\Phi$ is true if and only if $\orole_1\in\TEQ(T_1)$.
\end{theorem}

Together with the normal form of \Cref{lem:normal} and the polynomial-space algorithm of \Cref{prop:pspace}, this yields \Cref{thm:main}. The proof follows the construction from the inside out. After collecting a few tools (\Cref{sec:hard-tools}), we show that the clause tournament evaluates $\psi$ (\Cref{sec:hard-matrix}), that the $\exists$-layer computes a disjunction (\Cref{sec:hard-or}), and that the $\forall$-layer computes a conjunction (\Cref{sec:hard-and}). Each of these statements holds for arbitrary inputs to the gadget, so the final step is a short induction over the quantifiers (\Cref{sec:hard-reduction}). Proofs omitted from this section appear in \Cref{hard:app}.

\subsection{Tools}\label{sec:hard-tools}

Two special cases of the composition behavior of TEQ~\cite{LLL1996} are used repeatedly. In the first, a block that beats everything else is all that matters. In the second, three blocks that beat one another in a cycle each keep their selected vertices.

\begin{lemma}[Dominant set]\label{hard:dominant}
Let $A$ and $B$ be disjoint tournaments with $A\ne\varnothing$ and $A\succ B$. Then $\TEQ(A\cup B)=\TEQ(A)$.
\end{lemma}

\begin{lemma}[Three-block composition]\label{hard:cyclic}
Let $A$, $B$, and $C$ be nonempty, pairwise disjoint tournaments with $A\succ B\succ C\succ A$. Then $\TEQ(A\cup B\cup C)=\TEQ(A)\cup\TEQ(B)\cup\TEQ(C)$.
\end{lemma}

The layers repeatedly face the following situation: a query contains a part $A$ whose selected vertices we want to read off, together with a part $B$ that we would like to ignore, and we have no control over the edges between $A$ and $B$. Two extra vertices make $B$ irrelevant.

\begin{figure}[H]
\centering
\begin{tikzpicture}[>=Stealth, x=0.85cm, y=0.8cm]
\tikzset{arc/.style={->,thick},blk/.style={draw,ellipse,minimum width=1.7cm,minimum height=1cm}}
\node[blk] (A) at (0,0) {$A$};
\node[blk] (B) at (3.6,0) {$B$};
\node[circle,inner sep=1.5pt] (u) at (1.8,1.5) {$u$};
\node[circle,inner sep=1.5pt] (v) at (1.8,-1.5) {$v$};
\draw[arc] (u)--(A);
\draw[arc] (u)--(B);
\draw[arc] (A)--(v);
\draw[arc] (v)--(B);
\draw[arc] (v)--(u);
\draw[<->,densely dashed] (A)--node[above,font=\scriptsize]{arbitrary}(B);
\end{tikzpicture}
\caption{The two-vertex shield. The arrows are tournament edges between blocks. When $A\ne\varnothing$, the unique source component of the contestation graph is the cycle $\TEQ(A)\to v\to u\to\TEQ(A)$, and every vertex of $B$ lies downstream of $v$.}
\label{fig:shield}
\end{figure}
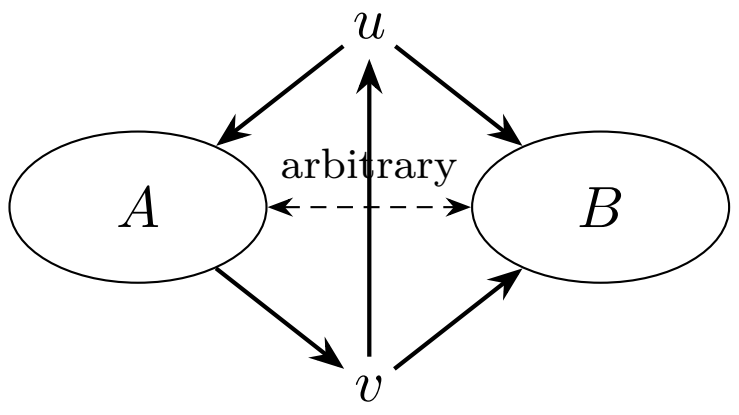

\begin{lemma}[Two-vertex shield]\label{hard:shield}\label{lem:shield}
Let $A$ and $B$ be disjoint tournaments with arbitrary edges between them, and add two vertices $u,v$ with
\[
u\succ A\cup B,\qquad A\succ v\succ u,\qquad v\succ B
\]
(\Cref{fig:shield}). Then
\[
\TEQ(A\cup B\cup\{u,v\})=
\begin{cases}
\{v\},&A=\varnothing,\\
\TEQ(A)\cup\{u,v\},&A\ne\varnothing.
\end{cases}
\]
\end{lemma}

\begin{proof}
We induct on $|B|$. If $A=\varnothing$, then $v$ is a Condorcet winner. Otherwise the query at $v$ is $A$, the query at $u$ is $\{v\}$, and every query at a vertex $x\in A$ has Condorcet winner $u$, which gives the cycle of blocks $\TEQ(A)\R v\R u\R\TEQ(A)$. For $y\in B$, the query $D(y)$ is a smaller instance of the same construction, with parts $A\cap D(y)$ and $B\cap D(y)$. Its first part may be empty, but $v$ is selected in both cases of the statement, so by induction $v\R y$. It remains to check that nothing outside the cycle points into it. A vertex of $B$ loses to $u$ and $v$, and it cannot point into $A$ because queries at vertices of $A$ select only $u$. A vertex of $A\setminus\TEQ(A)$ does not point to $v$. Hence the cycle is the unique source component.
\end{proof}

Finally, we record how the gadgets behave under the deletions that encode assignments. For this purpose, and for the lemmas below, we need the gadgets in a more general form than \Cref{def:filter,def:or,def:and}, in which the tournament built so far is replaced by an arbitrary tournament. We drop the indices $\ell$ and $i$ from the names of the new vertices.

\emph{The filter.} The four rules of \Cref{def:filter} do not use anything about clauses: they can be applied to any two disjoint tournaments $H$ and $K$ with outputs $\dout$ and $\cout$, in place of $\Cle{\ell-1}$ and the gadget of $C_\ell$. We write $\Filt(H,\dout;K,\cout)$ for the resulting tournament, with test $\test$, relay $\relay$ and output $\dout$. Thus $\Cle{\ell}=\Filt(\Cle{\ell-1},\cout_1;K,\cout_\ell)$, where $K$ is the gadget of $C_\ell$.

\emph{The layers.} Let $H$ be a tournament with output $\dout$; we call its vertices the old vertices. For two sets $\Hs{1},\Hs{0}\subseteq V(H)$ that contain $\dout$, the tournament $\Wex(H,\dout;\Hs{1},\Hs{0})$ extends $H$ by the six vertices $\rone,\rzero,\broot,\tri,\shw,\shz$ with the dominators in the table on the left, and its output is $\dout$. For a nonempty set $M\subseteq V(H)\setminus\{\dout\}$ and a vertex $\anc\in M$, the tournament $\Wall(H,\dout;M,\anc)$ extends $H$ by the six vertices $\eout,\slink,\groot,\rzero,\king,\wanc$ with the dominators in the table on the right, where $\Hs{1}=V(H)$ and $\Hs{0}=V(H)\setminus M$, and its output is $\eout$.
\begin{center}
\small
\renewcommand{\arraystretch}{1.12}
\begin{tabular}{@{}c@{\quad}l@{}}
\toprule
\multicolumn{2}{@{}l}{$\Wex(H,\dout;\Hs{1},\Hs{0})$}\\
\midrule
$\rone $ & $\Hs{1}\cup\{\shw ,\shz ,\tri \}$\\
$\rzero $ & $\Hs{0}\cup\{\shw ,\shz ,\rone \}$\\
$\broot $ & $\{\rone ,\rzero ,\tri \}$\\
$\tri $ & $\{\rzero ,\shw ,\shz \}$\\
$\shw $ & $\{\broot,\shz \}$\\
$\shz $ & $V(H)\cup\{\broot\}$\\
\bottomrule
\end{tabular}
\qquad\qquad
\begin{tabular}{@{}c@{\quad}l@{}}
\toprule
\multicolumn{2}{@{}l}{$\Wall(H,\dout;M,\anc)$}\\
\midrule
$\eout $ & $(V(H)\setminus\{\dout \})\cup\{\groot ,\slink ,\wanc \}$\\
$\slink $ & $(V(H)\setminus\{\dout \})\cup\{\groot ,\rzero ,\wanc \}$\\
$\groot $ & $V(H)$\\
$\rzero $ & $\Hs{0}\cup\{\eout ,\king ,\groot \}$\\
$\king $ & $\{\eout ,\slink ,\groot \}$\\
$\wanc $ & $\{\anc ,\king ,\groot ,\rzero \}$\\
\bottomrule
\end{tabular}
\end{center}
We abbreviate $\Wex(H,\dout;M)=\Wex\bigl(H,\dout;\,V(H),\,V(H)\setminus M\bigr)$. With this notation, the layers of \Cref{def:or,def:and} are $T_i=\Wex(T_{i+1},\orole_{i+1};M_i)$ and $T_i=\Wall(T_{i+1},\orole_{i+1};M_i,\anc_i)$. The $\exists$-layer is stated for two arbitrary sets $\Hs{1},\Hs{0}$ because this costs nothing and is used again in \Cref{sec:d2}.

Each definition orients the edges between a new vertex and an old vertex $h$ according to whether $h$ belongs to certain sets of old vertices, such as $\Hs{0}$, $\{\dout \}$, or $\{\anc \}$. Deleting old vertices only shrinks these sets, so as long as the individually named vertices survive, the deletion commutes with the construction.

\begin{observation}[Deletions inside gadgets]\label{obs:deletion}
Let $Z$ be a set of vertices of $H$, or of $H$ and $K$ in part~(a).
\begin{enumerate}[label=\textup{(\alph*)}]
\item If $\dout ,\cout \notin Z$, then $\Filt(H,\dout ;K,\cout )-Z=\Filt(H-Z,\dout ;K-Z,\cout )$.
\item If $\dout \notin Z$, then $\Wex(H,\dout ;\Hs{1},\Hs{0})-Z=\Wex(H-Z,\dout ;\Hs{1}\setminus Z,\Hs{0}\setminus Z)$.
\item If $\dout ,\anc \notin Z$, then $\Wall(H,\dout ;M,\anc )-Z=\Wall(H-Z,\dout ;M\setminus Z,\anc )$.
\end{enumerate}
\end{observation}

\subsection{Evaluating the clauses}\label{sec:hard-matrix}

Recall from \Cref{sec:hard-ideas} that an occurrence vertex is present exactly when its variable has value $1$, so that a clause is false exactly when its surviving occurrence vertices are those of its negative literals and no others. For example, the clause $(\neg x_1\lor\neg x_2\lor x_3)$ is false only for $x_1=x_2=1$ and $x_3=0$, that is, only when the occurrence vertices of $\neg x_1$ and $\neg x_2$ survive and that of $x_3$ is deleted. We first verify that each clause gadget recognizes this falsifying pattern of its clause. For $S\subseteq\{\oc{1},\oc{2},\oc{3}\}$, write $\Gad_j^S$ for the subtournament of $\Gad_j$ induced by $\{\cout,\ax{1},\ax{2}\}\cup S$, that is, the gadget in which exactly the occurrence vertices in $S$ survive.

\begin{lemma}[Clause gadgets]\label{lem:clause}\label{hard:clause}
For every $j\in\{0,1,2\}$ and every $S\subseteq\{\oc{1},\oc{2},\oc{3}\}$,
\[
\cout\notin\TEQ(\Gad_j^S)\quad\Longleftrightarrow\quad S=\{\oc{1},\ldots,\oc{j}\}.
\]
For $j=0$ the set on the right is empty.
\end{lemma}

\begin{proof}[Proof idea]
In $\Gad_0$ the output beats the occurrence vertices, which beat the auxiliary vertices, which in turn beat the output. If $S=\varnothing$, the auxiliary vertex $\ax{2}$ beats every other vertex and the output is not selected. If $S\ne\varnothing$, then $\{\cout\}\succ S\succ\{\ax{1},\ax{2}\}\succ\{\cout\}$, and \Cref{hard:cyclic} selects $\cout$. In $\Gad_1$ and $\Gad_2$ the negative occurrence vertices beat the output, and the remaining edges ensure that exactly the falsifying choice of $S$ excludes it. \Cref{hard:clause-proof} checks all cases.
\end{proof}

The filter joins the gadget of a clause to the clause tournament built from the earlier clauses, and its name describes what it does: the old output passes the filter, and remains selected, only if the output of the new part is selected.

\begin{lemma}[Filter]\label{lem:filter}\label{hard:filter}
Let $H$ and $K$ be disjoint tournaments with outputs $\dout$ and $\cout$. Then
\[
\dout \in\TEQ(\Filt(H,\dout ;K,\cout ))\quad\Longleftrightarrow\quad \dout \in\TEQ(H)\ \text{and}\ \cout \in\TEQ(K).
\]
\end{lemma}

\begin{proof}[Proof idea]
The query at the test $\test $ is $K$, so $\cout \R \test $ holds exactly when $\cout \in\TEQ(K)$, and this is the only possible contestation edge leaving $\cout $. A vertex $h\in H$ can leave $H$ only through $h\R \cout $, and the query at $\cout $ selects exactly $\TEQ(H)$ within $H$. If $\cout \in\TEQ(K)$, every $h\in\TEQ(H)$ follows $h\R \cout \R \test \R \relay $ and from there reaches every vertex. Otherwise $\cout $ has no exit, and $K\setminus\{\cout \}$ is nonempty because the only vertex of a one-vertex tournament is selected; every vertex of $H$ then receives a contestation edge from $K\setminus\{\cout \}$ and can leave $H$ only towards $\cout $, so no vertex of $H$ is selected. \Cref{hard:clause-proof} gives the details.
\end{proof}

Combining the two lemmas gives the answer of the clause tournament under every complete assignment.

\begin{proposition}[Clause tournament]\label{prop:matrix}
For every assignment $\alpha\in\{0,1\}^n$,
\[
\cout_1\in\TEQ(T_{n+1}-Z_\alpha)\quad\Longleftrightarrow\quad\psi(\alpha)=1.
\]
\end{proposition}

\begin{proof}
The set $Z_\alpha$ contains only occurrence vertices, so by \Cref{obs:deletion}(a), $T_{n+1}-Z_\alpha$ is built from the reduced clause gadgets, that is, the gadgets of $C_1,\ldots,C_m$ without the vertices of $Z_\alpha$, by the same chain of filters. The surviving occurrence vertices of the gadget of $C_\ell$ are those whose variables have value $1$ under $\alpha$. Since the three variables of $C_\ell$ are distinct, the clause is false exactly when the survivors are the occurrence vertices $\ocl{\ell}{1},\ldots,\ocl{\ell}{j_\ell}$ of its negative literals, and so by \Cref{lem:clause}, $\cout_\ell$ is selected in the reduced gadget of $C_\ell$ exactly when $C_\ell$ is true under $\alpha$. Applying \Cref{lem:filter} to the filters in the order of construction, $\cout_1$ is selected exactly when all clauses are true.
\end{proof}

\subsection{The \texorpdfstring{$\exists$}{exists}-layer}\label{sec:hard-or}

\begin{figure}[H]
\centering
\begin{tikzpicture}[x=1cm,y=1cm,font=\footnotesize,>=Stealth]
\tikzset{
  occi/.style={circle,draw,fill=blue!65!black,inner sep=0pt,minimum size=4pt},
  outv/.style={rectangle,draw,fill=black,inner sep=0pt,minimum size=4.5pt},
  reader/.style={circle,draw,fill=white,inner sep=1pt,minimum size=15pt,font=\small},
  blk/.style={->,very thick},
  lab/.style={font=\scriptsize,align=center}
}
\fill[blue!10,even odd rule] (0,0) rectangle (7.0,2.2) (4.8,0.25) rectangle (6.75,1.45);
\draw (0,0) rectangle (7.0,2.2);
\node[anchor=north west] at (0.05,2.17) {$H=\Hs{1}$};
\node[anchor=south west,text=blue!55!black] at (0.05,0.03) {$\Hs{0}=H-M_i$ (shaded)};
\node[outv] (d) at (2.3,1.15) {};
\node[anchor=west,inner sep=1.5pt] at (d.east) {$\dout $};
\draw[densely dashed,fill=white] (4.8,0.25) rectangle (6.75,1.45);
\node[anchor=north east] at (6.73,1.43) {$M_i$};
\node[occi] at (5.2,0.62) {}; \node[occi] at (5.6,0.62) {}; \node[occi] at (6.0,0.62) {};
\node[reader] (L) at (1.0,3.7) {};
\node[anchor=south] at (L.north) {reader $B^1$};
\draw[blk] (1.0,2.2) -- (L);
\node[lab,anchor=east] at (0.85,2.95) {every vertex\\of $H$ beats it};
\node[reader] (S) at (4.6,3.7) {};
\node[anchor=south] at (S.north) {reader $B^0$};
\draw[blk,blue!55!black] (3.5,2.2) -- (S);
\node[lab,text=blue!55!black,anchor=east] at (3.75,3.05) {every vertex\\of $\Hs{0}$ beats it};
\draw[blk] (S) -- (5.6,1.45);
\node[lab,anchor=west] at (5.35,2.95) {it beats every\\vertex of $M_i$};
\end{tikzpicture}
\caption{Reading both states of a single copy of $H$. The arrows are tournament edges between a new vertex and a set of old vertices. The reader $\Rd{1}$ loses to all of $H=\Hs{1}$. The reader $\Rd{0}$ loses to $\Hs{0}=H-M_i$ (shaded) and beats $M_i$, so its dominators from $H$ are exactly $\Hs{0}$, while $M_i$ remains in the tournament.}
\label{fig:layer}
\end{figure}
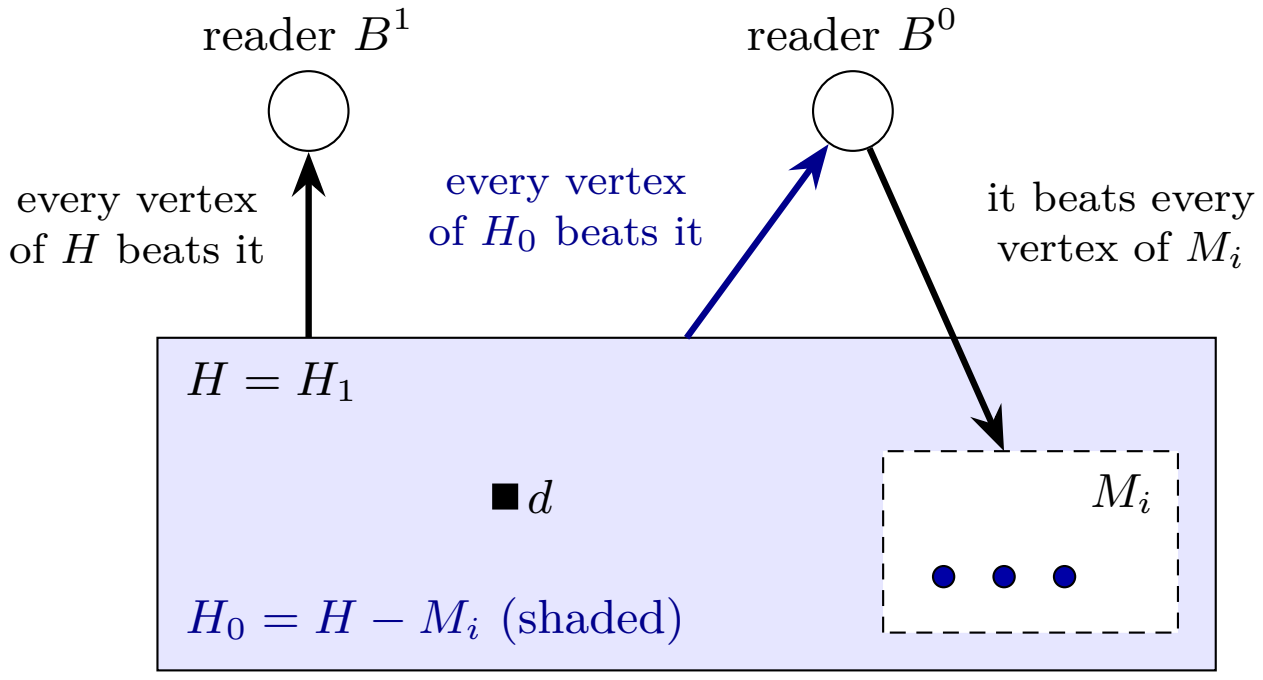

A layer must obtain two answers from a single copy of the inner tournament $H$, namely
\[
\bet{1}=\iverson{\dout \in\TEQ(\Hs{1})}\quad\text{and}\quad\bet{0}=\iverson{\dout \in\TEQ(\Hs{0})},
\]
where $\iverson{\cdot}$ denotes the truth value, $1$ or $0$, of the statement inside; in the reduction, $\Hs{1}=V(H)$ stands for $x_i=1$ and $\Hs{0}=V(H)\setminus M_i$ for $x_i=0$. The two readers provide them (\Cref{fig:layer}). The reader $\Rd{b}$ has all of $\Hs{b}$ among its dominators, together with up to three vertices of its own layer. These extra vertices are chosen so that the query at $\Rd{b}$ still reveals whether $\dout$ is selected in $\Hs{b}$; this is the purpose of the shield lemma. As a result, the contestation edge $\dout \R\Rd{1}$ is present exactly when $\bet{1}=1$, and the edge $\dout \R\Rd{0}$ exactly when $\bet{0}=1$.

\begin{figure}[H]
\centering
\begin{tikzpicture}[>=Stealth, every node/.style={inner sep=2pt}]
\tikzset{rel/.style={->,thick,blue!60!black}, lab/.style={font=\scriptsize,sloped}}
\begin{scope}
\node at (1.9,1.75) {\small $\exists$-layer: $\bet{0}\lor\bet{1}$};
\node (od) at (0,0) {$\dout $};
\node (op) at (1.9,.75) {$\rone $};
\node (oq) at (1.9,-.75) {$\rzero $};
\node (ob) at (3.8,0) {$\broot $};
\draw[rel] (od)--node[lab,above] {$\bet{1}$}(op);
\draw[rel] (od)--node[lab,below] {$\bet{0}$}(oq);
\draw[rel] (op)--(ob);
\draw[rel] (oq)--(ob);
\node[font=\scriptsize,left=1pt of od] {output};
\node[font=\scriptsize,above=0pt of op] {reader};
\node[font=\scriptsize,below=0pt of oq] {reader};
\node[font=\scriptsize,right=1pt of ob] {root};
\draw[rel,dashed] (ob) .. controls (4.9,-2.3) and (-1.1,-2.3) .. (od);
\end{scope}
\begin{scope}[xshift=6.6cm]
\node at (3.2,1.75) {\small $\forall$-layer: $\bet{0}\land\bet{1}$};
\node (ae) at (0,0) {$\eout $};
\node (aq) at (1.6,0) {$\rzero $};
\node (as) at (3.2,0) {$\slink $};
\node (ad) at (4.8,0) {$\dout $};
\node (ag) at (6.4,0) {$\groot $};
\draw[rel] (ae)--node[above,font=\scriptsize] {$\bet{0}$}(aq);
\draw[rel] (aq)--(as);
\draw[rel] (as)--(ad);
\draw[rel] (ad)--node[above,font=\scriptsize] {$\bet{1}$}(ag);
\node[font=\scriptsize,above=1pt of ae,align=center] {new\\output};
\node[font=\scriptsize,above=1pt of aq,align=center] {reader};
\node[font=\scriptsize,above=1pt of as] {link};
\node[font=\scriptsize,above=1pt of ad,align=center] {old\\output};
\node[font=\scriptsize,above=1pt of ag,align=center] {root,\\reader};
\draw[rel,dashed] (ag) to[bend left=24] (ae);
\end{scope}
\end{tikzpicture}
\caption{The intended return paths in the contestation graph. An edge labelled $\bet{b}$ is present exactly when $\bet{b}=1$, unlabelled solid edges are always present, and the dashed edge indicates that the root reaches every vertex. The role of each vertex is written next to its name. In the $\forall$-layer one vertex serves both as the root and as the reader $\Rd{1}$.}
\label{fig:quantifier-paths}
\end{figure}
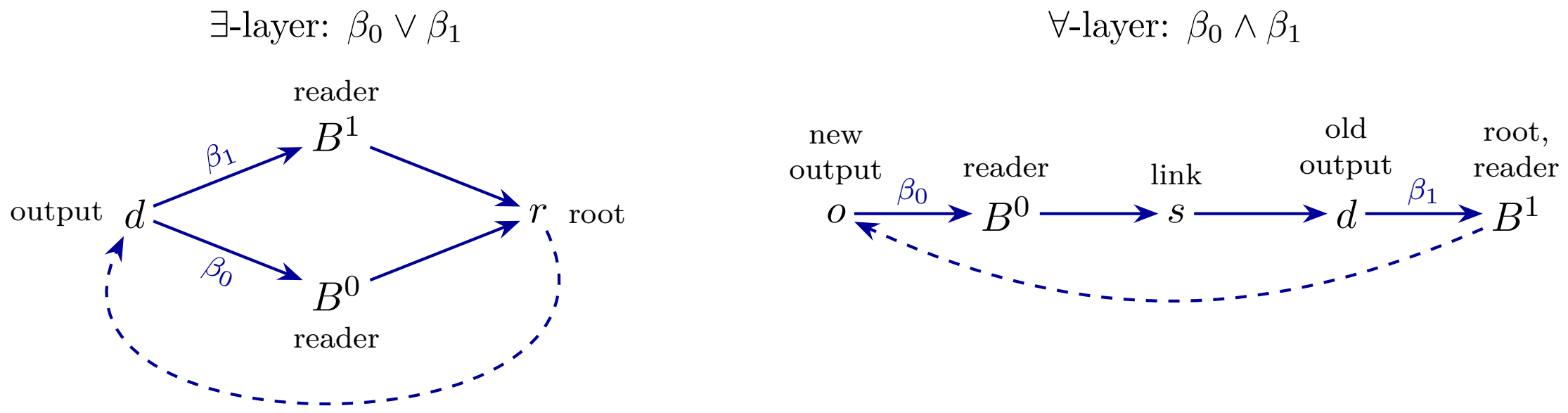

It remains to turn these two edges into the selection of a single vertex, and the return-path criterion (\Cref{lem:return}) does this. Each layer has a root that reaches every vertex of the contestation graph, so a vertex is selected exactly when it can reach the root. In an $\exists$-layer the output remains $\dout $, and it has two paths back to the root $\broot$, one through each reader (\Cref{fig:quantifier-paths}, left). The second half of each path is always present: the readers and the guard $\tri$ form a cycle $\rone\succ\rzero\succ\tri\succ\rone$, this cycle is the query at the root, and all three vertices of a cycle are selected, so both readers have a contestation edge to the root. Hence $\dout$ is selected exactly when $\bet{0}\lor\bet{1}$.

\begin{lemma}[$\exists$-layer]\label{lem:or}\label{hard:or}
Let $H$ be a tournament with output $\dout$, let $\Hs{0},\Hs{1}\subseteq V(H)$ contain $\dout $, and put $\bet{b}=\iverson{\dout \in\TEQ(\Hs{b})}$ for $b\in\{0,1\}$. Then
\[
\dout \in\TEQ(\Wex(H,\dout ;\Hs{1},\Hs{0}))\quad\Longleftrightarrow\quad\bet{0}\lor\bet{1}.
\]
\end{lemma}

\begin{proof}[Proof outline]
Let $W=\Wex(H,\dout ;\Hs{1},\Hs{0})$. We determine the contestation graph of $W$ query by query and then apply the return-path criterion with root $\broot $. The shield lemma does most of the work, in two different roles.

\emph{The readers read their sets.} The query $D(\rone )=\Hs{1}\cup\{\shw ,\shz ,\tri \}$ is a shield instance (\Cref{lem:shield}) with guards $u=\shw $ and $v=\shz $, kept part $\Hs{1}$, and hidden part $\{\tri \}$. Since $\Hs{1}$ contains $\dout $, it is nonempty, so $\TEQ(D(\rone ))=\TEQ(\Hs{1})\cup\{\shw ,\shz \}$. In the same way, $\TEQ(D(\rzero ))=\TEQ(\Hs{0})\cup\{\shw ,\shz \}$. Hence $\dout \R \rone $ holds exactly when $\bet{1}=1$, and $\dout \R \rzero $ exactly when $\bet{0}=1$.

\emph{The old vertices are silenced.} For an old vertex $h$, the query $D(h)$ consists of $D_H(h)$, the vertices $\broot $ and $\shw $, and those of $\rone ,\rzero ,\tri $ that beat $h$. It is again a shield instance, now with guards $u=\shw $ and $v=\broot$, kept part $D(h)\cap\{\rone ,\rzero ,\tri \}$, which always contains $\tri $, and hidden part $D_H(h)$. Consequently $\broot \R h$, and no old vertex has a contestation edge to another old vertex.

\emph{The root reaches every vertex.} The remaining queries are small. The query $D(\broot )=\{\rone ,\rzero ,\tri \}$ is the cycle $\rone \succ \rzero \succ \tri \succ \rone $ and selects all three vertices; the queries $D(\shw )=\{\broot,\shz \}$ and $D(\shz )=V(H)\cup\{\broot\}$ have Condorcet winner $\broot $; and $D(\tri )=\{\rzero ,\shw ,\shz \}$ has Condorcet winner $\shz $. Thus $\broot \R \shw $, $\broot \R \shz $, $\shz \R \rone ,\rzero ,\tri $, and $\broot \R h$ for every old vertex $h$, so $\broot $ reaches every vertex.

\emph{The output has only the two intended exits.} The out-neighbors of $\dout $ are $\shz $, the two readers, and some old vertices. The query at $\shz $ selects only $\broot $, and old vertices are silenced, so the only possible exits from $\dout $ are $\dout \R \rone $ and $\dout \R \rzero $, both of which continue to $\broot $. By \Cref{lem:return}, $\dout $ is selected exactly when $\bet{0}\lor\bet{1}$. \Cref{hard:wrapper-proofs} records the remaining details.
\end{proof}

\subsection{The \texorpdfstring{$\forall$}{forall}-layer}\label{sec:hard-and}

In a $\forall$-layer the output passes to the new vertex $\eout $, whose only path back to the root $\groot$ is $\eout\R\rzero\R\slink\R\dout\R\groot$ (\Cref{fig:quantifier-paths}, right). The path contains two tests in series, so that $\eout $ is selected exactly when $\bet{0}\land\bet{1}$. The second test is the edge from $\dout$ to the reader $\groot$. The first test is an edge to the reader $\rzero$ that starts at $\eout$ rather than at $\dout$. It nevertheless carries the answer $\bet{0}$: the vertices $\eout $ and $\slink $ both beat $\dout $ and lose to every other old vertex, and this lets $\eout $ stand in for $\dout$ inside the query at $\rzero $ and lets $\slink $ connect $\rzero $ back to $\dout $.

\begin{lemma}[$\forall$-layer]\label{lem:and}\label{hard:and}
Let $H$ be a tournament with output $\dout$, let $M\subseteq V(H)\setminus\{\dout \}$ be nonempty, and let $\anc \in M$. Put $\Hs{1}=V(H)$, $\Hs{0}=V(H)\setminus M$, and $\bet{b}=\iverson{\dout \in\TEQ(\Hs{b})}$ for $b\in\{0,1\}$. Then
\[
\eout \in\TEQ(\Wall(H,\dout ;M,\anc ))\quad\Longleftrightarrow\quad\bet{0}\land\bet{1}.
\]
\end{lemma}

The proof establishes each edge of this path, excludes every other path from $\eout $ back to $\groot $, and shows that $\groot $ reaches every vertex.

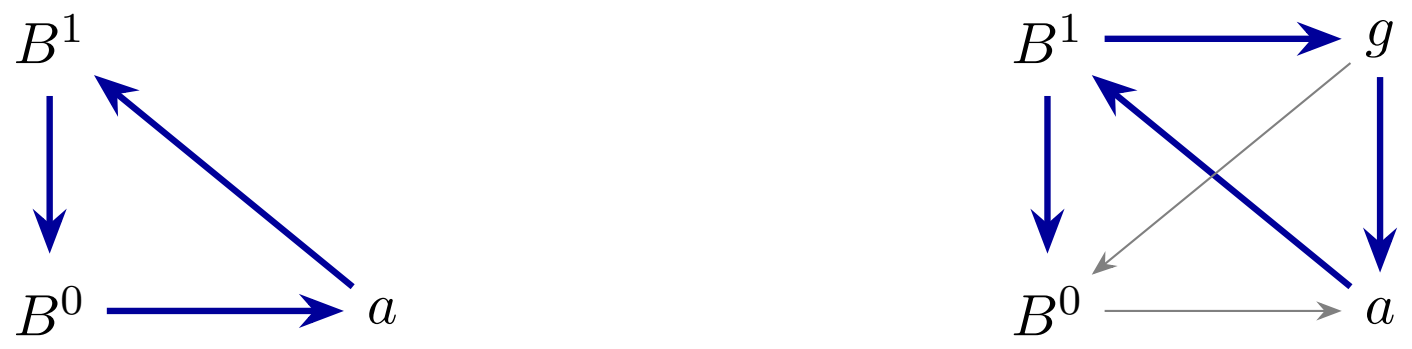
\begin{figure}[H]
\centering
\begin{tikzpicture}[>=Stealth, every node/.style={circle,inner sep=1pt,minimum size=12pt}]
\tikzset{arc/.style={->,gray,thin,shorten >=1pt,shorten <=1pt},
         rel/.style={->,very thick,blue!60!black,shorten >=1pt,shorten <=1pt}}
\begin{scope}
\node (g) at (0,1.8) {$\groot $};
\node (q) at (0,0) {$\rzero $};
\node (m) at (2.2,0) {$\anc $};
\draw[rel] (q) -- (m); \draw[rel] (m) -- (g); \draw[rel] (g) -- (q);
\node[draw=none,rectangle,align=center,font=\small] at (1.1,-0.95) {(a) dominators of $\wanc $ inside $D(\slink )$: $\{\rzero ,\anc ,\groot \}$\\[1pt] selected: $\{\rzero ,\anc ,\groot \}$, so $\rzero \R \wanc $ inside $D(\slink )$};
\end{scope}
\begin{scope}[xshift=6.6cm]
\node (g) at (0,1.8) {$\groot $};
\node (r) at (2.2,1.8) {$\king $};
\node (q) at (0,0) {$\rzero $};
\node (m) at (2.2,0) {$\anc $};
\draw[rel] (g) -- (r); \draw[rel] (r) -- (m); \draw[rel] (m) -- (g); \draw[rel] (g) -- (q);
\draw[arc] (q) -- (m); \draw[arc] (r) -- (q);
\node[draw=none,rectangle,align=center,font=\small] at (1.1,-0.95) {(b) $D(\wanc )=\{\anc ,\king ,\groot ,\rzero \}$\\[1pt] selected: $\{\anc ,\king ,\groot \}$, so not $\rzero \R \wanc $};
\end{scope}
\end{tikzpicture}
\caption{The dominators of $\wanc $ (a) inside the query $D(\slink )$ and (b) in the whole layer. Arrows are tournament edges. Thick blue arrows are also contestation edges of the displayed query, and thin gray arrows are not. In (a) the reader $\rzero $ is selected, which starts the route giving $\rzero \R \slink $. In (b) it is not, so $\rzero \R \wanc $ fails.}
\label{fig:anchor}
\end{figure}

\begin{proof}[Proof outline]
Let $W=\Wall(H,\dout ;M,\anc )$, and let $R$ be its contestation graph.

\emph{Old vertices other than $\dout $.} By the table of $\Wall$ in \Cref{sec:hard-tools}, the new dominators of an old vertex $h\ne \dout $ are $\king $ and $\wanc $ if $h\in \Hs{0}$, and $\king $ and $\rzero $, together with $\wanc $ unless $h=\anc $, if $h\in M$. The guard $\king $ beats every old vertex as well as $\rzero $ and $\wanc $, so it is the Condorcet winner of $D(h)$. Hence the only contestation edges into old vertices other than $\dout $ come from $\king $; in particular, none of $\eout ,\rzero ,\slink ,\dout $ can leave the intended route through an old vertex.

\emph{The $1$-test.} The query at $\groot $ is $D(\groot )=V(H)=\Hs{1}$, so $\dout \R \groot $ holds exactly when $\bet{1}=1$.

\emph{The $0$-test.} The edge $\eout \R \rzero $ means that $\eout $ is selected in the query $X=D(\rzero )=\Hs{0}\cup\{\eout ,\king ,\groot \}$, which we decide by the return-path criterion inside $X$, with root $\groot $. In $X$, the query at $\dout $ is $D_{\Hs{0}}(\dout )\cup\{\eout ,\king \}$, and it selects $\eout $: either $D_{\Hs{0}}(\dout )$ is empty and $\eout $ beats $\king $, or $\king \succ D_{\Hs{0}}(\dout )\succ \eout \succ \king $ and \Cref{hard:cyclic} applies. So $\eout \RX{X}\dout $. The query at $\groot $ in $X$ is $\Hs{0}$, so $\dout \RX{X}\groot $ holds exactly when $\bet{0}=1$. All other out-neighbors of $\eout $ and $\dout $ in $X$ have queries won by $\king $ or by $\groot $, so these are the only exits from $\eout $ and $\dout $. Finally, $\groot $ reaches every vertex of $X$: $\groot \RX{X}\king $, the vertex $\king $ reaches every vertex of $\Hs{0}\setminus\{\dout \}$, and these vertices together with $\groot $ form the query at $\eout $, so $\groot $ reaches $\eout $ and then $\dout $. By \Cref{lem:return}, $\eout $ is selected in $X$ exactly when $\bet{0}=1$, that is, $\eout \R \rzero $ holds exactly when $\bet{0}=1$.

\emph{The link $\rzero \R \slink \R \dout $.} In the query $D(\dout )$, the vertex $\slink $ reaches every vertex, so it is selected there and $\slink \R \dout $. The edge $\rzero \R \slink $ is where the anchor enters. In the query $D(\slink )=(V(H)\setminus\{\dout \})\cup\{\groot ,\rzero ,\wanc \}$, the reader $\rzero $ loses to every vertex of $\Hs{0}\setminus\{\dout \}$ and must nevertheless reach all of them. Its first step goes through $\wanc $: inside $D(\slink )$, the dominators of $\wanc $ are $\rzero $, $\anc $, and $\groot $, which form the cycle $\rzero \succ \anc \succ \groot \succ \rzero $ (\Cref{fig:anchor}(a)), so $\rzero $ is selected in that query and $\rzero $ reaches $\wanc $. From $\wanc $, an induction given in \Cref{hard:and-proof} shows that $\rzero $ reaches every vertex of $D(\slink )$, so $\rzero $ is selected in $D(\slink )$ and $\rzero \R \slink $. This is why the anchor must lie outside $\Hs{0}$: an anchor from $\Hs{0}$ would beat $\rzero $, and the triangle would become transitive with the anchor as its winner.

\emph{No shortcuts.} Two tempting shortcuts fail. In the query $D(\dout )$, the vertex $\eout $ has no exit, so $\eout \notin\TEQ(D(\dout ))$ and $\eout \R \dout $ is absent. In the query $D(\wanc )=\{\anc ,\king ,\groot ,\rzero \}$, the vertex $\king $ turns $\anc ,\king ,\groot $ into a cycle that beats $\rzero $ (\Cref{fig:anchor}(b)), so $\rzero \R \wanc $ is absent. Together with the first step, this gives the exact out-neighborhoods
\begin{equation}\tag{$*$}\label{hard:series}
\begin{gathered}
\Out_R(\eout)=\{\rzero:\bet{0}=1\},\qquad\Out_R(\rzero)=\{\slink\},\qquad\Out_R(\slink)=\{\dout\},\\
\Out_R(\dout)=\{\rzero:\bet{0}=1\}\cup\{\groot:\bet{1}=1\},
\end{gathered}
\end{equation}
where $\{\rzero :\bet{0}=1\}$ means $\{\rzero \}$ if $\bet{0}=1$ and $\varnothing$ otherwise.

\emph{The root.} The query $D(\king )=\{\eout ,\slink ,\groot \}$ has Condorcet winner $\groot $, so $\groot \R \king $. The vertex $\king $ reaches every old vertex other than $\dout $ and also $\wanc $, and $\groot \R \rzero \R \slink \R \dout $. Thus $\groot $ reaches every dominator of $\eout $, hence $\eout $ itself, and so every vertex. By \Cref{lem:return} and~\eqref{hard:series}, $\eout $ is selected exactly when $\bet{0}\land\bet{1}$. \Cref{hard:wrapper-proofs} checks each query in full.
\end{proof}

\subsection{Proof of correctness}\label{sec:hard-reduction}

An outer quantifier deletes occurrence vertices inside the layers that are already built, so we prove a statement about every intermediate tournament $T_i$ under every assignment to the variables outside it. For $\alpha\in\{0,1\}^{i-1}$, recall that $Z_\alpha$ is the set of occurrence vertices of the variables among $x_1,\ldots,x_{i-1}$ that $\alpha$ sets to $0$.

\begin{lemma}[Invariant]\label{lem:invariant}
For every $i\in\{1,\ldots,n+1\}$ and every $\alpha\in\{0,1\}^{i-1}$,
\[
\orole_i\in\TEQ(T_i-Z_\alpha)\quad\Longleftrightarrow\quad Q_ix_i\cdots Q_nx_n\ \psi(\alpha,x_i,\ldots,x_n)\ \text{is true}.
\]
\end{lemma}

\begin{proof}
We induct downward on $i$. For $i=n+1$ the quantifier prefix is empty, and the claim is \Cref{prop:matrix}. Let $i\le n$. The set $Z_\alpha$ consists of occurrence vertices of $x_1,\ldots,x_{i-1}$, so it avoids the output $\orole_{i+1}$, all of $M_i$ including the anchor $\anc_i$, and every vertex of a filter or a layer. By \Cref{obs:deletion}, $T_i-Z_\alpha$ is therefore the layer of type $Q_i$ around the smaller tournament $H=T_{i+1}-Z_\alpha$, with output $\orole_{i+1}$ and occurrence set $M_i$. Writing $\alpha b$ for the extension of $\alpha$ by $x_i=b$, so that $Z_{\alpha1}=Z_\alpha$ and $Z_{\alpha0}=Z_\alpha\cup M_i$, the two states of this layer are
\[
\Hs{1}=T_{i+1}-Z_{\alpha1}\qquad\text{and}\qquad \Hs{0}=T_{i+1}-Z_{\alpha0}.
\]
By the induction hypothesis, $\bet{b}=\iverson{\orole_{i+1}\in\TEQ(\Hs{b})}$ is the truth value of
\[
Q_{i+1}x_{i+1}\cdots Q_nx_n\ \psi(\alpha,b,x_{i+1},\ldots,x_n).
\]
By \Cref{lem:or} if $Q_i=\exists$ and \Cref{lem:and} if $Q_i=\forall$, the output $\orole_i$ is selected in $T_i-Z_\alpha$ exactly when $\bet{0}\lor\bet{1}$, respectively $\bet{0}\land\bet{1}$, which is the truth value of $Q_ix_i\cdots Q_nx_n\,\psi(\alpha,x_i,\ldots,x_n)$.
\end{proof}

Note that the induction compares answers in different tournaments only through the layer lemmas; it never assumes that deleting vertices preserves an answer.

\begin{proof}[Proof of \Cref{thm:correct}]
Apply \Cref{lem:invariant} with $i=1$ and the empty assignment, for which $Z_\alpha=\varnothing$.
\end{proof}

\begin{proof}[Proof of \Cref{thm:main}]
Membership in PSPACE is \Cref{prop:pspace}. For hardness, given a quantified $3$-CNF formula, apply \Cref{lem:normal} to obtain an equivalent formula satisfying~\ref{F1}--\ref{F3}, and then apply \Cref{def:reduction}. Both steps take polynomial time, and by \Cref{thm:correct} the result is a yes-instance of \prob{TEQ Membership} exactly when the formula is true.
\end{proof}

\section{Hereditary uniqueness}\label{sec:uniqueness}

The reduction of \Cref{sec:hardness,sec:correctness} proves that TEQ membership is PSPACE-hard. This section proves that the hardness does not depend on competing minimal retentive sets: the tournaments produced by the reduction have a unique minimal retentive set in every induced subtournament (\Cref{thm:structured-hardness}). The proof has two independent parts. The first is a general theorem, which does not refer to the reduction: a single vertex with at most four dominators already forces uniqueness (\Cref{sec:local}). The second shows that the tournaments of the reduction have such a vertex in every induced subtournament (\Cref{sec:hard-degree}).

\subsection{Local uniqueness}\label{sec:local}

\begin{restatable}[Local uniqueness]{theorem}{thmlocal}\label{thm:local}
Every tournament that has a vertex with at most four dominators has a unique minimal TEQ-retentive set.
\end{restatable}

Suppose that two minimal retentive sets coexist, and consider a vertex $v$ with at most four dominators. Neither set can be a singleton, since a singleton retentive set is the unique minimal retentive set. Each set that does not contain $v$ must contain at least two of those dominators. The two sets are disjoint, so this requirement leaves very little room. The remaining possibilities do fit within the bound of four, but their edges will turn out to be incompatible with retentiveness.

We write $D(v)=D_T(v)$, and say that a set $X$ \emph{dominates} a set $Y$ if every vertex of $Y$ is beaten by some vertex of $X$. Different vertices of $Y$ may be beaten by different vertices of $X$. The following facts make the counting argument precise. Their proofs, which do not assume uniqueness in smaller tournaments, appear in \Cref{uniq:app}.

\begin{lemma}[Basic facts]\label{uniq:basic}
Let $T$ be a tournament. Then the following hold.
\begin{enumerate}[label=\textup{(\arabic*)}]
\item Every retentive set $R$ of $T$ dominates its complement: every vertex outside $R$ is beaten by some vertex of $R$.
\item Distinct minimal retentive sets are disjoint, and every minimal retentive set induces a strongly connected tournament.
\item A singleton minimal retentive set consists of a Condorcet winner and is the unique minimal retentive set. No minimal retentive set has exactly two vertices.
\item If $R$ is a minimal retentive set of $T$ with $|R|\ge2$ and $v\in V(T)\setminus R$, then $|R\cap D_T(v)|\ge2$.
\end{enumerate}
\end{lemma}

We will also use a consequence inside dominator queries: if $X$ is nonempty and $\TEQ(X)$ is contained in a set of at most two vertices, then it consists of a single Condorcet winner of $X$. Indeed, every minimal retentive set in $X$ must then be a singleton, and a Condorcet winner is unique.

Now suppose that $R_1,R_2$ are distinct minimal retentive sets and $v$ has at most four dominators. By the lemma, both sets have at least three vertices. If $v$ lies outside both, each must supply two of the four dominators. If $v$ lies in one of the sets, strong connectivity supplies a dominator in that set, while the other set must still supply two. Thus only two configurations remain:
\begin{itemize}
\item $v\notin R_1\cup R_2$, and each $R_j$ contains exactly two dominators of $v$, or
\item $v\in R_1$, say. Then $R_2$ contains two or three dominators of $v$, and strong connectivity of $R_1$ supplies a dominator of $v$ in $R_1$, of which there are at most two.
\end{itemize}
The count has not yet produced a contradiction. When $v$ is in one set, the answer to its query is confined to at most two vertices of that set, so the query must have a Condorcet winner. To rule out this case, we will find a pair in the other set that cannot have a common dominator. When $v$ is outside both sets, we will find such a pair in each and show that the two pairs cannot coexist.

These pairs are given by \emph{captain} edges. If $a$ captains $b$, then $a\succ b$ and no vertex beats both. The query at $b$ therefore selects only $a$, so every retentive set containing $b$ contains $a$. To find these edges, we first show that an edge which is a captain inside a retentive set is also a captain in the whole tournament.

\begin{lemma}[Captain persistence]\label{uniq:persistence}
Let $R$ be a retentive set of a tournament $T$, and let $a,b\in R$. If $a$ captains $b$ in $T[R]$, then $a$ captains $b$ in $T$.
\end{lemma}

\begin{proof}
Put $H=T[D(b)]$ and $S=\TEQ(H)$. The first step is to show that $a$ is selected in this query. Retentiveness gives $S\subseteq R\cap D(b)$, and $a$ beats every other vertex of $R\cap D(b)$. By \Cref{uniq:basic}(1) applied in $H$, the set $S$ must beat $a$ unless it contains $a$. Since no vertex of $S\setminus\{a\}$ beats $a$, we have $a\in S$.

Now let $C\subseteq S$ be a minimal retentive set of $H$ containing $a$. A vertex beating both $a$ and $b$ would make $D_H(a)$ nonempty, so retentiveness of $C$ would require a selected dominator of $a$ in $C$. But $C\subseteq R\cap D(b)$, and $a$ beats every other vertex of this intersection. Formally, $D_H(a)\ne\varnothing$ would give $\varnothing\ne\TEQ(D_H(a))\subseteq C\cap D_H(a)\subseteq R\cap D(b)\cap D(a)=\varnothing$. So no vertex beats both $a$ and $b$.
\end{proof}

For minimal retentive sets this is~\cite[Lemma~5]{Yang2016}. The proof above covers all retentive sets. Now consider a retentive set $R$ that does not contain $v$. We seek a captain edge among the vertices of $R$ that beat $v$. Those vertices have to beat every member of $R$ outside the intersection. For each such member, its query must select a vertex of $R$ that beats $v$.

\begin{lemma}\label{uniq:intersection-dominates}
Let $R$ be a retentive set of a tournament $T$, let $v\in V(T)\setminus R$, and put $A=R\cap D_T(v)$. Then every vertex of $R\setminus A$ is beaten by some vertex of $A$.
\end{lemma}

\begin{proof}
Let $y\in R\setminus A$, so that $v\succ y$. The set $\TEQ(D(y))$ is contained in $R$, hence avoids $v$, and by \Cref{uniq:basic}(1) applied in $D(y)$ it contains a vertex beating $v$. That vertex lies in $A$ and beats $y$.
\end{proof}

When the intersection has two vertices, this domination property makes their edge a captain inside $R$, and captain persistence makes it global. Three vertices need more work: a failed captain edge creates a witness outside the intersection, and two such witnesses will force contradictory orientations.

\begin{lemma}[Two or three dominators]\label{uniq:two-three}
Let $R$ be a retentive set of a tournament $T$, and let $v\in V(T)\setminus R$. If $|R\cap D_T(v)|\in\{2,3\}$, then $R\cap D_T(v)$ contains a captain edge of $T$. If $R\cap D_T(v)$ induces a directed triangle, then at least two of its three edges are captain edges of $T$.
\end{lemma}

\begin{proof}[Proof idea]
Put $A=R\cap D(v)$. If $|A|=2$, \Cref{uniq:intersection-dominates} says that no vertex of $R\setminus A$ beats both vertices of $A$, so the edge inside $A$ is a captain edge in $T[R]$, and \Cref{uniq:persistence} lifts it to $T$. If $|A|=3$ and the pair $A\setminus\{a\}$ fails to be a captain edge, for a vertex $a$ that is not a Condorcet winner of $T[A]$, \Cref{uniq:persistence} yields a witness $x_a\in R\setminus A$ beating both vertices of the pair. Every minimal retentive set in the query $D(x_a)$ lies in $R$ and avoids $v$, but $a$ is the only vertex it can contain that beats $v$. By \Cref{uniq:basic}(4), it must therefore be a singleton, and by domination that singleton is $\{a\}$. Hence $\TEQ(D(x_a))=\{a\}$, so $a$ captains $x_a$. Two such failures, for distinct $a$ and $b$, force opposite orientations of the edge between $x_a$ and $x_b$. The full argument is in \Cref{uniq:app}.
\end{proof}

When $v$ lies outside both hypothetical minimal retentive sets, each contributes exactly two dominators, and the preceding lemma gives a captain edge in each pair. We must still explain why these two edges cannot coexist. In fact, the following stronger statement rules out any second captain edge among the other dominators of $v$. This extra generality will also be useful for finding a selected vertex.

\begin{lemma}[No disjoint captain pair]\label{uniq:disjoint}
Let $R$ be a retentive set of a tournament $T$, let $v\in V(T)\setminus R$, and suppose $R\cap D_T(v)=\{a,b\}$. Then $D_T(v)\setminus\{a,b\}$ contains no captain edge of $T$.
\end{lemma}

\begin{proof}[Proof idea]
Suppose $c$ captains $d$ with $c,d\in D(v)\setminus\{a,b\}$, and let $a\succ b$. The preceding lemma makes $a$ the captain of $b$. The two captain conditions force the four edges between the pairs into one of two alternating patterns. For example, $a\succ c$ forces $c\succ b\succ d\succ a$. In either pattern, domination inside $D(a)$ produces a vertex $x\in R$ whose query contains $v$ and in which $b$ is the only vertex of $R$ that beats $v$. \Cref{uniq:basic}(4) forces $\TEQ(D(x))=\{b\}$. This is impossible because the query also contains a vertex of $\{c,d\}$ that beats $b$. \Cref{uniq:app} gives both patterns and the witness argument.
\end{proof}

\Cref{uniq:disjoint} does not require the second pair to lie in a retentive set, and $D(v)$ may be arbitrarily large. Both features are used by the selector.

\begin{proof}[Proof of \Cref{thm:local}]
Suppose distinct minimal retentive sets $R_1,R_2$ exist and $|D(v)|\le4$. By \Cref{uniq:basic}, they are disjoint, strongly connected, and have at least three vertices each.

First suppose $v\in R_1$. Then $R_2$ contains two or three dominators of $v$, and these contain a captain edge $c\succ d$ by \Cref{uniq:two-three}. At most two dominators of $v$ lie in $R_1$, and retentiveness of $R_1$ gives $\TEQ(D(v))\subseteq R_1\cap D(v)$. By the consequence after \Cref{uniq:basic}, $\TEQ(D(v))$ is a single Condorcet winner of $D(v)$. It lies outside $R_2$ and therefore beats both $c$ and $d$, contradicting that $c$ captains $d$. The case $v\in R_2$ is symmetric.

Otherwise $v$ lies outside both sets, so each contains exactly two dominators of $v$. By \Cref{uniq:two-three}, the pair in $R_2$ contains a captain edge. This contradicts \Cref{uniq:disjoint} applied to $R_1$.
\end{proof}

A tournament in which every induced subtournament has such a vertex therefore has unique minimal retentive sets throughout. We name this condition for general $k$.

\begin{definition}[Hereditary dominator bound]\label{def:structural}
For $k\ge0$, let $\mD_k$ be the class of tournaments in which every nonempty induced subtournament has a vertex with at most $k$ dominators. An \emph{elimination order} of degree $k$ is an order $v_1,\ldots,v_N$ of the vertices in which $v_j$ has at most $k$ dominators in $\{v_j,\ldots,v_N\}$.
\end{definition}

A tournament lies in $\mD_k$ exactly when it has such an order: the first vertex of the order in any induced subtournament proves one direction, and repeatedly removing a vertex with at most $k$ current dominators proves the other. The condition bounds the minimum indegree in each subtournament, not the maximum; transitive tournaments of every size lie in $\mD_0$, for example, although their last vertex loses to every other vertex.

\begin{corollary}\label{cor:d4-unique}
Every nonempty induced subtournament of a tournament in $\mD_4$ has a unique minimal TEQ-retentive set.
\end{corollary}

We do not know whether a vertex with five dominators still forces uniqueness. \Cref{thm:local} applies to tournaments of unbounded order and is incomparable with the known guarantee for all tournaments of order at most $14$~\cite{Yang2016}.

\subsection{The hard instances}\label{sec:hard-degree}

We now show that the tournament $T_1$ of \Cref{def:reduction} lies in $\mD_3$, by exhibiting an elimination order in which each vertex has at most three remaining dominators when it is removed. Constant-size gadgets alone do not give such an order, because edges from many layers could enter the same old vertex and their contributions could add up. The order therefore begins by removing some guards so that this does not happen.

\begin{lemma}[Degree certificate]\label{lem:degree}
For every formula $\Phi$ satisfying~\ref{F1}--\ref{F3}, the tournament $T_1$ of \Cref{def:reduction} has an elimination order of degree three, computable in polynomial time.
\end{lemma}

\begin{proof}[Proof sketch]
The order has four phases; \Cref{hard:degree} gives the complete order and its degree counts.
\begin{enumerate}[label=(\arabic*)]
\item Process the layers from the outermost inward, removing $\shw,\tri,\broot$ from an $\exists$-layer and $\king,\wanc$ from a $\forall$-layer. Their remaining indegrees are $(2,2,2)$ and $(3,3)$. Afterwards, the surviving vertices of an $\exists$-layer beat old vertices only through $\rzero\succ M_i$, and those of a $\forall$-layer only through $\rzero\succ M_i$ and $\eout_i,\slink_i\succ\orole_{i+1}$. Hence each occurrence vertex has at most one dominator outside the clause tournament, since it belongs to exactly one set $M_i$, and each output has at most two. A $\forall$-layer replaces the output, so these contributions never add up at one vertex.
\item Process the filters from the last clause toward the first. Each non-output vertex of the right clause gadget has at most one further dominator from its filter, and the three clause gadgets admit orders of degree at most three under these budgets. Each filter leaves behind its relay $\relay$ and the output of its right clause gadget.
\item Remove these leftover filter vertices, from the first filter constructed to the last, each with indegree zero.
\item Remove the rest of each layer, from the innermost outward, in the orders $\shz,\rone,\rzero$ and $\groot,\rzero,\slink,\eout$.\qedhere
\end{enumerate}
\end{proof}

Together with the local uniqueness theorem, the degree certificate gives the second claim of the title.

\begin{theorem}[Structured hard instances]\label{thm:structured-hardness}
\prob{TEQ Membership} is PSPACE-complete under polynomial-time many-one reductions on the class $\mD_3$, even when a witnessing elimination order is supplied. Every nonempty induced subtournament of a tournament in $\mD_3$ has a unique minimal TEQ-retentive set and a vertex with at most three dominators.
\end{theorem}

\begin{proof}[Proof of \Cref{thm:structured-hardness}]
The reduction of \Cref{thm:main} proves hardness on the stated class. The witnessing elimination order is given by \Cref{lem:degree}. Its first vertex in any nonempty induced subtournament has at most three dominators there, so \Cref{thm:local} gives hereditary uniqueness (\Cref{cor:d4-unique}). Membership in PSPACE follows from \Cref{prop:pspace}.
\end{proof}

\section{Finding a member versus recognizing one}\label{sec:algorithms}

The uniqueness theorem tells us that a tournament with a vertex of indegree at most four has one minimal retentive set. It does not tell us which vertices belong to that set. We will use the same local constraints to find one of its members in quadratic time. The tournaments of \Cref{thm:structured-hardness} have such a vertex in every induced subtournament. On this class, therefore, a selected alternative can be found in quadratic time, although deciding whether a specified alternative is selected is PSPACE-complete.

We first ask what a known member tells us about the rest of TEQ. Under hereditary uniqueness, a procedure that finds one member in every induced subtournament lets us recover the entire set. We then give such a procedure for tournaments with a vertex of indegree at most four.

\begin{restatable}[Finding a member]{theorem}{thmselect}\label{thm:select}
\leavevmode
\begin{enumerate}[label=\textup{(\alph*)}]
\item In a tournament that has a vertex with at most four dominators, one member of TEQ can be found in $O(N^2)$ time.
\item If every nonempty induced subtournament of $T$ has a vertex with at most four dominators, then $\TEQ(T)$ can be computed deterministically in $O(2^tN^2)$ time, where $t=|\TEQ(T)|$. Neither $t$ nor any member of $\TEQ(T)$ needs to be known in advance.
\end{enumerate}
\end{restatable}

\subsection{Computing TEQ from a certified member}\label{sec:closure}

Suppose first that we know one vertex in the unique minimal retentive set. Retentiveness forces that set to contain the selected dominators of the vertex, then the selected dominators of those vertices, and so on. Starting from a known member and repeatedly adding these required vertices cannot take us outside the answer. Once no more vertices need to be added, the resulting set is itself retentive and therefore contains the unique minimal retentive set.

To carry out this process, we must answer a TEQ query at each vertex we add. Each query is a smaller tournament, so we can apply the same process recursively if we can find a selected vertex there too. This is why the algorithm below assumes both a selector and uniqueness throughout a hereditary class, rather than only in the original tournament.

The starting vertex matters. In \Cref{ex:four}, closure started at the unselected vertex $d$ adds $a,b,c$ but never removes $d$. A \emph{certified} member, one known to be selected, avoids this. \Cref{alg:closure} starts from such a member in every recursive call and caches completed answers by their vertex sets. Even with a certified seed, the recursive queries might appear to visit exponentially many subsets of the tournament. The key is that every pivot used anywhere in the recursion belongs to the final answer. Each domain is obtained by intersecting the dominator sets of some of those pivots, so caching limits the algorithm to at most $2^t$ domains, where $t=|\TEQ(T)|$. We prove this bound without knowing the answer in advance.

\begin{algorithm}[H]
\caption{$\textsc{Closure}(X)$: TEQ on a hereditary class with unique minimal retentive sets}
\label{alg:closure}
\noindent\textbf{Input:} A set $X\subseteq V$ and a selector $\sigma$ returning a member of $\TEQ(T[Y])$ for every nonempty $Y\subseteq V$.
\begin{enumerate}[label=\arabic*.,leftmargin=*,itemsep=1pt,topsep=2pt]
\item If $X=\varnothing$, return $\varnothing$.
\item If $X$ is in the cache, return the cached answer.
\item Set $b\gets\sigma(T[X])$ and $R\gets\{b\}$, and initialize $Q$ as a queue containing $b$.
\item While $Q\ne\varnothing$, remove some $r$ from $Q$ and, for each $c\in\textsc{Closure}(X\cap D(r))\setminus R$, add $c$ to $R$ and to $Q$.
\item Store $R$ in the cache under the key $X$ and return $R$.
\end{enumerate}
\end{algorithm}

\begin{theorem}[Closure from a certified seed]\label{uniq:general}
Let $\mC$ be a hereditary class of tournaments whose nonempty members each have a unique minimal TEQ-retentive set. Suppose that a deterministic algorithm $\sigma$ returns a member of $\TEQ(H)$ for every nonempty $H\in\mC$ in time $p(|V(H)|)$, where $p$ is nondecreasing. Then, for every $T\in\mC$, \Cref{alg:closure} called on $X=V(T)$ returns $\TEQ(T)$ in deterministic time $O(2^t(p(N)+N^2))$, where $N=|V(T)|$ and $t=|\TEQ(T)|$.
\end{theorem}

\begin{proof}
\emph{Correctness.} Induct on $|X|$ and write $W_X=\TEQ(T[X])$. Every recursive call is on a proper subset, because $X\cap D(r)$ excludes $r$, and returns $\TEQ(T[X\cap D(r)])$ by induction. Initially $R\subseteq W_X$ because the seed is certified. If $r\in R\subseteq W_X$, retentiveness of $W_X$ puts the answer of the call at $r$ inside $W_X$, so $R\subseteq W_X$ is maintained. At termination every member of $R$ has been processed, so $R$ is nonempty and retentive in $T[X]$. It contains a minimal retentive set, which by uniqueness equals $W_X$. Hence $R=W_X$. Heredity is used because the uniqueness and seed hypotheses are needed in every recursive call.

\emph{Number of domains.} If $r\in W_X$, retentiveness gives $\TEQ(T[X\cap D(r)])\subseteq W_X$. Hence, along any chain of recursive calls, the answer of each call is contained in the answer of its parent. Every pivot $r$ used anywhere therefore belongs to $W=\TEQ(T)$. Each domain reached is therefore of the form
\[
X_B=\bigcap_{b\in B}D(b),\qquad B\subseteq W,\qquad X_\varnothing=V,
\]
so at most $2^t$ distinct domains occur, although the algorithm never needs to know $W$. A call on a domain still being evaluated cannot recur on the current stack, because domains strictly shrink along it, so each domain is evaluated once.

\emph{Time.} Each domain calls $\sigma$ once and processes at most $N$ pivots. Using membership vectors for sets and a binary trie on the $N$ bits of a domain for the cache, each pivot costs $O(N)$ time, so each domain costs $O(p(N)+N^2)$. \Cref{uniq:closure-details} gives the implementation details.
\end{proof}

\begin{corollary}[Finding several winners]\label{cor:partial-teq}
    Under the hypotheses of \Cref{uniq:general}, for every integer $k\ge 1$ there is a deterministic algorithm that finds $k$ distinct members of $\TEQ(T)$, or returns the entire set if it has fewer than $k$ members, in time $O(2^{k-1}(p(N)+N^2))$. On $\mD_4$, the running time is $O(2^{k-1}N^2)$.
\end{corollary}

\begin{proof}
    Run \Cref{alg:closure}, reporting each distinct seed the first time the selector chooses it anywhere in the recursion. Stop at the $k$th report. By retentiveness, the winners of each recursive call belong to the winner set of its parent, so every reported seed belongs to $\TEQ(T)$. Every vertex in a returned set originates as a seed. Thus, if the run finishes with fewer than $k$ reports, it has reported all of $\TEQ(T)$.
    
    Let $S$ be the set of pivots used to generate recursive calls. Each such pivot has already been reported, since it is either the seed of its call or a vertex returned by a completed subcall. The algorithm stops before a $k$th distinct seed can be used as a pivot, so $|S|\le k-1$. Every domain reached has the form $X_B$ for some $B\subseteq S$, in the notation of \Cref{uniq:general}. With caching, at most $2^{k-1}$ domains are evaluated, each in $O(p(N)+N^2)$ time. On $\mD_4$, \Cref{cor:d4-unique,thm:select} give the required uniqueness and a selector with $p(N)=O(N^2)$.
\end{proof}

\subsection{Finding a member next to a vertex with at most four dominators}\label{sec:selector}

The closure algorithm leaves us with a smaller task: find one selected vertex in each tournament encountered. We now do this for any tournament with a vertex $v$ having at most four dominators, without evaluating the recursive queries on the full tournament. Put $A=D(v)$. By \Cref{thm:local}, $R=\TEQ(T)$ is the unique minimal retentive set. We do not know $R$, but the lemmas of \Cref{sec:uniqueness} restrict its intersection $I=A\cap R$.

The tournament on $A$ is small enough to solve directly. Its selected vertices are useful candidates, but selection within $A$ does not by itself certify selection in $T$. To narrow the possibilities for $I$, we test which edges inside $A$ are captain edges in the whole tournament. The selector then returns vertices that belong to every intersection allowed by those tests and the constraints from \Cref{sec:uniqueness}.

If $A=\varnothing$, then $v$ is a Condorcet winner. If $T[A]$ has a Condorcet winner $c$, then $c$ is selected. To see this, first suppose that $R$ is a singleton. Its vertex is a global Condorcet winner, which must lie in $A$ and equal $c$. If $v\in R$, retentiveness gives $c\in R$. Finally, suppose that neither $v$ nor $c$ belongs to $R$. The set $I$ has at least two vertices by \Cref{uniq:basic}(4), and at most three because it excludes $c$. By \Cref{uniq:two-three}, it contains a captain edge. But $c$ beats both endpoints of that edge, a contradiction.

It remains to handle the case in which $T[A]$ has no Condorcet winner. The possible captain edges then have a particularly simple form.

\begin{lemma}[The local captain graph]\label{uniq:four-vertices}
Let $A$ be a nonempty set of at most four vertices of a tournament $T$ such that $T[A]$ has no Condorcet winner. Then $E=\TEQ(T[A])$ is a directed triangle, each edge of $E$ is a captain edge of $T[A]$, and the only other possible captain edge of $T[A]$ points from a vertex of $E$ to the vertex of $A\setminus E$, if there is one. (Proof in \Cref{uniq:selector-details}.)
\end{lemma}

\paragraph{The constraints.}
Assume from now on that $T[A]$ has no Condorcet winner. Then $T$ has no Condorcet winner either, since it would lie in $A$ and win $T[A]$, so $R$ has at least three vertices. Let $C$ be the set of edges inside $A$ that are captain edges of the whole tournament $T$. Every such edge is a captain edge of $T[A]$, so by \Cref{uniq:four-vertices} its tail lies in $E$. Testing whether an edge is a captain edge takes one scan for a common dominator, so $C$ can be computed in $O(N)$ time. The set $C$ constrains the unknown set $I$ as follows:
\begin{enumerate}[label=(C\arabic*)]
\item if $v\in R$, then $E\subseteq I$, by retentiveness.
\item if $v\notin R$, then $|I|\ge2$ by \Cref{uniq:basic}(4), and unless $I=A$, the set $I$ contains an edge of $C$ by \Cref{uniq:two-three}.
\item $I$ is closed under captain predecessors: if $a\succ b$ is an edge of $C$ and $b\in I$, then $a\in I$.
\item if $|I|=2$, then $v\notin R$ by (C1), since $|E|=3$, and $C$ has no edge disjoint from $I$ by \Cref{uniq:disjoint}.
\end{enumerate}

\paragraph{A worked case.}
Suppose $C$ contains two disjoint edges $a\succ b$ and $c\succ d$. We show that $a$ and $c$ are selected, although we do not know $R$. Both captains lie in $E$, so by (C1) they are selected if $v\in R$. Suppose $v\notin R$ and $a\notin I$. By (C3), $b\notin I$ as well, so $I\subseteq\{c,d\}$, and (C2) forces $I=\{c,d\}$. But then $a\succ b$ is an edge of $C$ disjoint from $I$, which contradicts (C4). Hence $a\in R$, and symmetrically $c\in R$.

\paragraph{The selector.}
The other configurations of $C$ are handled in the same way, and \Cref{tab:selector} lists the resulting rule. For example, if $C$ is empty, then $E\subseteq I$. This follows from (C1) when $v\in R$. When $v\notin R$, (C2) forces $I=A$, since an intersection $I\ne A$ would contain an edge of $C$. At the other extreme, suppose $C$ contains the three edges of $E$. Each vertex of $E$ is then captained in $T$ by its predecessor on the triangle. Thus $E$ is retentive, and every nonempty retentive subset of $E$ is all of $E$. Uniqueness gives $R=E$.

\begin{table}[H]
\centering
\small
\begin{tabular}{@{}ll@{}}
\toprule
First applicable condition & Certified set $B\subseteq\TEQ(T)$\\
\midrule
$A=\varnothing$ & $\{v\}$\\
$T[A]$ has a Condorcet winner $c$ & $\{c\}$\\
$C=\varnothing$ & $E$\\
$C$ contains the three edges of $E$ & $E$\\
$C$ contains disjoint edges $a\succ b$ and $c\succ d$ & $\{a,c\}$\\
$C$ consists of a single edge $a\succ b$ & $E\cap\{a,b\}$\\
otherwise & $\{s\}$, where $s$ is the common endpoint of all edges of $C$\\
\bottomrule
\end{tabular}
\caption{The selector for a vertex $v$ with $A=D(v)$ and $|A|\le4$. Here $E=\TEQ(T[A])$ and $C$ is the set of edges inside $A$ that are captain edges of the whole tournament.}
\label{tab:selector}
\end{table}

\begin{proposition}[Selector]\label{prop:selector}
Let $T$ be a tournament with a vertex $v$ such that $|D(v)|\le4$. Put $A=D(v)$ and $E=\TEQ(T[A])$, and let $C$ be the set of edges with both endpoints in $A$ that are captain edges of $T$. Then the set $B$ given by the first applicable row of \Cref{tab:selector} is well defined, nonempty, and contained in $\TEQ(T)$, and it can be computed in $O(N^2)$ time without any recursive TEQ call and without any hypothesis on proper subtournaments.
\end{proposition}

The first two rows were verified above and the third to fifth in the preceding paragraphs. \Cref{uniq:selector-details} verifies the last two rows and the running time.

\begin{proof}[Proof of \Cref{thm:select}]
Part~(a) is \Cref{prop:selector}. Choosing the least-labelled vertex of the certified set makes the selector deterministic. For part~(b), $\mD_4$ is hereditary, every nonempty member has a unique minimal retentive set by \Cref{cor:d4-unique}, and every nonempty member has a vertex with at most four dominators. Apply \Cref{uniq:general} with $\mC=\mD_4$ and $p(N)=O(N^2)$.
\end{proof}

\section{The boundary of the dominator bound}\label{sec:boundaries}

The PSPACE-hard instances of \Cref{thm:structured-hardness} have an elimination order in which at most three dominators remain at each step. What happens if we allow fewer? With at most one remaining dominator, the recursive computation has a particularly simple form and takes quadratic time. With two, membership is already NP-hard. We give the detailed constructions and proofs in \Cref{rest:details}.

\begin{theorem}[The degree boundary]\label{thm:boundary}
\leavevmode
\begin{enumerate}[label=\textup{(\alph*)}]
\item For every tournament $T\in\mD_1$, the set $\TEQ(T)$ can be computed in $O(N^2)$ time.
\item \prob{TEQ Membership} is NP-hard under polynomial-time many-one reductions on $\mD_2$, even when a witnessing elimination order is supplied.
\end{enumerate}
\end{theorem}

\subsection{One dominator: a polynomial case}\label{sec:d1}

Suppose $a$ has just one dominator $b$. Any retentive set containing $a$ must also contain $b$. If $b$ has no dominators, it wins the tournament. Otherwise, the query at $b$ selects vertices from its dominator set $I$. Every vertex of $I$ is beaten by $a$, and $a$ beats every vertex except $b$. Thus $a$ captains each vertex of $I$. The selected vertices in $I$ bring us back to $a$, closing the recursion around one smaller core. The lemma identifies the whole answer from that core.

\begin{lemma}[One dominator]\label{lem:one-dominator}\label{uniq:one-dominator}
Let $T$ be a tournament with vertices $a,b$ such that $D_T(a)=\{b\}$, and put $I=D_T(b)$. Then
\[
\TEQ(T)=
\begin{cases}
\{b\},& I=\varnothing,\\
\{a,b\}\cup\TEQ(T[I]),& I\ne\varnothing,
\end{cases}
\]
and in both cases the minimal retentive set is unique.
\end{lemma}

\begin{proof}
If $I=\varnothing$, then $b$ is a Condorcet winner. Otherwise there is no Condorcet winner, since the only candidate would be $b$. Put $E=\TEQ(T[I])$. The query at $a$ selects $b$, the query at $b$ selects $E$, and the query at every $e\in E$ selects $a$: indeed $e\succ b$, and $a$ beats $e$ and every other vertex except $b$, so $a$ captains $e$. Thus $S=\{a,b\}\cup E$ is retentive. A nonempty retentive subset of $S$ containing any of its vertices must contain $a$, then $b$, then all of $E$, so $S$ is minimal. This uses no uniqueness assumption on $T[I]$. Every minimal retentive set has at least two vertices and must contain $a$, since otherwise $a$ would need two dominators in it (\Cref{uniq:basic}(4)). It then contains $b$ and $E$. This proves uniqueness and the formula.
\end{proof}

In particular, a vertex with one dominator is selected exactly when its dominator has a dominator. A target with at most one dominator can therefore be decided immediately. On $\mD_1$, the lemma yields a single chain of subproblems. Repeatedly pick a vertex $x$ of indegree at most one in the current set $X$. If $x$ has no dominator, it is a Condorcet winner. Record $x$ and end the process. Otherwise let $y$ be its unique dominator. If $y$ has no dominator, it is a Condorcet winner. Record $y$ and end the process. If its dominator set $I$ is nonempty, record $x,y$ as selected and replace $X$ by $I$. Maintaining indegree counters implements this in $O(N^2)$ time, which proves \Cref{thm:boundary}(a). \Cref{alg:d1} gives the details.

\subsection{Two dominators: NP-hardness}\label{sec:d2}

The PSPACE reduction uses three remaining dominators when it processes a $\forall$-layer. For satisfiability, we need only existential choices, and the $\exists$-layer can combine the two values of each variable. We can therefore reuse that layer, but the clause tournament and its elimination order must fit within the smaller degree budget. This calls for a different clause tournament, built from a gadget we already have.

We change how the clause tournament represents an assignment. In the PSPACE reduction, an occurrence vertex recorded the value of its variable, and the clause gadget had to account for the sign of the literal. Here, an occurrence vertex is present exactly when its literal is true. Every clause can therefore use the same gadget, namely $\Gad_0$ of \Cref{def:clause}, which selects its output exactly when at least one occurrence remains. A different filter joins the clauses while leaving room in the degree bound for the variable layers (\Cref{rest:matrix}).

For a variable $x_i$, let $M_i^+$ and $M_i^-$ be its positive and negative occurrence vertices. Setting $x_i=1$ deletes $M_i^-$, whereas setting $x_i=0$ deletes $M_i^+$. For example, if both $x_i$ and $\neg x_i$ occur, setting $x_i=1$ retains the positive occurrence and deletes the negative one. Setting $x_i=0$ does the reverse. Unlike the two states used in the PSPACE reduction, these states need not be nested. The $\exists$-layer accepts arbitrary vertex sets $H_1$ and $H_0$ containing the output, so it still reads both states from one copy (\Cref{hard:or}).

\begin{proposition}[The degree-two reduction]\label{rest:d2hard}
There is a polynomial-time algorithm that maps every $3$-CNF formula with $n\ge1$ variables and $m\ge1$ clauses to a tournament $T$ on $6n+8m-2$ vertices, a vertex $d$ of $T$, and an elimination order of $T$ of degree two whose last vertex is $d$, such that $d\in\TEQ(T)$ if and only if the formula is satisfiable. 
\end{proposition}

\begin{proof}[Proof of \Cref{thm:boundary}]
Part~(a) was shown in \Cref{sec:d1}, with details in \Cref{alg:d1}. Part~(b) follows from \Cref{rest:d2hard}, proved in \Cref{rest:details}.
\end{proof}

\subsection{A consequence for preference profiles}

Our reduction specifies pairwise comparisons directly. It also yields a consequence for preference profiles, where each voter gives a linear order on the alternatives. A profile is \emph{two-crossing} if its voters can be ordered so that the relative preference between each pair of alternatives changes at most twice. Constantinescu and Wattenhofer~\cite[Theorem~7]{CW2022} realize every tournament as the majority relation of such a profile, with $O(N^2)$ voters. Thus this restriction on profiles does not exclude the hard tournaments.

\begin{corollary}\label{cor:two-crossing}
Given an explicitly listed two-crossing profile whose majority relation is a tournament, deciding TEQ membership is PSPACE-complete under polynomial-time many-one reductions. The witnessing voter order may be supplied.
\end{corollary}

The proof in \Cref{app:consequences} applies the realization theorem to our reduction. Since the number of voters grows with $N$, this does not establish PSPACE-hardness for a fixed number of voters. For seven voters, NP-hardness is known~\cite{BBGHKPS2019}.

\section{Discussion and open problems}\label{sec:conclusion}\label{sec:discussion}

TEQ membership is PSPACE-complete even when every recursive subproblem has a unique minimal retentive set. This separates the computational difficulty of the recursion from the failure of Schwartz's uniqueness conjecture. At the same time, whenever a tournament has an alternative with at most four dominators, one TEQ winner can be found in quadratic time (it is not necessarily the alternative that satisfies the dominator condition). Our hard instances satisfy this condition, so finding one winner is efficient on a class where deciding whether a specified alternative is a winner is PSPACE-complete.

The hereditary dominator bound (intuitively, where the dominator condition is true recursively) also lets us compare the hard instances with more restricted tournaments. On $\mD_1$, TEQ can be computed in quadratic time. On $\mD_2$, membership is NP-hard. On $\mD_3$ and $\mD_4$, it is PSPACE-complete, even with a witnessing elimination order supplied. All four classes have hereditary uniqueness and a quadratic-time selector. The selector gives a starting point for the certified-closure algorithm, which computes the entire set in $O(2^tN^2)$ time, where $t=|\TEQ(T)|$. In particular, when $t=O(\log N)$, the entire set can be computed in polynomial time even though membership on the class is PSPACE-complete.
Regardless of the total number $t$ of winners, we can find $\min\{k,t\}$ distinct winners in $O(2^{k-1}N^2)$ time, giving quadratic time for fixed $k$ and cubic time for $k=\lfloor\log_2 N\rfloor$ (\Cref{cor:partial-teq}, applied to $\mD_4$).


Four questions remain unresolved.
\begin{itemize}
    \item \emph{Degree two.} Is \prob{TEQ Membership} on $\mD_2$ PSPACE-hard, in NP, or neither? Our NP-hardness reduction uses only $\exists$-layers. Extending the quantifier construction to $\mD_2$ would require a $\forall$-layer whose elimination steps have at most two remaining dominators.
    \item \emph{Degree five.} Does a vertex with five dominators still force a unique minimal retentive set? The proof for four uses the fact that two competing sets nearly exhaust the available dominators. A fifth dominator leaves configurations not covered by that argument.
    \item \emph{Few voters.} Is TEQ membership PSPACE-complete for majority tournaments of a constant number of voters? The two-crossing transfer uses a number of voters that grows with the tournament. For seven voters, NP-hardness is known~\cite{BBGHKPS2019}.
    \item \emph{Finding a member.} Can some member of TEQ be found in polynomial time in every tournament? Our selector needs a vertex with at most four dominators, and every tournament on at most ten vertices has one. For the Banks set, a member can be found in polynomial time in every tournament even though membership is NP-complete~\cite{Hudry2004,Woeginger2003}. It would be interesting to study what further properties the winners found by our selector satisfy and whether those properties can help find a TEQ winner in arbitrary tournaments.
\end{itemize}


TEQ's role as the common endpoint of repeated retentiveness makes its complexity relevant to the broader study of tournament solutions. Together with the PSPACE-completeness of Stable Voting and Simple Stable Voting~\cite{SV2026}, our result shows that several prominent recursive rules for pairwise comparison can encode polynomial-space computation. TEQ shows that this complexity can arise from the directions of the pairwise comparisons alone, even when every recursive subproblem has a unique minimal retentive set. At the same time, on the structured tournaments in our construction, finding one TEQ winner is efficient while recognizing a specified winner is PSPACE-complete. Understanding what can be guaranteed of efficiently found winners may therefore help make recursive tournament solutions useful in settings where computing the full choice set is difficult.

\clearpage
\appendix
\section{Proofs for Sections~\ref{sec:hardness} and~\ref{sec:correctness}}\label{hard:app}

All tournaments in this appendix are explicit and unweighted. The notation $A\succ B$ prescribes all edges between $A$ and $B$, while $R_X$ denotes the contestation graph of the specific induced tournament $X$. Passing to a query changes the contestation graph, and none of the arguments below identifies $R_X$ with the restriction of a larger contestation graph without checking the relevant dominator queries.

\subsection{The normal form}\label{hard:normal}

TQBF remains PSPACE-complete for closed prenex formulas whose quantifier-free part is a conjunction of clauses with at most three literals~\cite{SM1973}. The following lemma puts such formulas into the form~\ref{F1}--\ref{F3} assumed in \Cref{sec:hardness}. Each assumption serves the construction in one way. \ref{F1} spares us clauses in which a variable occurs twice, so that the occurrence sets $M_i$ contain at most one vertex of each clause gadget. \ref{F2} is forced by our clause gadgets, which handle at most two negative literals per clause. \ref{F3} guarantees that every variable is represented by at least one vertex, which the $\forall$-layer needs for its anchor. Listing the negative literals of a clause first is only a convention, since the order of the literals within a clause is immaterial. For two formulas $\chi$ and $\chi'$ over the same variables, we say that $\chi$ implies $\chi'$ pointwise if every assignment satisfying $\chi$ satisfies $\chi'$. In that case $Q_1x_1\cdots Q_nx_n\,\chi$ implies $Q_1x_1\cdots Q_nx_n\,\chi'$, by induction on $n$.

\begin{lemma}[Normal form]\label{lem:normal}
Given a closed prenex formula whose quantifier-free part is a conjunction of clauses with at most three literals, one can compute in linear time an equivalent closed prenex formula with at least one clause that satisfies~\ref{F1}--\ref{F3}.
\end{lemma}

\begin{proof}
We apply four steps. Each new variable is fresh, so every new clause has distinct variables.

\emph{Clean.} Delete every clause that contains a literal and its negation, since it is always true, and delete repeated literals within a clause. If no clause remains, the formula is true, and we output the true formula $\exists x_1\exists x_2\exists x_3\,(x_1\lor x_2\lor x_3)$.

\emph{Pad.} Let $C$ be a clause with $k<3$ literals, now on $k$ distinct variables. Take $3-k$ fresh variables and replace $C$ by the $2^{3-k}$ clauses $C\lor\pi$, where $\pi$ ranges over the disjunctions of the fresh variables with all choices of signs. An empty clause becomes the eight clauses on three fresh variables. Under each assignment exactly one $\pi$ is false, so the conjunction of the new clauses is equivalent to $C$ under every assignment. Quantify the fresh variables existentially at the end of the prefix.

\emph{Positive literals.} Let $\psi$ be the quantifier-free part after padding, and let $\psi^+$ be the conjunction of its clauses that contain a positive literal. If $\psi=\psi^+$, skip this step. Otherwise take one fresh variable $f$, quantified universally at the front of the prefix, and for each all-negative clause $C_\ell=(\neg y_1\lor\neg y_2\lor\neg y_3)$ a fresh variable $p_\ell$, quantified existentially at the end of the prefix. Replace $C_\ell$ by
\[
(\neg y_1\lor\neg y_2\lor p_\ell)\land(\neg p_\ell\lor\neg y_3\lor f).
\]
Call the new quantifier-free part $\psi'$, and fix an assignment to all variables other than $f$ and the $p_\ell$. If $f=0$, the two new clauses can be satisfied by a choice of $p_\ell$ if and only if $C_\ell$ is true. Since each $p_\ell$ occurs only in its own two clauses, the existential quantifiers over the $p_\ell$ turn $\psi'$ into $\psi$. If $f=1$, choosing $p_\ell=1$ satisfies both new clauses, so these quantifiers turn $\psi'$ into $\psi^+$. Writing $\mathcal{Q}$ for the original prefix together with the padding variables, the new formula is therefore equivalent to $(\mathcal{Q}\,\psi)\land(\mathcal{Q}\,\psi^+)$. Since $\psi$ implies $\psi^+$ pointwise, the second conjunct follows from the first, and the new formula is equivalent to $\mathcal{Q}\,\psi$. Every new clause contains $p_\ell$ or $f$ positively.

\emph{Unused variables.} Delete the quantifiers of variables that occur in no clause. This changes no truth value.

Each step keeps the earlier properties, and the result satisfies~\ref{F1}--\ref{F3}. Every clause produces at most nine clauses and at most four fresh variables, and $f$ is added once, so the transformation takes linear time.
\end{proof}

\subsection{Composition tools, clause gadgets, and the filter}\label{hard:clause-proof}

\begin{proof}[Proof of \Cref{hard:dominant}]
Induct on $|A\cup B|$. For $a\in A$, the dominator set of $a$ is unchanged, so the contestation edges into $A$ agree with those of $R_A$. For $b\in B$, the query is $A\cup D_B(b)$ with $A\succ D_B(b)$. This query is smaller than the whole tournament, so its TEQ is $\TEQ(A)$ by induction. Thus every vertex of $B$ receives contestation edges from $\TEQ(A)$, and no vertex of $B$ sends a contestation edge into $A$. The source components are therefore exactly those of $R_A$.
\end{proof}

\begin{proof}[Proof of \Cref{hard:cyclic}]
For $a\in A$, the query is $C\cup D_A(a)$, and $C$ dominates the other part, so \Cref{hard:dominant} gives the answer $\TEQ(C)$. The analogous identities hold cyclically for $B$ and $C$. The three selected sets are strongly connected to one another and send contestation edges to every remaining vertex, while the remaining vertices send no contestation edges. Hence the union of the three selected sets is the unique source component.
\end{proof}

For the case analysis of the clause gadgets, let $u\in\{0,1\}^3$ and write $\Gad_j[u]=\Gad_j^S$, where $S=\{\oc{k}:k\in\{1,2,3\},\ u_k=1\}$. The entries of $u$ record variable values, not the truth of literals. In $\Gad_1$ the only falsifying pattern is $100$, and in $\Gad_2$ it is $110$. Before checking all cases, we explain why the output of each gadget is not selected in that pattern and is selected in every other one. The mechanism for $\Gad_0$ is described in the proof idea of \Cref{lem:clause}.
\begin{itemize}
\item In $\Gad_1$ the fixed vertices form the directed triangle $\cout\succ\ax{2}\succ\ax{1}\succ\cout$, so the output is selected when no occurrence vertex remains. The negative occurrence vertex $\oc{1}$ beats all three fixed vertices, so on its own it is a Condorcet winner and the output is not selected. Each positive occurrence vertex is beaten by the output and beats vertices $\ax{1},\ax{2},\oc{1}$. If at least one is present, the output is selected on the source cycle $\cout\R h\R\ax{1}\R\cout$ or $\cout\R h\R\oc{1}\R\cout$, according as $\oc{1}$ is absent or present, where $h$ is the retained positive occurrence vertex of largest index.
\item In $\Gad_2$ the fixed vertices are transitive with the output on top. Each negative occurrence vertex beats the output. On its own, it lies with the output on the source cycle $\cout\R\ax{2}\R\oc{1}\R\cout$ or $\cout\R\ax{1}\R\oc{2}\R\cout$, for $\oc{1}$ and $\oc{2}$ respectively, and the output is selected. If both negative occurrence vertices are present and $\oc{3}$ is absent, they form the source cycle $\ax{1}\R\oc{2}\R\oc{1}\R\ax{1}$ with vertex $\ax{1}$, and the output is not selected. The positive occurrence vertex $\oc{3}$ is beaten only by the output, and whenever it is present the output is selected, either as a Condorcet winner or on the source cycle $\cout\R\oc{3}\R\oc{1}\R\cout$ or $\cout\R\oc{3}\R\oc{2}\R\cout$.
\end{itemize}

\begin{proof}[Proof of \Cref{lem:clause}]
Recall that $\Gad_j[u]$ retains the occurrence vertex $\oc{k}$ exactly when $u_k=1$.

\emph{The gadget $\Gad_0$.} If no occurrence vertex remains, vertex $\ax{2}$ is a Condorcet winner. Otherwise the nonempty blocks $\{\cout\}$, the retained occurrence vertices, and $\{\ax{1},\ax{2}\}$ form a cyclic composition in that order, and the latter two blocks are transitive. \Cref{hard:cyclic} gives $\TEQ(\Gad_0[u])=\{\cout,\ax{2},h\}$, where $h$ is the retained occurrence vertex of largest index.

\emph{The gadget $\Gad_1$.} First suppose that neither $\oc{2}$ nor $\oc{3}$ is retained. If $\oc{1}$ is absent, the fixed vertices form a directed triangle and all are selected. If $\oc{1}$ is present, it is a Condorcet winner. Now suppose that at least one of $\oc{2},\oc{3}$ is retained, and let $h$ be the retained member of $\{\oc{2},\oc{3}\}$ of largest index. Each query at a retained vertex of $\{\oc{2},\oc{3}\}$ has Condorcet winner $\cout$, so $\cout\R x$ for these vertices. If $\oc{1}$ is absent, the queries at $\cout,\ax{1},\ax{2}$ have Condorcet winners $\ax{1},h,\cout$, respectively. The source cycle is $\cout\R h\R\ax{1}\R\cout$, and the other vertices have no path back to it. If $\oc{1}$ is present, the query at $\cout$ has winner $\oc{1}$, the queries at $\oc{1}$ and at $\ax{1}$ have winner $h$, and the query at $\ax{2}$ consists of $\{\cout,\oc{1}\}$ and the retained vertices of $\{\oc{2},\oc{3}\}$, a cyclic composition with selected set $\{\cout,\oc{1},h\}$. The source cycle is now $\cout\R h\R\oc{1}\R\cout$. Vertex $\ax{1}$, vertex $\ax{2}$, and a retained vertex of $\{\oc{2},\oc{3}\}$ of smaller index have no contestation edge back into it.

\emph{The gadget $\Gad_2$.} All eight selected sets are as follows:
\[
\setlength{\arraycolsep}{3.5pt}
\begin{array}{c|cccccccc}
u&000&001&010&011&100&101&110&111\\\hline
\TEQ(\Gad_2[u])&\{\cout\}&\{\cout\}&\{\cout,\ax{1},\oc{2}\}&\{\cout,\oc{2},\oc{3}\}&
\{\cout,\ax{2},\oc{1}\}&\{\cout,\oc{1},\oc{3}\}&\{\ax{1},\oc{1},\oc{2}\}&\{\cout,\oc{2},\oc{3}\}.
\end{array}
\]
If neither $\oc{1}$ nor $\oc{2}$ remains, vertex $\cout$ is a Condorcet winner, whether or not $\oc{3}$ remains. Without $\oc{3}$, retaining only $\oc{2}$ gives the source cycle $\cout\R\ax{1}\R\oc{2}\R\cout$ with vertex $\ax{2}$ downstream, and retaining only $\oc{1}$ gives $\cout\R\ax{2}\R\oc{1}\R\cout$ with vertex $\ax{1}$ downstream. When both $\oc{1}$ and $\oc{2}$ remain but $\oc{3}$ does not, the queries at $\cout,\ax{1},\ax{2},\oc{1},\oc{2}$ select $\{\oc{2}\},\{\oc{1}\},\{\cout,\ax{1},\oc{2}\},\{\oc{2}\},\{\ax{1}\}$, respectively: the middle query is a directed triangle and the others have Condorcet winners. The nonempty contestation out-neighborhoods are therefore
\[
\cout:\{\ax{2}\},\qquad\ax{1}:\{\ax{2},\oc{2}\},\qquad\oc{1}:\{\ax{1}\},\qquad\oc{2}:\{\cout,\ax{2},\oc{1}\},
\]
and the unique source cycle is $\ax{1}\R\oc{2}\R\oc{1}\R\ax{1}$.

Finally, suppose that $\oc{3}$ remains and at least one of $\oc{1},\oc{2}$ does. If $\oc{2}$ is absent, the queries at $\cout,\ax{1},\ax{2},\oc{1},\oc{3}$ select $\{\oc{1}\},\{\cout,\oc{1},\oc{3}\},\{\cout\},\{\oc{3}\},\{\cout\}$, respectively. The source cycle is $\cout\R\oc{3}\R\oc{1}\R\cout$, and $\ax{1},\ax{2}$ are sinks of the contestation graph. If $\oc{2}$ remains, the query at $\cout$ selects $\oc{2}$, the query at $\oc{2}$ selects $\oc{3}$, and the query at $\oc{3}$ selects $\cout$. The query at $\ax{1}$ selects $\cout$ if $\oc{1}$ is absent and the triangle $\{\cout,\oc{1},\oc{3}\}$ otherwise. The query at $\ax{2}$ is the four-vertex tournament $\{\cout,\ax{1},\oc{2},\oc{3}\}$, whose contestation graph consists of $\oc{2}\R\cout$, $\cout\R\ax{1}$, $\oc{3}\R\oc{2}$, and $\cout\R\oc{3}$. It selects $\{\cout,\oc{2},\oc{3}\}$. If present, vertex $\oc{1}$ has query winner $\oc{3}$. Thus $\cout\R\oc{3}\R\oc{2}\R\cout$ is the unique source cycle, vertices $\ax{1},\ax{2}$ are sinks, and $\oc{1}$ can point only to $\ax{1}$.

This establishes every entry of the table, and the falsifying vectors $1^j0^{3-j}$ follow.
\end{proof}

\begin{proof}[Proof of \Cref{lem:filter}]
Write $F=\Filt(H,\dout ;K,\cout )$ and $K^-=K\setminus\{\cout \}$. For $h\in H$, the query is $D_H(h)\cup K^-\cup\{\test \}$. If $K^-$ is nonempty, it dominates the other vertices of this query, so by \Cref{hard:dominant} the selected dominators of $h$ are exactly $\TEQ(K^-)$. If $K^-$ is empty, $\test $ is the Condorcet winner of the query. Hence no vertex of $H$ sends a contestation edge to another vertex of $H$. Every $k\in K^-$ has query winner $\relay $, and $\relay $ has query winner $\test $, giving $\test \R \relay \R k$. The query at $\test $ is $K$, so $\cout \R \test $ holds exactly when $\cout \in\TEQ(K)$. This is the only possible contestation edge out of $\cout $: the queries at vertices of $K^-$ select $\relay $, and $\cout $ loses to all of $H$ and to $\relay $.

Put $S=D_K(\cout )$. The query at $\cout $ is $S\cup H\cup\{\relay \}$. If $S$ is empty, $H$ dominates $\relay $ in this query. Otherwise $S\succ H\succ \relay \succ S$ is a cyclic composition. In either case its selected set meets $H$ in exactly $\TEQ(H)$. Thus $h\in H$ has a contestation edge to $\cout $ exactly when $h\in\TEQ(H)$, and this is its only possible contestation edge out.

If $\cout \in\TEQ(K)$, every $h\in\TEQ(H)$ follows $h\R \cout \R \test \R \relay $ and reaches all of $K^-$. If $K^-\ne\varnothing$, it has reached the nonempty set $\TEQ(K^-)$, whose members point to every vertex of $H$. If $K^-=\varnothing$, the vertex $\test $ points to every vertex of $H$. Either way $h$ reaches every vertex. Vertices of $H\setminus\TEQ(H)$ have no contestation edge out and have an incoming one, so they are not selected. If $\cout \notin\TEQ(K)$, then $\cout $ has no contestation edge out, and $K^-\ne\varnothing$. Every vertex of $H$ receives a contestation edge from $K^-$ but can reach outside $H$ at most the sink $\cout $, so no vertex of $H$ is selected. This proves the equivalence, including for one-vertex tournaments $H$ or $K$.
\end{proof}

\subsection{Supporting lemmas for the \texorpdfstring{$\forall$}{forall}-layer}\label{hard:and-proof}

The $\forall$-layer needs the unconditional edge $\rzero \R \slink $ to connect its two tests. Since this edge means that $\rzero $ is selected in $D(\slink )$, we prove that $\rzero $ reaches every vertex of the contestation graph of that query and then apply the return-path criterion. The following two lemmas supply the reachability argument. The first is the elementary step of the anchored route: vertices that beat the anchor are reached through a new vertex $\wanc $, and vertices that lose to it have smaller queries in which the anchor is selected.

\begin{lemma}[Completing a cycle]\label{hard:complete-cycle}
Let $Y\cup\{\anc \}$ be a tournament with $\anc \notin Y$, and add a vertex $\wanc $ with $\anc \succ \wanc \succ Y$. Then $\anc $ reaches every vertex in the contestation graph of the new tournament.
\end{lemma}

\begin{proof}
Induct on $|Y|$. Always $\anc \R \wanc $. If $x\in Y$ beats $\anc $, then $\anc $ is absent from the query at $x$, and $\wanc $ is a Condorcet winner of that query, so $\wanc \R x$. If $\anc \succ x$, the query at $x$ contains $\anc $ and $\wanc $, together with a proper subset of $Y$, with $\anc \succ \wanc \succ D_Y(x)$. By induction $\anc $ reaches every vertex of that query. The return-path criterion therefore selects $\anc $ in the query, giving $\anc \R x$.
\end{proof}

The second lemma handles arbitrary edges between the old vertices. We prove that $\rzero $ reaches the entire query, because the return-path criterion then selects it regardless of the edges among the old vertices. In the proof of \Cref{lem:and} it is applied to $D(\slink )$ with $C=V(H)\setminus\{\dout \}$, $L=\Hs{0}\setminus\{\dout \}$, and the given $M$ and $\anc $. Only the anchor $\anc $, not all of $M$, is needed.

\begin{lemma}[An anchored return path]\label{hard:anchor}
Let $C=L\mathbin{\dot\cup}M$ be a tournament with $M\ne\varnothing$, let $\anc \in M$, and add vertices $\rzero ,\wanc $ with
\[
L\succ \rzero \succ M,\qquad \rzero \succ \wanc ,\qquad \anc \succ \wanc \succ C\setminus\{\anc \}.
\]
Then $\rzero $ reaches every vertex in the contestation graph. The conclusion still holds after adding a vertex $\groot $ with $C\succ \groot \succ\{\rzero ,\wanc \}$.
\end{lemma}

\begin{proof}
First consider the simpler construction on $A=A_L\mathbin{\dot\cup}A_M$ with $A_L\succ \rzero \succ A_M$ and $\rzero \succ \wanc \succ A$. By induction on $|A|$, the vertex $\rzero $ reaches everything: $\rzero \R \wanc $, every query at a vertex of $A_L$ has winner $\wanc $, and a query at a vertex of $A_M$ is a smaller instance of the same construction. This includes the case $A=\varnothing$.

For the anchored construction, induct on $|C|$. The query at $\wanc $ is $\{\rzero ,\anc \}$ with winner $\rzero $, so $\rzero \R \wanc $. Let $U=\{x\in C:x\succ \anc \}$. For $x\in U\cap L$, the query at $x$ has winner $\wanc $. For $x\in U\cap M$, the anchor is absent from the query at $x$, which is an instance of the simpler construction, so $\rzero \R x$. Thus $\rzero $ reaches all of $U$. The nonempty query at $\anc $ is $U\cup\{\rzero \}$, all of whose vertices have been reached, so following a selected dominator gives a path to $\anc $. For $x\in L$ with $\anc \succ x$, the query at $x$ contains $\anc $ and $\wanc $ with $\anc \succ \wanc \succ D_C(x)\setminus\{\anc \}$, and \Cref{hard:complete-cycle} gives $\anc \R x$. For $x\in M$ with $\anc \succ x$, the query at $x$ is a smaller anchored instance, and induction gives $\rzero \R x$. Hence $\rzero $ reaches all of $C\cup\{\rzero ,\wanc \}$.

Adding $\groot $ changes no query at a vertex of $C$, since $\groot$ loses to all of $C$. It does change the query at $\rzero$, but that query plays no role: the argument only follows contestation edges out of $\rzero$, never into it. The query at $\wanc $ becomes the triangle $\rzero \succ \anc \succ \groot \succ \rzero $, so $\rzero \R \wanc $ still holds, and all paths into $C$ remain. Finally $D(\groot )=C$, so a reached vertex of $\TEQ(C)$ points to $\groot $.
\end{proof}

\subsection{Complete proofs of the layer lemmas}\label{hard:wrapper-proofs}

The outlines in \Cref{sec:hard-or,sec:hard-and} explain how the readers and return paths combine the two answers. We verify here that every query has the claimed answer and that no other edge changes the computation. The proof for the $\forall$-layer uses the two lemmas of \Cref{hard:and-proof}.

\begin{proof}[Proof of \Cref{lem:or}]
Let $W=\Wex(H,\dout ;\Hs{1},\Hs{0})$. We determine the contestation edges query by query and then apply \Cref{lem:return} with root $\broot $. The shield plays two roles. It preserves the answer in each of $\Hs{1}$ and $\Hs{0}$ despite the additional guards, and it prevents old vertices from creating further contestation edges among themselves.

\emph{The readers query $\Hs{1}$ and $\Hs{0}$.} The query at $\rone $ is $D(\rone )=\Hs{1}\cup\{\shw ,\shz ,\tri \}$. It is a shield instance (\Cref{lem:shield}) with guards $u=\shw $ and $v=\shz $, kept part $\Hs{1}$, and hidden part $\{\tri \}$: indeed $\shw \succ \Hs{1}\cup\{\tri \}$, $\Hs{1}\succ \shz \succ \shw $, and $\shz \succ \tri $. The kept part contains $\dout $, so $\TEQ(D(\rone ))=\TEQ(\Hs{1})\cup\{\shw ,\shz \}$. In the same way, $D(\rzero )=\Hs{0}\cup\{\shw ,\shz ,\rone \}$ is a shield instance with guards $(\shw ,\shz )$, kept part $\Hs{0}$, and hidden part $\{\rone \}$, since $\shw \succ \Hs{0}\cup\{\rone \}$, $\Hs{0}\succ \shz \succ \shw $, and $\shz \succ \rone $. Hence $\TEQ(D(\rzero ))=\TEQ(\Hs{0})\cup\{\shw ,\shz \}$. Consequently
\[
\dout \R \rone \iff\bet{1}=1,\qquad \dout \R \rzero \iff\bet{0}=1,\qquad \shw ,\shz \R \rone ,\qquad \shw ,\shz \R \rzero .
\]

\emph{The old vertices are silenced.} For $h\in V(H)$, let $P_h$ be the set of dominators of $h$ among $\rone ,\rzero ,\tri $. It always contains $\tri $, because $\tri \succ H$. The query at $h$ is $D_H(h)\cup\{\broot,\shw \}\cup P_h$. It is a shield instance with guards $u=\shw $ and $v=\broot$, kept part $P_h$, and hidden part $D_H(h)$, since $\shw \succ P_h\cup V(H)$, $P_h\succ \broot \succ \shw $, and $\broot \succ V(H)$. Hence $\TEQ(D(h))=\TEQ(P_h)\cup\{\shw ,\broot \}$. Two facts follow: $\broot \R h$ for every old vertex $h$, and no old vertex has a contestation edge to another old vertex.

\emph{The root reaches every vertex.} The remaining queries are $D(\broot )=\{\rone ,\rzero ,\tri \}$, which is the directed triangle $\rone \succ \rzero \succ \tri \succ \rone $ and selects all three vertices, the queries $D(\shw )=\{\broot,\shz \}$ and $D(\shz )=V(H)\cup\{\broot\}$, which both have Condorcet winner $\broot $, and the query $D(\tri )=\{\rzero ,\shw ,\shz \}$, which has Condorcet winner $\shz $. They give $\rone ,\rzero ,\tri \R \broot $, $\broot \R \shw $, $\broot \R \shz $, and $\shz \R \tri $. Hence $\broot $ reaches $\shw $ and $\shz $ directly, $\rone $, $\rzero $, and $\tri $ through $\shz $, and every old vertex directly.

\emph{The output has no other exit.} The out-neighbors of $\dout $ are $\shz $, the two readers, and the old vertices it beats. The query at $\shz $ selects only $\broot $, and we showed that no old vertex has a contestation edge from $\dout $. So the only possible contestation edges out of $\dout $ are $\dout \R \rone $ and $\dout \R \rzero $, and both continue to $\broot $. By \Cref{lem:return}, $\dout $ is selected if and only if one of these two conditional edges is present, that is, if and only if $\bet{0}\lor\bet{1}$.
\end{proof}

\begin{proof}[Proof of \Cref{lem:and}]
Let $W=\Wall(H,\dout ;M,\anc )$, and write $R$ for its contestation graph. We first list the queries at old vertices, which show that the old vertices cannot create a shortcut. We then verify the two tests and the link, exclude the remaining shortcuts, and show that $\groot $ reaches every vertex. The return-path criterion is used twice: first inside the query $D(\rzero )$ to establish the $0$-test, and then in $W$ itself.

The table of $\Wall$ in \Cref{sec:hard-tools} gives the following dominator sets for old vertices:
\begin{equation}\tag{$\dagger$}\label{hard:oldqueries}
\begin{array}{c|l}
\dout&D_H(\dout)\cup\{\eout,\king,\slink,\wanc\}\\
x\in \Hs{0}\setminus\{\dout\}&D_H(x)\cup\{\king,\wanc\}\\
\anc&D_H(\anc)\cup\{\king,\rzero\}\\
y\in M\setminus\{\anc\}&D_H(y)\cup\{\king,\rzero,\wanc\}.
\end{array}
\end{equation}
In each of these sets other than the first, $\king $ is a Condorcet winner, since $\king $ beats all of $H$ and beats $\rzero $ and $\wanc $. Thus every old vertex other than $\dout $ receives a contestation edge from $\king $ and from no other vertex. None of $\eout ,\rzero ,\slink ,\dout $ can therefore enter an old vertex other than $\dout $, so these vertices cannot leave the intended route through an old vertex.

\emph{The $1$-test.} The query at $\groot $ is $D(\groot )=V(H)=\Hs{1}$, so $\dout \R \groot $ if and only if $\bet{1}=1$.

\emph{The $0$-test.} We analyze the contestation graph $R_X$ of the query $X=D(\rzero )=\Hs{0}\cup\{\eout ,\king ,\groot \}$, which differs from $R$. The additional vertices mean that selecting $\eout $ in $X$ is not literally the question whether $\dout $ is selected in $\Hs{0}$, so we apply the return-path criterion a second time, inside $X$. The intended route from $\eout $ to $\groot $ is $\eout \RX{X}\dout \RX{X}\groot $. Note that the edge from $\dout $ to $\groot $ tests $\bet{0}$ inside $X$ but $\bet{1}$ in $W$. Restricting \Cref{def:and} and~\eqref{hard:oldqueries} to $X$ gives
\[
D_X(\groot )=\Hs{0},\quad D_X(\king )=\{\eout ,\groot \},\quad D_X(\eout )=(\Hs{0}\setminus\{\dout \})\cup\{\groot \},\quad D_X(\dout )=D_{\Hs{0}}(\dout )\cup\{\eout ,\king \},
\]
and $D_X(x)=D_{\Hs{0}}(x)\cup\{\king \}$ for $x\in \Hs{0}\setminus\{\dout \}$. In $R_X$, therefore, $\groot \RX{X}\king $ because $\groot \succ \eout $, $\king \RX{X}x$ for every $x\in \Hs{0}\setminus\{\dout \}$ because $\king \succ \Hs{0}$, and $\dout \RX{X}\groot $ exactly when $\bet{0}=1$. The query $D_X(\dout )$ selects $\eout $: either $D_{\Hs{0}}(\dout )=\varnothing$ and $\eout $ beats $\king $, or the blocks $\{\king \}\succ D_{\Hs{0}}(\dout )\succ\{\eout \}\succ\{\king \}$ form a cyclic composition (\Cref{hard:cyclic}). So $\eout \RX{X}\dout $. The out-neighbors of $\eout $ in $X$ are $\dout $ and $\king $, and the query at $\king $ selects only $\groot $, so the only contestation edge out of $\eout $ goes to $\dout $. The out-neighbors of $\dout $ in $X$ are $\groot $ and some vertices of $\Hs{0}\setminus\{\dout \}$, whose queries select only $\king $. So the only possible contestation edge out of $\dout $ goes to $\groot $, and it is present exactly when $\bet{0}=1$.

It remains to verify that $\groot $ is a root of $R_X$. It reaches every vertex of $\Hs{0}\setminus\{\dout \}$ through $\king $. These vertices and $\groot $ itself form the nonempty query $D_X(\eout )$, so some selected dominator of $\eout $ has been reached, and $\groot $ reaches $\eout $ and then $\dout $. By \Cref{lem:return}, both $\eout $ and $\dout $ are selected in $X$ exactly when $\eout \RX{X}\dout \RX{X}\groot $ is a path, that is, exactly when $\bet{0}=1$, and $\groot $ is always selected. Since $X=D(\rzero )$, selection in $X$ gives contestation edges into $\rzero $ in $W$:
\[
\eout \R \rzero \iff\bet{0}=1,\qquad \dout \R \rzero \iff\bet{0}=1,\qquad \groot \R \rzero .
\]

\emph{The link from $\rzero $ to $\dout $.} First, $\slink \R \dout $. In $U=D(\dout )=D_H(\dout )\cup\{\eout ,\king ,\slink ,\wanc \}$, the query at $\king $ is $\{\eout ,\slink \}$, with Condorcet winner $\slink $. The queries at the vertices of $D_H(\dout )$, and the query at $\wanc $, which is $\{\king \}$ or $\{\king ,\anc \}$, all have Condorcet winner $\king $. So $\slink $ reaches $\king $, every vertex of $D_H(\dout )$, and $\wanc $. In particular, $\slink $ reaches every vertex of the nonempty query $D_U(\eout )=D_H(\dout )\cup\{\slink ,\wanc \}$, so $\slink $ also reaches $\eout $. Thus $\slink $ reaches every vertex of $U$, and it is selected there by \Cref{lem:return}. Second, $\rzero \R \slink $. Inside the query $D(\slink )=(V(H)\setminus\{\dout \})\cup\{\groot ,\rzero ,\wanc \}$ the dominators of $\wanc $ form the triangle $\rzero \succ \anc \succ \groot \succ \rzero $, so $\rzero \R \wanc $ inside $D(\slink )$. \Cref{hard:anchor}, applied with $C=V(H)\setminus\{\dout \}$ and $L=\Hs{0}\setminus\{\dout \}$, extends this route: $\rzero $ reaches every vertex of $D(\slink )$ and is therefore selected there.

\emph{No shortcuts.} Two edges would break the serial route, and neither is present. The edge $\eout \R \dout $ would skip the $0$-test. In $U=D(\dout )$, the only out-neighbor of $\eout $ is $\king $, and the query at $\king $ in $U$ selects only $\slink $. So $\eout $ has no contestation edge out in $R_U$. Its own query $D_U(\eout )$ is nonempty, so $\eout $ also has an incoming contestation edge. A vertex with an incoming edge and no outgoing edge is not in a source component, so $\eout \notin\TEQ(U)$ and $\eout \R \dout $ fails. The edge $\rzero \R \wanc $ would let $\rzero $ leave the route. The query $D(\wanc )=\{\anc ,\king ,\groot ,\rzero \}$ has the contestation edges $\king \R \anc $, $\anc \R \groot $, $\groot \R \king $, and $\groot \R \rzero $, because inside it the queries at $\anc ,\groot ,\king ,\rzero $ have Condorcet winners $\king ,\anc ,\groot ,\groot $ (\Cref{fig:anchor}(b)). Compared with the triangle inside $D(\slink )$, the presence of $\king $, which beats $\anc $ and $\rzero $ and loses to $\groot $, turns $\{\anc ,\king ,\groot \}$ into the source cycle and leaves $\rzero $ downstream. So $\TEQ(D(\wanc ))=\{\anc ,\king ,\groot \}$ and $\rzero \R \wanc $ fails.

Collecting the out-neighbors gives the exact contestation out-neighborhoods in~\eqref{hard:series}. Indeed, $\eout $ beats only $\dout ,\king ,\rzero $, and the query at $\king $, which is $\{\eout ,\slink ,\groot \}$, selects only $\groot $. The vertex $\rzero $ beats only $M\cup\{\slink ,\wanc \}$, and the queries at vertices of $M$ select only $\king $ by~\eqref{hard:oldqueries}. The vertex $\slink $ beats only $\eout ,\king ,\dout $, and in $D(\eout )=(V(H)\setminus\{\dout \})\cup\{\groot ,\slink ,\wanc \}$ the vertex $\slink $ loses to every other vertex, so it has no contestation edge out there while its own query there is nonempty, and it is not selected. The vertex $\dout $ beats only $\rzero $, $\groot $, and some old vertices, whose queries select only $\king $.

\emph{The root.} The query at $\king $ selects $\groot $, so $\groot \R \king $. Then $\king $ reaches every vertex of $V(H)\setminus\{\dout \}$ by~\eqref{hard:oldqueries}, and $\king \R \wanc $ because $\king \in\TEQ(D(\wanc ))$. Also $\groot \R \rzero \R \slink \R \dout $. Thus $\groot $ reaches every vertex of the nonempty query $D(\eout )=(V(H)\setminus\{\dout \})\cup\{\groot ,\slink ,\wanc \}$, hence some selected dominator of $\eout $, and therefore $\eout $. So $\groot $ reaches every vertex. By \Cref{lem:return}, $\eout $ is selected if and only if it reaches $\groot $, which by~\eqref{hard:series} happens if and only if $\bet{0}\land\bet{1}$.
\end{proof}

\subsection{The degree certificate of Section~\ref{sec:hard-degree}}\label{hard:degree}

\begin{proof}[Proof of \Cref{lem:degree}]
The risk is that guards from many layers could all dominate the same vertex. We first remove some guards from each layer so that the remaining edges into old vertices have only two controlled forms: one surviving layer vertex may beat each occurrence vertex of its own variable, and at most two may beat an output. Because the occurrence sets are disjoint and a $\forall$-layer replaces its input output, these contributions do not accumulate. We then remove the clause tournament, the leftover filter vertices, and the rest of each layer. All counts below are current indegrees. To name vertices globally, we give the vertices of layer $i$ the subscript $i$, use the names $\cout_\ell,\axl{\ell}{1},\axl{\ell}{2},\ocl{\ell}{1},\ocl{\ell}{2},\ocl{\ell}{3}$ of \Cref{def:reduction} for the vertices of the gadget of $C_\ell$, and write $\relay_\ell,\test_\ell$ for the relay and test of the filter of $C_\ell$, for $\ell\ge2$.

\emph{Phase 1: outer guards.} Process the layers from the outermost toward the clause tournament. At an $\exists$-layer, where $\Hs{1}=V(H)$ and $\Hs{0}=V(H)\setminus M_i$, delete $\shw,\tri,\broot$ in that order. At a $\forall$-layer delete $\king,\wanc$. The vertices removed in this phase are neither occurrence vertices nor outputs, and surviving outer layers beat old vertices only at vertices of those two kinds, so they do not increase the counts below. Within each active layer and its old vertices, the remaining dominators are
\[
\begin{array}{c|c|l}
\text{Layer}&\text{Removed vertex}&\text{Remaining dominators}\\\hline
\exists&\shw&\broot,\shz\\
\exists&\tri&\rzero,\shz\\
\exists&\broot&\rone,\rzero\\
\forall&\king&\eout,\slink,\groot\\
\forall&\wanc&\anc_i,\groot,\rzero
\end{array}
\]
The retained vertices of an $\exists$-layer are $\shz,\rone,\rzero$, whose only edges into old vertices are $\rzero\succ M_i$. The retained vertices of a $\forall$-layer are $\eout,\slink,\groot,\rzero$, whose only edges into old vertices are $\eout_i,\slink_i\succ\orole_{i+1}$ and $\rzero\succ M_i$. Inductively, none of the guards deleted in this phase has an additional dominator from an earlier layer, so every deletion in this phase has degree at most three.

After this phase, an occurrence vertex has at most one dominator outside the clause tournament, namely the reader $\rzero$ of its own variable. An output has at most two such dominators, the vertices $\eout,\slink$ of the next $\forall$-layer, which replaces that output. Intervening $\exists$-layers add no incoming edges to it, and once an output has been replaced it is never designated again, so this budget does not accumulate. Outputs and occurrence vertices are disjoint, so the two budgets never add up at a single vertex. Every other vertex of the clause tournament has no dominator among the retained layer vertices.

\emph{Phase 2: the filters.} The right child of a filter sends many edges into its left child, so we remove it first. Process the filters in the reverse order of construction, from the last clause toward the first. In a filter $\Filt(H,\dout;K,\cout)$, the right child $K$ is a clause gadget with output $\cout$, and the left child $H$ is the clause tournament built so far. First delete the five vertices of $K\setminus\{\cout\}$. Each has one possible dominator $\relay$ from its filter, and an occurrence vertex can additionally have the $0$-reader of its variable as a dominator. Previously retained filter pairs lose to this entire active part, so no further retained vertex dominates it. The following orders account for these worst-case contributions:
\[
\begin{array}{c|c|c}
\text{Gadget}&\text{Order on }K\setminus\{\cout\}&\text{Current indegrees, at most}\\\hline
\Gad_0&\oc{3},\ax{2},\ax{1},\oc{2},\oc{1}&3,3,3,3,3\\
\Gad_1&\oc{3},\oc{1},\ax{1},\ax{2},\oc{2}&3,3,3,3,3\\
\Gad_2&\oc{3},\ax{1},\oc{2},\ax{2},\oc{1}&3,3,2,2,2
\end{array}
\]
These numbers follow directly from the adjacency lists of the gadgets, with the output $\cout$ retained throughout. Now $\test$ has only $\cout$ as a remaining dominator. Delete $\test$. Leave $\relay$ and $\cout$ for later: they lose to every vertex of the remaining left child and add no incoming edge to it. Repeat until the first clause is reached.

The first clause has no remaining incoming edges from filter vertices. Its output can have up to two dominators from an enclosing $\forall$-layer. Each occurrence vertex can have one reader as a dominator, and the other vertices have none. Use the complete orders
\[
\begin{array}{c|c|c}
\text{Gadget}&\text{Complete order}&\text{Current indegrees, at most}\\\hline
\Gad_0&\oc{3},\ax{2},\ax{1},\cout,\oc{2},\oc{1}&2,2,2,2,1,1\\
\Gad_1&\oc{3},\oc{1},\ax{1},\cout,\ax{2},\oc{2}&2,2,2,2,1,1\\
\Gad_2&\oc{3},\ax{1},\oc{2},\ax{2},\oc{1},\cout&2,2,1,1,1,2
\end{array}
\]

\emph{Phase 3: leftover filter vertices.} All clause vertices except the retained outputs of the right children are now gone. Process the leftover filters from the first constructed to the last, removing $\relay_\ell$ and then $\cout_\ell$. Their left child has already been removed, and both have indegree zero when removed.

\emph{Phase 4: the rest of each layer.} Process the layers from the innermost to the outermost. The rest of an $\exists$-layer is removed in the order $\shz,\rone,\rzero$. It is transitive in that direction, and no outer layer sends an edge into it, so each removal has degree zero. The rest of a $\forall$-layer is removed in the order $\groot,\rzero,\slink,\eout$. Vertex $\groot$ beats the other three. After its removal, the remaining triangle is $\eout\succ\rzero\succ\slink\succ\eout$, so $\rzero$ has degree one and then $\slink$ degree zero. The output $\eout$ has at most its two dominators from an enclosing $\forall$-layer. This completes an elimination order of maximum current indegree three.


\emph{The order as a list.} Concatenate: for $i=1,\ldots,n$, the list $(\shw_i,\tri_i,\broot_i)$ if $Q_i=\exists$ and $(\king_i,\wanc_i)$ if $Q_i=\forall$; for $\ell=m,\ldots,2$, the vertices of the gadget of $C_\ell$ other than $\cout_\ell$, in the order of the first table, followed by $\test_\ell$; the vertices of the gadget of $C_1$ in the order of the second table; for $\ell=2,\ldots,m$, the pair $(\relay_\ell,\cout_\ell)$; and for $i=n,\ldots,1$, the list $(\shz_i,\rone_i,\rzero_i)$ if $Q_i=\exists$ and $(\groot_i,\rzero_i,\slink_i,\eout_i)$ if $Q_i=\forall$. Every vertex occurs exactly once, and the list is computed in linear time from the construction.
\end{proof}

\section{Proofs for Sections~\ref{sec:uniqueness} and~\ref{sec:algorithms}}\label{uniq:app}

All tournaments in this appendix are finite, and $D(v)=D_T(v)$ when the ambient tournament is clear. As in \Cref{sec:uniqueness}, a set \emph{dominates} another if every vertex of the latter is beaten by some vertex of the former. The structural arguments in the first two subsections do not assume uniqueness in smaller tournaments. The closure argument below uses the hereditary uniqueness hypothesis of its theorem.

\subsection{Basic facts and captains}

We first show that a retentive set cannot be beaten in its entirety by an outsider. If it were, the query at any of its vertices would contain the outsider but would select only members of the set. Applying the same observation to that smaller query gives a contradiction. This domination property then lets us count the dominators that a minimal retentive set must contribute.

\begin{proof}[Proof of \Cref{uniq:basic}]
(1) Induct simultaneously over all tournaments of a given order and all their retentive sets. The order-one case is immediate. Suppose that an outsider $x$ beats every vertex of a retentive set $R$. For $r\in R$, the smaller nonempty tournament $H=T[D(r)]$ contains $x$, whereas $S=\TEQ(H)\subseteq R$ avoids $x$. The set $S$ is nonempty and retentive in $H$, because $H$ has a minimal retentive set and unions of retentive sets are retentive. By induction, some vertex of $S$ beats $x$, a contradiction.

(2) A nonempty intersection of retentive sets is retentive, so distinct minimal retentive sets are disjoint. Let $R$ be minimal retentive. The condensation of the tournament $T[R]$ is an acyclic tournament and therefore has a unique source strongly connected component $C$. For $c\in C$ we have $D(c)\cap R\subseteq C$, so retentiveness of $R$ implies retentiveness of $C$, and minimality gives $C=R$.

(3) If $R=\{r\}$ is retentive, then $D(r)=\varnothing$, since otherwise its nonempty TEQ would lie in $\{r\}\cap D(r)=\varnothing$. Thus $r$ is a Condorcet winner, and by (1) every retentive set contains $r$, so $\{r\}$ is the unique minimal retentive set. A two-vertex set is not strongly connected, so by (2) it is not minimal retentive.

(4) By (1), some vertex of $R$ beats $v$. Suppose $a$ is the only one. For every $b\in R\setminus\{a\}$, the query $T[D(b)]$ contains $v$, and its TEQ lies in $R$. By (1) applied in that query, it contains a vertex beating $v$, necessarily $a$. Hence $a\succ b$ for every $b\in R\setminus\{a\}$, contradicting the strong connectivity of $T[R]$, which has at least two vertices.
\end{proof}

The remark after \Cref{uniq:basic} follows: if $X$ is nonempty and $\TEQ(X)$ lies in a set of size at most two, then its minimal retentive sets are singletons by~(3), so $\TEQ(X)$ is a single Condorcet winner. To see why captain edges constrain retentive sets, every retentive set is closed under captain predecessors in $T$: if $a$ captains $b$, then $\TEQ(D(b))=\{a\}$, so every retentive set containing $b$ contains $a$. The selector uses this closure.

\subsection{Small intersections}

The main proof uses captain edges inside small intersections with $D(v)$. For two vertices, the domination lemma already supplies one. With three, we turn each alleged failure into a witness outside the intersection. A dominator query at that witness forces a singleton answer, which makes one intersection vertex its captain. Two failures then demand opposite orientations between their witnesses.

\begin{proof}[Proof of \Cref{uniq:two-three}]
Put $A=R\cap D(v)$. When $A=\{a,b\}$ with $a\succ b$, \Cref{uniq:intersection-dominates} says that no vertex of $R\setminus A$ beats both $a$ and $b$, and no vertex of $A$ can. Thus $a$ captains $b$ in $T[R]$, and \Cref{uniq:persistence} lifts this to $T$.

Now let $|A|=3$. Call $a\in A$ \emph{eligible} if it is not a Condorcet winner of $T[A]$. At least two vertices are eligible, and all three are eligible when $T[A]$ is a directed triangle. Suppose that, for an eligible $a$, the edge between the two vertices of $A\setminus\{a\}$ is not a captain edge of $T$. By the contrapositive of \Cref{uniq:persistence}, some $x_a\in R$ beats both vertices of $A\setminus\{a\}$. The witness is not one of these two vertices, and eligibility excludes $x_a=a$. Thus $x_a\notin A$, so $v\succ x_a$, and \Cref{uniq:intersection-dominates} forces $a\succ x_a$.

We now use $x_a$ to force a captain edge out of $a$. Its query contains $v$, but retentiveness confines every selected vertex to $R$. Among the vertices of $R$ in the query, only $a$ beats $v$. The two-dominator bound therefore rules out a nonsingleton minimal retentive set in this query. Formally, let $H=T[D(x_a)]$. Every minimal retentive set of $H$ lies in $R$ and avoids $v$, and
\[
R\cap D(x_a)\cap D(v)=\{a\}.
\]
A minimal retentive set of $H$ with at least two vertices would contain two dominators of $v$ by \Cref{uniq:basic}(4), which this intersection does not allow. So the minimal retentive set of $H$ is a singleton that dominates $v$, namely $\{a\}$. Consequently $a$ is a Condorcet winner of $H$, that is, $a$ captains $x_a$.

Two distinct eligible vertices $a,b$ cannot both have failing complementary pairs. Their witnesses would satisfy $x_a\succ b$ and $x_b\succ a$, and they are distinct, since $a\succ x_a$ but $x_b\succ a$. Because $a$ captains $x_a$, the edge $x_b\succ a$ forces $x_a\succ x_b$. Because $b$ captains $x_b$, the edge $x_a\succ b$ forces $x_b\succ x_a$. Hence at most one eligible vertex has a failing complementary pair, which proves both assertions.
\end{proof}

Eligibility is essential: if $T[A]$ is transitive, its top vertex may itself witness the failure of the other pair. This is why the assertion about two captain edges is restricted to directed triangles.

For the next lemma, the obstacle is a captain edge outside the two-vertex intersection. The two captain conditions constrain all edges between the pairs. In either resulting orientation, a witness from a retentive query forces a singleton answer that is beaten by a vertex in that query.

\begin{proof}[Proof of \Cref{uniq:disjoint}]
Write $R\cap D(v)=\{a,b\}$ with $a\succ b$. By \Cref{uniq:two-three}, $a$ captains $b$. Suppose $c$ captains $d$ with distinct $c,d\in D(v)\setminus\{a,b\}$. Both $c$ and $d$ lie outside $R$. The two captain conditions leave two cases.

If $a\succ c$, the captain conditions force $c\succ b$, $b\succ d$, and $d\succ a$. Since $d\in D(a)$, the set $S=\TEQ(T[D(a)])\subseteq R$ dominates $d$ inside this query by \Cref{uniq:basic}(1). Choose $x\in S$ with $x\succ d$. Thus $x$ beats $a$ and $d$, and the captain edges force $b\succ x$ and $c\succ x$, since otherwise $x$ would beat both endpoints of one of them. In particular $x\notin\{a,b,c,d\}$, and $x\in R\setminus\{a,b\}$ gives $v\succ x$. In $H=T[D(x)]$, all minimal retentive sets lie in $R$ and avoid $v$, and the only dominator of $v$ they can contain is $b$. By \Cref{uniq:basic}(4) the minimal retentive set of $H$ is a singleton dominating $v$, namely $\{b\}$. This is impossible, because $c\in V(H)$ and $c\succ b$.

If $c\succ a$, the captain conditions force $a\succ d$, $d\succ b$, and $b\succ c$. Now $c\in D(a)$, and domination by $\TEQ(T[D(a)])\subseteq R$ supplies $x$ beating $a$ and $c$. This time $b,d\succ x$, and $v\succ x$ as before. In $T[D(x)]$ the forced singleton is again $\{b\}$, which is contradicted by $d\succ b$.
\end{proof}

\subsection{Implementing the closure algorithm}\label{uniq:closure-details}

The closure proof in \Cref{sec:closure} bounds the number of distinct query domains. To obtain the claimed deterministic running time, we must also store and process each domain efficiently. Sets are stored as $N$-entry membership vectors, and the cache is a binary trie keyed by the $N$ bits of a domain, so each lookup or insertion takes $O(N)$ time deterministically. No constant-time operation on an arbitrary $N$-bit integer is assumed. For each new domain $X$, the selector costs at most $p(N)$, and constructing the induced adjacency matrix, if needed, costs $O(N^2)$. At most $N$ queue entries are processed per domain, and each requires $O(N)$ work for the intersection $X\cap D(r)$, the union with the answer, and the cache lookup. The calls are made in depth-first order, and completed answers suffice for memoization: a child domain has smaller cardinality than every active ancestor, so an unfinished domain is never reentered. With at most $2^t$ domains, the total time is $O(2^t(p(N)+N^2))$. The trie has $O(N2^t)$ nodes, so word addresses need $O(t+\log N)$ bits.

\subsection{The selector}\label{uniq:selector-details}

Let $v$ have at most four dominators, put $A=D(v)$, and let $R=\TEQ(T)$, the unique minimal retentive set by \Cref{thm:local}. We use $R$ only in the correctness proof: the algorithm inspects $T[A]$ and tests captain edges in $T$. The first two rows of \Cref{tab:selector} were verified in \Cref{sec:selector}.

\begin{proof}[Proof of \Cref{uniq:four-vertices}]
Necessarily $|A|\in\{3,4\}$. Every local indegree is positive and the indegrees sum to $\binom{|A|}{2}$, so some $u$ has local indegree one. Let $p$ be its sole dominator in $A$. The local dominator set of $p$ is nonempty, has at most two vertices, and therefore has a Condorcet winner $q$. Then
\[
p\succ u\succ q\succ p.
\]
The edges $p\succ u$ and $q\succ p$ are captain edges of $T[A]$ by construction. The edge $u\succ q$ is also one: $p$ does not beat $q$, and the possible fourth vertex is beaten by $u$. Each vertex of $E=\{p,u,q\}$ has its predecessor in the cycle as a captain, so the query at that vertex selects exactly its predecessor. Thus $E$ is retentive. Any nonempty retentive subset of $E$ must contain the predecessor of each of its vertices and hence all three vertices. Therefore $E$ is minimal retentive in $T[A]$. Any other minimal retentive set would be disjoint from $E$ and therefore a singleton, contradicting the absence of a Condorcet winner. Thus $\TEQ(T[A])=E$. A vertex has at most one captain, since two captains of the same vertex would have to beat each other. Each vertex of $E$ already has a captain in $E$, so the fourth vertex captains none of them, and at most one captain edge enters the fourth vertex.
\end{proof}

\begin{proof}[Proof of \Cref{prop:selector}]
The first five rows were verified in \Cref{sec:selector}. For the last two rows, assume that $T[A]$ has no Condorcet winner, so that $R$ has at least three vertices, and use the constraints (C1)--(C4) of \Cref{sec:selector} on $I=A\cap R$.

\emph{$C$ consists of one edge $a\succ b$.} Its tail lies in $E$, so the returned set $E\cap\{a,b\}$ is nonempty. If $v\in R$, then $E\subseteq R$. If $v\notin R$, then either $I=A$, or $I$ contains the unique edge of $C$. In both cases $\{a,b\}\subseteq R$.

\emph{Otherwise.} Now $C$ has at least two edges, no two of which are disjoint, and $C$ does not contain the triangle on $E$. A family of pairwise intersecting edges is a star or the three edges of a triangle, and the only triangle available in the local captain graph is the excluded one on $E$. So all edges of $C$ share an endpoint $s$. It lies in $E$, because the fourth vertex is incident to at most one captain edge. Hence $s\in R$ if $v\in R$. If $v\notin R$ and $I\ne A$, then $I$ contains an edge of $C$ and hence $s$. If $I=A$, then $s\in I$ trivially.

\emph{Running time.} The indegrees and a suitable $v$ can be found in $O(N^2)$ time from the adjacency matrix. The tournament $T[A]$ has constant size, and testing a proposed captain edge $a\succ b$ requires scanning all vertices for a common dominator, in $O(N)$ time. At most six edges need to be tested.
\end{proof}

\section{Proofs for Section~\ref{sec:boundaries}}\label{rest:details}

\subsection{Quadratic time on \texorpdfstring{$\mD_1$}{D1}}\label{alg:d1}

\begin{proof}[Proof of \Cref{thm:boundary}(a)]
Each application of \Cref{lem:one-dominator} identifies part of the answer and leaves at most one smaller TEQ query. We follow this chain until a Condorcet winner is reached, recording the vertices already identified. This gives an $O(N^2)$-time algorithm from an adjacency matrix, without a witnessing elimination order.

Return $\varnothing$ on an empty input. Otherwise, maintain an induced set $X$, initially $V$, and an accumulated answer $A$, initially empty. The recorded vertices and the answer still to be found satisfy
\[
\TEQ(T)=A\cup\TEQ(T[X]),\qquad A\cap X=\varnothing.
\]
Choose $x\in X$ of current indegree at most one, which exists by the hereditary promise. If $x$ has indegree zero, return $A\cup\{x\}$. Otherwise let $D_X(x)=\{y\}$ and $I=D_X(y)$. If $I=\varnothing$, return $A\cup\{y\}$. If not, add $x,y$ to $A$, replace $X$ by $I$, and repeat. \Cref{lem:one-dominator} preserves the invariant, because the selected vertices outside $I$ are exactly $x$ and $y$. The current set strictly shrinks and remains nonempty until a terminal case is reached.

For the running time, initialize the indegrees and a queue of vertices of indegree at most one. When $X$ is replaced by $I$, remove every vertex of $X\setminus I$, not only $x$ and $y$. For each removed vertex, decrement the indegree of each remaining out-neighbor and enqueue a vertex when it becomes eligible. Each vertex is removed once, and each ordered pair is inspected a constant number of times. Scanning for $y$ and $I$ in each iteration adds $O(N^2)$ work in total, and stale queue entries are ignored. The promise itself can be checked in $O(N^2)$ time by the same greedy deletion on a separate copy of the indegree counters.
\end{proof}

Every nonempty induced subtournament of a member of $\mD_1$ therefore has a unique minimal retentive set. Its selected subtournament arises from a single vertex by repeatedly adding two vertices that form a directed triangle $x\succ I\succ y\succ x$ with the previous selected set, so the number of selected vertices is odd. The class $\mD_1$ is not contained in the packing-equals-covering condition of Mnich, Shrestha, and Yang~\cite[Theorem~1]{MSY2015}, where packing counts vertex-disjoint directed triangles and covering is the minimum number of vertices meeting all directed triangles. To see this, let $C=\{a,b,c\}$ induce the triangle $a\succ b\succ c\succ a$, and add $u,v$ with $C\succ u\succ v\succ C$. The order $v,a,b,c,u$ has degree one. This five-vertex tournament has packing number one but covering number two: deleting either outer vertex leaves $C$, deleting any vertex of $C$ leaves an outer triangle through a surviving vertex of $C$, and deleting one outer and one inner vertex suffices. This separates the input conditions, without asserting that the example cannot be handled by composition consistency.

\subsection{Covered-vertex deletion}

The degree-two gadgets below are easier to analyze after deleting certain vertices. A vertex $x$ \emph{covers} $y$ if $x\succ y$ and every vertex beaten by $y$ is also beaten by $x$. Covered vertices are never selected~\cite{BFHM2010,BBH2016}. We need more than this exclusion: deleting a vertex can change the queries that determine the rest of TEQ. The following lemma shows that a covered vertex can be deleted without changing those answers.

\begin{lemma}[Covered-vertex deletion]\label{alg:coverdelete}
Let $T$ be a tournament in which $x$ covers $y$. Then $y\notin\TEQ(T)$ and $\TEQ(T-y)=\TEQ(T)$. Moreover, $R_{T-y}$ is the restriction of $R_T$ to $V\setminus\{y\}$.
\end{lemma}

\begin{proof}
Induct on $N$, proving both statements together. Let $z\ne y$. If $y\succ z$, then $x\succ z$ because $x$ covers $y$, so the query $D(z)$ contains both $x$ and $y$, and $x$ still covers $y$ inside it. By induction, $\TEQ(D(z))$ excludes $y$ and equals $\TEQ(D(z)\setminus\{y\})$. If $z\succ y$, the query $D(z)$ does not contain $y$ at all. Hence $y$ has no outgoing contestation edge, and the relation on the remaining vertices is unchanged. The query $D(y)$ contains $x$, so it has nonempty TEQ and $y$ has an incoming contestation edge. Thus $\{y\}$ is a non-source component, and deleting it changes no source component.
\end{proof}

\subsection{A positive-clause tournament with a degree-two certificate}

The reduction of \Cref{rest:d2hard} uses a different meaning for occurrence vertices than \Cref{sec:hardness}: each occurrence is present exactly when its literal is true under the chosen assignment. Thus a clause is satisfied precisely when at least one of its three occurrence vertices survives. We first construct a tournament whose output records this condition. We then combine these tests into a clause tournament while reserving enough indegree for the variable layers.

Define the six-vertex positive-clause tournament $G$ on $\cout ,\ax{1},\ax{2},\oc{1},\oc{2},\oc{3}$ by the out-neighbor lists
\[
\begin{array}{c|l}
 \cout &\oc{1},\oc{2},\oc{3}\\ \ax{1}&\cout \\ \ax{2}&\cout ,\ax{1}\\
 \oc{1}&\ax{1},\ax{2}\\ \oc{2}&\ax{1},\ax{2},\oc{1}\\ \oc{3}&\ax{1},\ax{2},\oc{1},\oc{2}.
\end{array}\tag{R1}\label{rest:g0}
\]
This is the clause gadget $\Gad_0$ of \Cref{def:clause}, written as out-neighbor lists; we call it $G$ here because every clause uses it, whatever the signs of its literals. The vertices $\oc{1},\oc{2},\oc{3}$ are optional and $\cout $ is the output. If no optional vertex remains, $\ax{2}$ is a Condorcet winner. Otherwise, let $\oc{*}$ be the last retained optional vertex in the order $\oc{1},\oc{2},\oc{3}$. It covers every other retained optional vertex, and $\ax{2}$ covers $\ax{1}$. Exact covered deletion (\Cref{alg:coverdelete}) leaves the triangle $\cout \succ \oc{*}\succ \ax{2}\succ \cout $. Consequently $\cout $ is selected exactly when at least one optional vertex is present.

The clause tournament must accept only when every clause is satisfied. To append a clause, we keep the old output and make its selection depend on the new clause. If an occurrence survives, the filter preserves the selected vertices of the old tournament. If none survives, it excludes them all. The guards $\relay ,\test $ produce these two behaviors through different source components of the contestation graph. Later assignment choices can delete occurrences from the old tournament as well, so the filter must continue to work on the surviving old tournament.

The filter from the PSPACE reduction already computes conjunction, but its elimination order does not leave the indegree budget needed here. Its relay would beat each optional occurrence vertex. Together with the clause output $\cout $, this would leave no room for the variable layer's additional dominator. We reverse those edges and check that the modified filter still computes conjunction.

For a nonempty old tournament $H$, take a disjoint copy of $G$, add $\relay ,\test $, and put $K=G-\cout $. In addition to the edges of $H$ and of~\eqref{rest:g0}, orient
\begin{equation}\label{rest:positivefilter}
 K\succ H\succ \cout ,\qquad H\succ \relay ,\qquad
 G\succ \test \succ H,\qquad \test \succ \relay ,\qquad
 \{\oc{1},\oc{2},\oc{3}\}\succ \relay \succ\{\cout ,\ax{1},\ax{2}\}.
\end{equation}
These rules specify every pair. Optional vertices may be absent.

\begin{lemma}\label{rest:positive}
Let $H$ be a nonempty tournament, let $G$ be a disjoint copy of the tournament of~\eqref{rest:g0} with any subset of its optional vertices $\oc{1},\oc{2},\oc{3}$ deleted, put $K=V(G)\setminus\{\cout \}$, and let $F$ be the tournament on $V(H)\cup V(G)\cup\{\relay ,\test \}$, for two new vertices $\relay ,\test $, obtained by keeping the edges of $H$ and $G$ and orienting~\eqref{rest:positivefilter}. Then
\[
\TEQ(F)\cap V(H)=
\begin{cases}
\TEQ(H),&\text{if some optional vertex is present},\\
\varnothing,&\text{if no optional vertex is present}.
\end{cases}
\]
Both conclusions remain valid after arbitrary deletions in $H$ that leave $H$ nonempty.
\end{lemma}

\begin{proof}
The rules refer only to the surviving core, so it suffices to prove the unrestricted statement. Suppose first that no optional vertex remains. The queries at $\ax{2},\relay ,\test $ are $D(\ax{2})=\{\relay \}$, $D(\relay )=H\cup\{\test \}$ with Condorcet winner $\test $, and $D(\test )=\{\cout ,\ax{1},\ax{2}\}$ with Condorcet winner $\ax{2}$, so they have answers $\relay ,\test ,\ax{2}$. Thus $\{\ax{2},\relay ,\test \}$ is a closed source cycle. It reaches every old vertex $h$, because $\ax{2}$ is the Condorcet winner of $D(h)$, and it reaches $\ax{1}$, because $\relay $ is the Condorcet winner of $D(\ax{1})=\{\ax{2},\relay \}$. Finally $D(\cout )=H\cup\{\ax{1},\ax{2},\relay \}$. Here $\ax{2}$ covers $\ax{1}$, and deleting $\ax{1}$ leaves $\ax{2}\succ H\succ \relay \succ \ax{2}$, so $\ax{2},\relay \in\TEQ(D(\cout ))$. The source cycle therefore reaches $\cout $ as well and is the unique source component. It contains no old vertex.

Now suppose some optional vertex remains, and let $\oc{*}$ be the last one. It covers every other optional vertex in the entire construction, and $\ax{2}$ covers $\ax{1}$. Delete these covered vertices. The remaining right-hand core is $\cout \succ \oc{*}\succ \ax{2}\succ \cout $, with queries
\[
D(\oc{*})=\{\cout \},\quad D(\ax{2})=\{\oc{*},\relay \},\quad
D(\relay )=H\cup\{\test ,\oc{*}\},\quad D(\test )=\{\cout ,\ax{2},\oc{*}\}
\]
and answers $\{\cout \},\{\oc{*}\},\{\oc{*}\},\{\cout ,\ax{2},\oc{*}\}$. For every $h\in H$, the vertex $\oc{*}$ is the Condorcet winner of $D(h)$. The remaining query is $D(\cout )=H\cup\{\ax{2},\relay \}$, where $\ax{2}\succ H\succ \relay \succ \ax{2}$, so $\TEQ(D(\cout ))=\TEQ(H)\cup\{\ax{2},\relay \}$. Hence $\cout $ reaches all vertices through $\oc{*}$. An old vertex has an outgoing contestation edge exactly when it lies in $\TEQ(H)$, in which case it points to $\cout $. By \Cref{lem:return}, exactly these old vertices lie in the unique source component.
\end{proof}

The clause tournament must also leave room for the variable layers in a degree-two elimination order. Once the layers are partly removed, each occurrence vertex can still have one incoming edge from its own variable layer. The other surviving layer vertices do not dominate it. We therefore require an order ending at the output $d$ in which each occurrence has at most one remaining internal dominator and every other vertex has at most two. The output is removed last, when it has no dominator left. Equivalently, give each optional occurrence an allowance of one external dominator and give every other vertex an allowance of zero. The remaining internal indegree plus the allowance must be at most two at every step. These allowances count only possible additional dominators, and the order restricts to every context in which optional vertices are deleted.

The order $\oc{3},\oc{2},\oc{1},\ax{2},\ax{1},\cout $ has this property for $G$ with output $\cout $: each optional vertex has one internal dominator when removed, and the last three steps have none. Given an old order ending at $d$, the positive filter preserves the property with the order
\begin{equation}\label{rest:matrixorder}
 \oc{3},\oc{2},\oc{1},\ax{2},\ax{1},\test ,\quad
 (\text{old order without }d),\quad \relay ,\cout ,d.
\end{equation}
Each optional vertex has only $\cout $ as a remaining internal dominator. The next three vertices $\ax{2},\ax{1},\test $ have only $\relay $, $\relay $, and $\cout $, respectively. Neither residual vertex $\relay $ nor $\cout $ beats an old vertex. After the old non-output vertices are removed, $\relay $ and then $\cout $ have only $d$ as a dominator, and the output is last.

\begin{proposition}\label{rest:matrix}
For every $m\ge1$ there is a tournament $P$ on $8m-2$ vertices, constructible in polynomial time without any TEQ queries, with a vertex $d$ and pairwise disjoint three-element sets $O_1,\ldots,O_m\subseteq V(P)\setminus\{d\}$ of occurrence vertices, such that for every set $Z$ of occurrence vertices, $d\in\TEQ(P-Z)$ if and only if $O_j\not\subseteq Z$ for every $j$. Moreover, $P$ has an elimination order ending at $d$ in which, at its removal, every occurrence vertex has at most one remaining dominator, $d$ has none, and every other vertex has at most two.
\end{proposition}

\begin{proof}
Use $G$ for the first clause and append the others by~\eqref{rest:positivefilter}. \Cref{rest:positive} proves the conjunction by induction. The vertex count is $6+8(m-1)$, and~\eqref{rest:matrixorder} gives the certificate inductively.
\end{proof}

\subsection{NP-hardness on \texorpdfstring{$\mD_2$}{D2}}

\begin{proof}[Proof of \Cref{rest:d2hard}]
If the input has an empty clause, it is unsatisfiable. Return a transitive tournament on $6n+8m-2$ vertices, designate its last vertex $d$, and use the transitive order. The designated vertex is not selected, and the order has degree zero and ends at $d$, and it allows two additional surviving dominators at $d$ without changing the other removal steps.

Otherwise every clause is nonempty. Give each literal occurrence its own optional vertex and build the clause tournament of \Cref{rest:matrix}. Short clauses are padded by repeated literals, represented by distinct vertices. Write $M_i^+$ and $M_i^-$ for the positive and negative occurrence sets of $x_i$. All these sets are pairwise disjoint, and some may be empty.

For the current tournament $H$ with output $\dout$, put $\Hs{1}=V(H)\setminus M_i^-$ and $\Hs{0}=V(H)\setminus M_i^+$. Both sets contain $\dout$ and all fixed old guards. Apply the $\exists$-layer $\Wex(H,\dout;\Hs{1},\Hs{0})$ in the general form of \Cref{sec:hard-tools}. By \Cref{lem:or}, its unchanged output $\dout$ is selected exactly when it is selected in $\Hs{1}$ or in $\Hs{0}$, the branches $x_i=1$ and $x_i=0$. Build the layers from the last variable to the first. An outer variable deletes occurrences inside every inner layer already built, but disjointness of the occurrence sets preserves the inner output and every fixed guard, so by \Cref{obs:deletion}(b) the deletion leaves the same layer around the smaller tournament, with both sets restricted. Induction as in \Cref{lem:invariant} shows that the final output is selected exactly when the formula is satisfiable. An empty sign class causes no difficulty, since the corresponding set is then all of $V(H)$. Each variable adds six vertices.

It remains to find an elimination order for the full construction. If guards from many layers remain while we remove a vertex of the clause tournament, their incoming edges could exceed the degree budget. We first remove enough guards that each occurrence vertex receives only the contribution of its own variable, and the other vertices of the clause tournament receive none. Process the layers from the outermost to the innermost, removing $\shw,\tri,\broot$ in each. Their remaining dominator sets are $\{\broot,\shz\}$, $\{\rzero,\shz\}$, and $\{\rone,\rzero\}$. Residual outer guards lose to all fixed guards of an inner layer and contribute no dominators to these steps. Afterwards, a residual $\shz$ loses to every old vertex, the reader $\rone$ of $x_i$ beats exactly $M_i^-$ among the vertices of the clause tournament, and the reader $\rzero$ of $x_i$ beats exactly $M_i^+$. Thus each occurrence vertex has exactly one external dominator and every other vertex of the clause tournament has none. Now apply the weighted order of the clause tournament, except for its output $d$.

Finally, process the residual triples $\shz,\rone,\rzero$ from the innermost to the outermost. Every residual inner guard beats every residual outer guard, because all fixed guards lie in both masks. A triple has internal order $\shz\succ\rone\succ\rzero$ with $\shz\succ\rzero$, and its only remaining old vertex $\dout$ beats the entire triple. Removing $\shz,\rone,\rzero$ therefore takes indegree one at each step. Remove $\dout$ last, when its indegree is zero. All steps, the tournament, and the order are computable in polynomial time.
\end{proof}

\subsection{Two-crossing preference profiles}\label{app:consequences}

\begin{proof}[Proof of \Cref{cor:two-crossing}]
Constantinescu and Wattenhofer~\cite[Theorem~7]{CW2022} realize every weighted tournament with same-parity integer margins as the majority tournament of a two-crossing profile. If the largest absolute margin is $W$, the construction uses $O(N^2W)$ voters and runs in time polynomial in the size of this explicit output.

Apply the reduction of \Cref{thm:structured-hardness}. Only its orientation matters, so assign margin $1$ to every winning edge and $-1$ to its reverse. This satisfies the same-parity hypothesis with $W=1$. The realization gives $O(N^2)$ voters, each with a ballot of length $N$, in polynomial time, preserves the majority orientation exactly, and supplies a two-crossing order. TEQ depends only on the orientation, so the designated answer is preserved. Conversely, compute the majority tournament of the profile and apply \Cref{prop:pspace}.
\end{proof}

This is a direct consequence of the realization theorem, whose original complexity corollary established NP-hardness of TEQ on the same domain~\cite[Corollary~8]{CW2022}. Since the number of voters grows with $N$, the corollary says nothing about profiles with a constant number of voters, such as the seven-voter setting of~\cite[Theorem~3]{BBGHKPS2019}.

\clearpage
\bibliographystyle{alpha}
\bibliography{references}

@inproceedings{BFHM2008,
  author = {Felix Brandt and Felix Fischer and Paul Harrenstein and Maximilian Mair},
  title = {A Computational Analysis of the Tournament Equilibrium Set},
  booktitle = {Proceedings of the Twenty-Third AAAI Conference on Artificial Intelligence},
  pages = {38--43},
  year = {2008},
  publisher = {AAAI Press},
  url = {https://aaai.org/Papers/AAAI/2008/AAAI08-006.pdf}
}

@article{Schwartz1990,
  author = {Thomas Schwartz},
  title = {Cyclic tournaments and cooperative majority voting: A solution},
  journal = {Social Choice and Welfare},
  volume = {7},
  number = {1},
  pages = {19--29},
  year = {1990},
  doi = {10.1007/BF01832917},
  note = {\url{https://doi.org/10.1007/BF01832917}}
}

@article{BFHM2010,
  author = {Felix Brandt and Felix Fischer and Paul Harrenstein and Maximilian Mair},
  title = {A computational analysis of the tournament equilibrium set},
  journal = {Social Choice and Welfare},
  volume = {34},
  number = {4},
  pages = {597--609},
  year = {2010},
  doi = {10.1007/s00355-009-0419-z},
  note = {\url{https://doi.org/10.1007/s00355-009-0419-z}}
}

@incollection{BBH2016,
  author = {Felix Brandt and Markus Brill and Paul Harrenstein},
  title = {Tournament solutions},
  booktitle = {Handbook of Computational Social Choice},
  editor = {Felix Brandt and Vincent Conitzer and Ulle Endriss and J{\'e}r{\^o}me Lang and Ariel D. Procaccia},
  publisher = {Cambridge University Press},
  chapter = {3},
  pages = {57--84},
  year = {2016},
  doi = {10.1017/CBO9781107446984.004},
  note = {\url{https://doi.org/10.1017/CBO9781107446984.004}}
}

@inproceedings{BBS2011,
  author = {Felix Brandt and Markus Brill and Hans Georg Seedig},
  title = {On the fixed-parameter tractability of composition-consistent tournament solutions},
  booktitle = {Proceedings of the Twenty-Second International Joint Conference on Artificial Intelligence (IJCAI)},
  pages = {85--90},
  year = {2011},
  doi = {10.5591/978-1-57735-516-8/IJCAI11-026},
  url = {https://www.ijcai.org/Proceedings/11/Papers/026.pdf},
  note = {\url{https://doi.org/10.5591/978-1-57735-516-8/IJCAI11-026}}
}

@inproceedings{MSY2015,
  author = {Matthias Mnich and Yash Raj Shrestha and Yongjie Yang},
  title = {When does {Schwartz} conjecture hold?},
  booktitle = {Proceedings of the Twenty-Fourth International Joint Conference on Artificial Intelligence (IJCAI)},
  pages = {603--609},
  year = {2015},
  url = {https://www.ijcai.org/Proceedings/15/Papers/091.pdf},
  note = {\url{https://www.ijcai.org/Proceedings/15/Papers/091.pdf}}
}

@misc{Yang2016,
  author = {Yongjie Yang},
  title = {A further step towards an understanding of the tournament equilibrium set},
  year = {2016},
  eprint = {1611.03991},
  archivePrefix = {arXiv},
  primaryClass = {math.CO},
  doi = {10.48550/arXiv.1611.03991},
  note = {arXiv:1611.03991v1. \url{https://doi.org/10.48550/arXiv.1611.03991}},
  url = {https://arxiv.org/abs/1611.03991v1}
}

@article{LLL1993,
  author = {Gilbert Laffond and Jean-Fran{\c c}ois Laslier and Michel {Le Breton}},
  title = {More on the tournament equilibrium set},
  journal = {Math{\'e}matiques informatique et sciences humaines},
  volume = {123},
  pages = {37--44},
  year = {1993},
  url = {https://www.numdam.org/item/MSH_1993__123__37_0/},
  note = {\url{https://www.numdam.org/item/MSH_1993__123__37_0/}}
}

@article{LLL1996,
  author = {Gilbert Laffond and Jean Lain{\'e} and Jean-Fran{\c c}ois Laslier},
  title = {Composition-consistent tournament solutions and social choice functions},
  journal = {Social Choice and Welfare},
  volume = {13},
  number = {1},
  pages = {75--93},
  year = {1996},
  doi = {10.1007/BF00179100},
  note = {\url{https://doi.org/10.1007/BF00179100}}
}

@article{BCKLNSST2013,
  author = {Felix Brandt and Maria Chudnovsky and Ilhee Kim and Gaku Liu and Sergey Norin and Alex Scott and Paul Seymour and Stephan Thomass{\'e}},
  title = {A counterexample to a conjecture of {Schwartz}},
  journal = {Social Choice and Welfare},
  volume = {40},
  number = {3},
  pages = {739--743},
  year = {2013},
  doi = {10.1007/s00355-011-0638-y},
  note = {\url{https://doi.org/10.1007/s00355-011-0638-y}}
}

@misc{BS2013,
  author = {Felix Brandt and Hans Georg Seedig},
  title = {A tournament of order 24 with two disjoint {TEQ}-retentive sets},
  year = {2013},
  eprint = {1302.5592},
  archivePrefix = {arXiv},
  primaryClass = {cs.MA},
  doi = {10.48550/arXiv.1302.5592},
  note = {arXiv:1302.5592v2. \url{https://doi.org/10.48550/arXiv.1302.5592}},
  url = {https://arxiv.org/abs/1302.5592v2}
}

@article{BBSS2018,
  author = {Felix Brandt and Markus Brill and Hans Georg Seedig and Warut Suksompong},
  title = {On the structure of stable tournament solutions},
  journal = {Economic Theory},
  volume = {65},
  number = {2},
  pages = {483--507},
  year = {2018},
  doi = {10.1007/s00199-016-1024-x},
  note = {\url{https://doi.org/10.1007/s00199-016-1024-x}}
}

@article{BBFH2014,
  author = {Felix Brandt and Markus Brill and Felix Fischer and Paul Harrenstein},
  title = {Minimal retentive sets in tournaments},
  journal = {Social Choice and Welfare},
  volume = {42},
  number = {3},
  pages = {551--574},
  year = {2014},
  doi = {10.1007/s00355-013-0740-4},
  note = {\url{https://doi.org/10.1007/s00355-013-0740-4}}
}

@article{BBGHKPS2019,
  author = {Georg Bachmeier and Felix Brandt and Christian Geist and Paul Harrenstein and Keyvan Kardel and Dominik Peters and Hans Georg Seedig},
  title = {{$k$}-majority digraphs and the hardness of voting with a constant number of voters},
  journal = {Journal of Computer and System Sciences},
  volume = {105},
  pages = {130--157},
  year = {2019},
  doi = {10.1016/j.jcss.2019.04.005},
  note = {\url{https://doi.org/10.1016/j.jcss.2019.04.005}}
}

@inproceedings{Suksompong2021,
  author = {Warut Suksompong},
  title = {Tournaments in computational social choice: Recent developments},
  booktitle = {Proceedings of the Thirtieth International Joint Conference on Artificial Intelligence (IJCAI)},
  pages = {4611--4618},
  year = {2021},
  doi = {10.24963/ijcai.2021/626},
  url = {https://www.ijcai.org/proceedings/2021/626},
  note = {\url{https://doi.org/10.24963/ijcai.2021/626}}
}

@inproceedings{CW2022,
  author = {Andrei Constantinescu and Roger Wattenhofer},
  title = {Voting in two-crossing elections},
  booktitle = {Proceedings of the Thirty-First International Joint Conference on Artificial Intelligence (IJCAI)},
  pages = {208--214},
  year = {2022},
  doi = {10.24963/ijcai.2022/30},
  url = {https://www.ijcai.org/proceedings/2022/30},
  note = {\url{https://doi.org/10.24963/ijcai.2022/30}}
}

@misc{SV2026,
  author = {Ethan Dickey and Alexandros Psomas and Athina Terzoglou},
  title = {{Stable Voting} is {PSPACE}-complete},
  year = {2026},
  eprint = {2607.14366},
  archivePrefix = {arXiv},
  primaryClass = {cs.GT},
  doi = {10.48550/arXiv.2607.14366},
  note = {arXiv:2607.14366v1. \url{https://doi.org/10.48550/arXiv.2607.14366}},
  url = {https://arxiv.org/abs/2607.14366v1}
}

@inproceedings{SM1973,
  author = {Larry J. Stockmeyer and Albert R. Meyer},
  title = {Word problems requiring exponential time: Preliminary report},
  booktitle = {Proceedings of the Fifth Annual ACM Symposium on Theory of Computing (STOC)},
  pages = {1--9},
  year = {1973},
  publisher = {ACM},
  doi = {10.1145/800125.804029},
  note = {\url{https://doi.org/10.1145/800125.804029}}
}

@book{Laslier1997,
  author = {Jean-Fran{\c c}ois Laslier},
  title = {Tournament Solutions and Majority Voting},
  series = {Studies in Economic Theory},
  volume = {7},
  publisher = {Springer},
  address = {Berlin, Heidelberg},
  year = {1997}
}

@article{Woeginger2003,
  author = {Gerhard J. Woeginger},
  title = {Banks winners in tournaments are difficult to recognize},
  journal = {Social Choice and Welfare},
  volume = {20},
  number = {3},
  pages = {523--528},
  year = {2003},
  doi = {10.1007/s003550200197},
  note = {\url{https://doi.org/10.1007/s003550200197}}
}

@article{Hudry2004,
  author = {Olivier Hudry},
  title = {A note on ``{Banks} winners in tournaments are difficult to recognize'' by {G. J. Woeginger}},
  journal = {Social Choice and Welfare},
  volume = {23},
  number = {1},
  pages = {113--114},
  year = {2004},
  doi = {10.1007/s00355-003-0241-y},
  note = {\url{https://doi.org/10.1007/s00355-003-0241-y}}
}

@article{Hudry2009,
  author = {Olivier Hudry},
  title = {A survey on the complexity of tournament solutions},
  journal = {Mathematical Social Sciences},
  volume = {57},
  number = {3},
  pages = {292--303},
  year = {2009}
}

@inproceedings{Yan2022,
  author = {Xinyi Yan and Chengxi Luo and Charles L. A. Clarke and Nick Craswell and Ellen M. Voorhees and Pablo Castells},
  title = {Human Preferences as Dueling Bandits},
  booktitle = {Proceedings of the 45th International ACM SIGIR Conference on Research and Development in Information Retrieval},
  pages = {567--577},
  year = {2022},
  doi = {10.1145/3477495.3531991},
  url = {https://doi.org/10.1145/3477495.3531991}
}

@inproceedings{Ramamohan2016,
  author = {Siddartha Ramamohan and Arun Rajkumar and Shivani Agarwal},
  title = {Dueling Bandits: Beyond {Condorcet} Winners to General Tournament Solutions},
  booktitle = {Advances in Neural Information Processing Systems 29},
  pages = {1253--1261},
  year = {2016},
  url = {https://papers.nips.cc/paper/2016/hash/fccb3cdc9acc14a6e70a12f74560c026-Abstract.html}
}

@article{Fishburn1977,
  author = {Peter C. Fishburn},
  title = {Condorcet social choice functions},
  journal = {SIAM Journal on Applied Mathematics},
  volume = {33},
  number = {3},
  pages = {469--489},
  year = {1977}
}

@article{LLL1993b,
  author = {Gilbert Laffond and Jean-Fran{\c c}ois Laslier and Michel {Le Breton}},
  title = {The bipartisan set of a tournament game},
  journal = {Games and Economic Behavior},
  volume = {5},
  number = {1},
  pages = {182--201},
  year = {1993}
}

@article{BF2008,
  author = {Felix Brandt and Felix Fischer},
  title = {Computing the minimal covering set},
  journal = {Mathematical Social Sciences},
  volume = {56},
  number = {2},
  pages = {254--268},
  year = {2008}
}

@article{Alon2006,
  author = {Noga Alon},
  title = {Ranking tournaments},
  journal = {SIAM Journal on Discrete Mathematics},
  volume = {20},
  number = {1},
  pages = {137--142},
  year = {2006}
}

@article{CTY2007,
  author = {Pierre Charbit and St{\'e}phan Thomass{\'e} and Anders Yeo},
  title = {The minimum feedback arc set problem is {NP}-hard for tournaments},
  journal = {Combinatorics, Probability and Computing},
  volume = {16},
  number = {1},
  pages = {1--4},
  year = {2007}
}

@inproceedings{KMS2007,
  author = {Claire Kenyon-Mathieu and Warren Schudy},
  title = {How to rank with few errors},
  booktitle = {Proceedings of the 39th Annual ACM Symposium on Theory of Computing (STOC)},
  pages = {95--103},
  year = {2007}
}

@inproceedings{VW2010,
  author = {Virginia {Vassilevska Williams}},
  title = {Fixing a tournament},
  booktitle = {Proceedings of the 24th AAAI Conference on Artificial Intelligence},
  pages = {895--900},
  year = {2010}
}

@article{YBKJ2012,
  author = {Yisong Yue and Josef Broder and Robert Kleinberg and Thorsten Joachims},
  title = {The {$K$}-armed dueling bandits problem},
  journal = {Journal of Computer and System Sciences},
  volume = {78},
  number = {5},
  pages = {1538--1556},
  year = {2012}
}

@inproceedings{ZKWR2015,
  author = {Masrour Zoghi and Zohar S. Karnin and Shimon Whiteson and Maarten de Rijke},
  title = {Copeland dueling bandits},
  booktitle = {Advances in Neural Information Processing Systems 28},
  pages = {307--315},
  year = {2015}
}

@inproceedings{DHSSZ2015,
  author = {Miroslav Dud{\'\i}k and Katja Hofmann and Robert E. Schapire and Aleksandrs Slivkins and Masrour Zoghi},
  title = {Contextual dueling bandits},
  booktitle = {Proceedings of the 28th Conference on Learning Theory (COLT)},
  series = {Proceedings of Machine Learning Research},
  volume = {40},
  pages = {563--587},
  year = {2015}
}

@inproceedings{Munos2024,
  author = {R{\'e}mi Munos and Michal Valko and Daniele Calandriello and Gheshlaghi Azar, Mohammad and Mark Rowland and Zhaohan Daniel Guo and Yunhao Tang and Matthieu Geist and Thomas Mesnard and C{\^o}me Fiegel and Andrea Michi and Marco Selvi and Sertan Girgin and Nikola Momchev and Olivier Bachem and Daniel J. Mankowitz and Doina Precup and Bilal Piot},
  title = {Nash learning from human feedback},
  booktitle = {Proceedings of the 41st International Conference on Machine Learning (ICML)},
  series = {Proceedings of Machine Learning Research},
  volume = {235},
  pages = {36743--36768},
  year = {2024}
}

@inproceedings{Lampis2022,
    author = {Michael Lampis},
    title = {Determining a {Slater} Winner Is Complete for Parallel Access to {NP}},
    booktitle = {39th International Symposium on Theoretical Aspects of Computer Science (STACS 2022)},
    series = {Leibniz International Proceedings in Informatics (LIPIcs)},
    volume = {219},
    pages = {45:1--45:14},
    year = {2022},
    editor = {Petra Berenbrink and Benjamin Monmege},
    publisher = {Schloss Dagstuhl -- Leibniz-Zentrum f{\"u}r Informatik},
    doi = {10.4230/LIPIcs.STACS.2022.45}
}

\clearpage
\section*{AI methodology statement}
\addcontentsline{toc}{section}{AI methodology statement}
We used AI to perform most of the research activities associated with this work. Our contribution is providing input as to the direction of the research, redirecting ideas unlikely to succeed, significant prose review and revision, direction of the flow of the paper, explicit direction in the generation of figures and intuition statements, and overall guidance of the tools. Our primary contributions are thus \textit{taste}, \textit{interpretation and translation of results}, and \textit{extracting what we, as a community, can actually learn from these results}. One example of this is deciding which results \textit{not} to include in the manuscript. Future versions of this article will include those omitted results as supplementary material.

OpenAI's ChatGPT (Astra 6.0, Terra 6.1) was used mainly to develop the reduction; in our experience it is stronger at finding reductions, which is why we used it for that. It worked under our guidance and from the ideas of our earlier result on Stable Voting~\cite{SV2026}, which suggested the reduction strategy. Anthropic's Claude (Fable 5.1, Fable 5.2 and Opus 5.5) was used mainly to check proofs and to help with the writing and the figures; in our experience, it produced the more meaningful and intuitive diagrams, which is why we used it for them. Both were also used to write verification programs and to search the literature. Computational checks served as supporting evidence only.

We have checked the construction and the proofs. Responsibility for the correctness, originality and attribution of the work rests with the authors.

\end{document}